\documentclass[11pt,a4paper]{article}

\usepackage[a4paper,margin=2.5cm]{geometry}

\usepackage[ruled,linesnumbered]{algorithm2e}
\usepackage{booktabs}
\usepackage{makecell}
\usepackage{multirow}
\usepackage{algorithmic}

\usepackage{ltablex} 
\usepackage[referable]{threeparttablex}
\usepackage[commandnameprefix=always]{changes}
\usepackage{graphicx}
\usepackage{amsmath,amssymb}
\usepackage{xcolor}
\usepackage{array}
\usepackage{calc}
\usepackage{float}
\usepackage{fancyvrb}
\usepackage[numbers,sort&compress]{natbib}
\providecommand{\real}[1]{#1}

\providecommand{\tightlist}{\setlength{\itemsep}{0pt}\setlength{\parskip}{0pt}}
\usepackage{hyperref}   

\begin{document}

\title{VISA: A Structured Description Protocol for Agent-Based Simulation Models Towards Machine Reproducibility}

\author{Zhou He\\[2pt]\small\texttt{hezhou@ucas.ac.cn}\\University of Chinese Academy of Sciences, Beijing, China}
\date{}

\maketitle

\begin{abstract}
Agent-based models (ABMs) are difficult to reproduce: their behavior is spread across prose narratives, platform-specific code, and implicit assumptions, so that two readers routinely reconstruct different models from the same documentation. We present VISA, a structured, symbol-based description protocol that specifies a model in eight interconnected tables---four at the agent level (Agent, Variable, Sensing, Internal Function) and four at the model level (Associated Data, Input/Output, Schedule, Validation)---under the principle of minimality with completeness. VISA makes a model machine-parseable and unambiguous via two artifacts: nineteen executable consistency rules that turn model validity into a checkable property, and three reusable LLM-executable skills (authoring, checking, and code generation) that operationalize the full author--check--code--reproduce loop. We validate the protocol on three external, independently authored ABMs spanning three platforms: we reproduce two cross-language (NetLogo to Python) directly from their VISA specifications, and we capture a third, an industrial AnyLogic model, in eight tables (passing all nineteen rules) while honestly demarcating where reproduction is blocked by a proprietary movement library and unavailable data---itself a transparency contribution. VISA moves the reproduction barrier from the model, where it is invisible, to a named, localized dependency, where it is actionable.
\end{abstract}


\raggedbottom  

\section{Introduction}

Agent-based modeling (ABM) has emerged as a powerful methodology for studying complex systems across diverse domains, including ecology~\cite{Grimm2005,Railsback2001}, economics~\cite{Dosi2006,Caiani2016}, epidemiology~\cite{Epstein2009,Tracy2018}, urban planning~\cite{Batty2007,Filatova2015}, supply chain management~\cite{Bonzelet2022,Tian2022}, and public policy~\cite{Mahmood2020,Guan2018}.
Unlike equation-based approaches, ABM enables the representation of heterogeneous agents, their adaptive behaviors, and the emergent macro-level phenomena arising from micro-level interactions~\cite{Railsback2001,Macal2014}.
This bottom-up paradigm has proven particularly valuable in complex management systems, where organizational actors exhibit diverse decision-making strategies and interact within institutional constraints~\cite{North2007,Siebers2010}.

Despite the widespread adoption of ABM, the field continues to face a fundamental reproducibility challenge~\cite{Fanelli2018,OpenScience2015,Squazzoni2020}.
A model that cannot be unambiguously understood and independently reproduced undermines the scientific credibility of simulation-based research~\cite{Axtell1996,Edmonds2003}.
This challenge is especially acute for ABM because, unlike analytical models, agent-based models involve numerous design choices regarding agent behaviors, interaction topologies, scheduling mechanisms, and parameter configurations that must be precisely communicated for faithful replication~\cite{Donkin2017,Zhang2021replication}.
The difficulty is further compounded by the diversity of modeling platforms---NetLogo~\cite{Wilensky1999}, Repast~\cite{North2007}, Mesa, and others---each with distinct programming paradigms and abstractions~\cite{Daly2022}.

Recognizing this challenge, the simulation community has developed several documentation protocols, reporting guidelines, and model-sharing platforms.
The \textbf{ODD} (Overview, Design concepts, Details) protocol, first proposed by Grimm et al.~\cite{Grimm2006} and subsequently updated~\cite{Grimm2010,Grimm2020}, has become the de facto standard for describing agent-based models.
ODD organizes model descriptions into seven elements across three blocks, providing a structured narrative that guides authors through model purpose, state variables, process scheduling, and submodels.
Several extensions have been proposed to address specific modeling aspects: \textbf{ODD+D} adds a decision-making module~\cite{Mueller2013}, \textbf{ODD+2D} incorporates data description for empirical ABMs~\cite{Laatabi2018}, the \textbf{Visual ODD} approach~\cite{Szangolies2024} aims to improve the visual communication of model structures, and \textbf{vODD-DD} provides visual documentation for data-driven models~\cite{Lee2026}.
Beyond the ODD family, the \textbf{MR POTATOHEAD} framework offers a conceptual design pattern for structuring ABM components~\cite{Parker2006,Parker2008}, the \textbf{TRACE} protocol focuses on model evaluation and testing~\cite{Grimm2014}, and the \textbf{STRESS} guidelines address empirical simulation reporting~\cite{Monks2019}.
UML-based representations~\cite{Bersini2012} offer an alternative visual formalism for specifying model structures.
On the infrastructure side, model repositories such as the \textbf{CoMSES} Computational Model Library (formerly OpenABM) have been established to host and share agent-based models, promoting code-level reproducibility and systematic archiving practices~\cite{Janssen2017}.
Despite these valuable efforts, the fundamental question of \emph{what information must be documented} for a model to be faithfully reproduced remains only partially addressed.

While these protocols have significantly advanced the state of model documentation, several fundamental limitations persist.
First, existing protocols are predominantly \emph{text-centric}---they rely on natural language narratives to describe model components, which inevitably introduce ambiguity and interpretation variability.
As noted by Grimm et al.~\cite{Grimm2020}, even with the ODD protocol, readers often struggle to extract precise specifications sufficient for independent model reimplementation.
Second, most protocols are not \emph{machine-readable}---they lack the structured, symbol-based representation that enables automated parsing and code generation.
This limitation becomes increasingly critical as large language models (LLMs) offer new opportunities for automated model construction and replication~\cite{Kazieva2026}.
Third, existing frameworks exhibit \emph{redundancy and overlap} among their descriptive elements, which can lead to inconsistent specifications and increased documentation burden.
Fourth, there is a disconnect between model descriptions and their \emph{computational implementation}---the gap between a narrative description and executable code remains wide, requiring substantial manual effort and domain expertise to bridge~\cite{Mueller2014}.

To address these limitations, we propose \textbf{VISA}, a table-based, symbol-driven description protocol for agent-based simulation models.
The name reflects the paired structure of the protocol across two levels: the \emph{Agent level} comprises four tables---Variable, Sensing, Internal Function, and Agent---that capture the micro-level structure and behavior of individual agents; the \emph{Model level} comprises another four tables---Input/Output, Schedule, Associated Data, and Validation---that describe the macro-level configuration, data flow, temporal dynamics, and credibility assessment of the simulation.
Each table employs mathematical notation, explicit cross-references, and standardized schemas that are both human-interpretable and machine-parseable.

At its core, this work addresses a fundamental scientific question: \emph{what are the essential elements of an agent-based model, and what information is indispensable for reproducing its simulation results?}
Existing protocols such as ODD cover a broad range of elements, yet for many classic models (e.g., Schelling's segregation model), numerous ODD items are either absent or redundant, suggesting that not all elements carry equal weight for reproducibility.
In contrast, VISA is designed under the principle of \emph{minimality with completeness}---every element in the eight tables is indispensable, and together they constitute the necessary and sufficient conditions for a model to be unambiguously understood, faithfully reproduced, critically evaluated, and productively extended by subsequent researchers.

The main contributions of this paper are as follows:
\begin{itemize}
\item We propose the VISA protocol, a structured, table-based description standard comprising eight interconnected tables---four at the agent level and four at the model level---built on symbol-based notation and the principle of minimality with completeness.
\item We demonstrate VISA through a running example based on a multi-newsvendor inventory management system, illustrating how each table captures essential model elements.
\item We define nineteen consistency rules that govern the relationships among the tables, turning model validity from a matter of judgment into a checkable property.
\item We operationalize the protocol through two artifacts: nineteen executable consistency rules and three reusable skills (authoring, checking, and code generation) that together advance transparency, reproducibility, and verifiability.
\item We validate the protocol empirically by running the full author--check--code loop on three external, independently authored ABMs spanning three platforms---reproducing two cross-language (NetLogo~$\to$~Python) and capturing the third (AnyLogic) as an expressiveness and honest-limit case.
\end{itemize}

The remainder of this paper is organized as follows.
Section~\ref{sec:related} reviews related work on ABM documentation protocols and reproducibility.
Section~\ref{sec:visa} presents the VISA protocol in detail, including its design principles, notation, the eight standardized tables illustrated through a multi-newsvendor example, and the nineteen consistency rules.
Section~\ref{sec:guidelines} operationalizes the protocol through its two artifacts---the nineteen consistency rules and the three skills---and states how evaluators and reproducers use them.
Section~\ref{sec:evaluation} empirically evaluates the protocol by reproducing three external agent-based models.
Section~\ref{sec:discussion} discusses visualization and the protocol's applicability to learned agents, together with limitations and future directions.
Section~\ref{sec:conclusion} concludes.

\section{Related Work}
\label{sec:related}

This section reviews the landscape of reproducibility research and documentation standards for agent-based modeling.
We organize the discussion along three dimensions: empirical evidence on the reproducibility challenge, existing documentation protocols and reporting standards, and efforts to bridge model descriptions with computational implementations.

\subsection{Reproducibility of Agent-Based Models}

The reproducibility crisis documented across scientific disciplines~\cite{Fanelli2018,OpenScience2015} poses particularly acute challenges for simulation-based research~\cite{Squazzoni2020}.
Unlike analytical models that can be fully specified by closed-form equations, agent-based models involve numerous interacting components---agent types, state variables, behavioral rules, interaction topologies, scheduling mechanisms, spatial representations, and stochastic elements---each of which must be precisely communicated for faithful replication~\cite{Railsback2001,Macal2014}, a task made harder by the stochastic, emergent character of agent-based dynamics, whose outcomes are intrinsically difficult to repeat~\cite{Li2017}.

Empirical replication studies have consistently demonstrated the severity of this challenge.
The seminal alignment study by Axtell et al.~\cite{Axtell1996} compared independent implementations of the same model and found that even when two implementations produced qualitatively similar outcomes, quantitative discrepancies frequently arose from subtle differences in scheduling order, numerical precision, and boundary condition handling.
Edmonds and Hales~\cite{Edmonds2003} reinforced these findings, concluding that successful alignment required far more implementation-level detail than was typically communicated in publications.
Donkin et al.~\cite{Donkin2017} attempted to replicate complex agent-based models following structured guidelines and characterized the task as ``formidable,'' identifying incomplete behavioral specifications and ambiguous interaction rules as primary sources of replication failure.
More recently, Zhang and Robinson~\cite{Zhang2021replication} proposed a formal replication standard and applied it to an existing model, finding that even with explicit guidelines, key information about scheduling, stochasticity handling, and agent activation order was often missing from original descriptions.

The diversity of ABM platforms further complicates reproducibility.
Daly et al.~\cite{Daly2022} and Abar et al.~\cite{Abar2017} surveyed agent-based modeling tools and documented substantial heterogeneity in their abstractions for scheduling, spatial representation, agent interaction, and data management.
Models described using the conceptual vocabulary of one platform (e.g., NetLogo's \texttt{ask} semantics~\cite{Wilensky1999}) may be difficult to translate into another (e.g., Repast's discrete-event scheduling~\cite{North2007}) without losing behavioral fidelity.
Mueller et al.~\cite{Mueller2014} identified this platform-dependence as a key barrier to standardized model documentation.

Analyses of publication practices have quantified the extent of the documentation gap.
Angus and Hassani-Mahmooei~\cite{Angus2015} surveyed agent-based modeling publications in JASSS and found that a significant proportion lacked sufficient information for independent model reimplementation, even when structured documentation guidelines were available.
These findings collectively underscore the need for documentation standards that are not only comprehensive in scope but also precise and unambiguous in specification.

\subsection{Documentation Protocols and Reporting Standards}

\subsubsection{The ODD Protocol and Its Extensions}

The ODD (Overview, Design concepts, Details) protocol, introduced by Grimm et al.~\cite{Grimm2006} and updated in 2010~\cite{Grimm2010} and 2020~\cite{Grimm2020}, has become the most widely adopted standard for describing agent-based models.
ODD organizes model descriptions into seven elements grouped into three blocks.
The \emph{Overview} block comprises Purpose, State Variables and Scales, and Process Overview and Scheduling.
The \emph{Design Concepts} block addresses theoretical considerations including emergence, adaptation, objectives, learning, prediction, sensing, interaction, stochasticity, and collectives.
The \emph{Details} block covers Initialization, Input Data, and Submodels.

The protocol has had substantial impact on documentation practices across ecology, social sciences, and economics~\cite{Polhill2008,Polhill2010,Vincenot2018}.
Its adoption has been credited with improving the completeness and comparability of model descriptions~\cite{Grimm2020}.
However, several limitations have been identified.
Mueller et al.~\cite{Mueller2014} observed that despite ODD's availability, many published descriptions remain incomplete, and the protocol's reliance on natural language narratives leaves considerable room for interpretation.
The 2020 update~\cite{Grimm2020} acknowledged that ``readers often struggle to extract precise specifications sufficient for independent model reimplementation,'' and introduced refinements to reduce ambiguity.
Furthermore, ODD's Design Concepts block contains elements (e.g., emergence, collectives) that are relevant primarily to certain theoretical perspectives and may be absent or redundant for many applied models.

Several extensions address specific aspects. ODD+D~\cite{Mueller2013} adds a module for human decision-making (decision algorithms, behavioral theories, learning); ODD+2D~\cite{Laatabi2018} adds a data-description module linking variables to empirical sources; Visual ODD~\cite{Szangolies2024} and vODD-DD~\cite{Lee2026} add standardized visualization for narratives and data flow; and Bersini~\cite{Bersini2012} advocates UML for representing ABM structures.

While these extensions address important gaps, their proliferation has introduced a new challenge: modelers must navigate multiple overlapping standards, and the combined documentation burden has grown without a corresponding improvement in reproducibility rates~\cite{Mueller2014}.

\subsubsection{Other Documentation and Evaluation Frameworks}

Several alternative frameworks address other parts of the modeling lifecycle. The MR POTATOHEAD pattern~\cite{Parker2006,Parker2008} decomposes ABMs into modular elements (agents, environment, rules, interactions) and aids interdisciplinary design~\cite{Innocenti2020}, but focuses on conceptual design rather than a complete replication-ready specification. The TRACE protocol~\cite{Grimm2014} and its ``evaludation'' extension~\cite{Augusiak2014} structure model evaluation and testing across the modeling cycle, ensuring credibility without specifying how a model itself is described. The STRESS guidelines~\cite{Monks2019} structure the reporting of simulation studies generally, and economics-focused proposals~\cite{Richiardi2006,Wolf2013,Groeneveld2017} call for standardized description of agent characteristics and interactions. None of these, however, fixes a machine-readable set of information elements sufficient to reproduce an ABM.

\subsubsection{Model Repositories and Infrastructure}

Complementing documentation protocols, infrastructure-level efforts have promoted code-level reproducibility through model sharing and archiving.
The CoMSES Computational Model Library (formerly OpenABM) provides a centralized repository for hosting agent-based models with standardized metadata~\cite{Janssen2017}.
Janssen's analysis of archiving practices revealed that while code availability improves reproducibility, source code alone is often insufficient for understanding a model's conceptual foundation---the same model implemented in different languages may differ in subtle but consequential ways.
Janssen et al.~\cite{Janssen2008} had earlier called for a community framework to support shared modeling practices.

Related standardization efforts include MIASE (Minimum Information About a Simulation Experiment)~\cite{Waltemath2011}, which defines the minimum information required to reproduce computational biology simulations, and the FAIR (Findable, Accessible, Interoperable, Reusable) principles~\cite{Wilkinson2016}, which provide general guidelines for scientific data stewardship.
While these initiatives address important aspects of reproducibility infrastructure, they operate at a different level of abstraction from model description protocols and do not specify the particular information elements required for an ABM to be faithfully reproduced.

\subsection{Bridging Description and Implementation}

A persistent and largely unresolved challenge in ABM research is the gap between model description and computational implementation.
Smaldino~\cite{Smaldino2020} analyzed the general process of translating verbal theories into formal models, identifying numerous implicit decisions and assumptions that arise during formalization.
For ABMs, Mueller et al.~\cite{Mueller2014} explicitly identified the disconnect between narrative descriptions and executable code as a major barrier to reproducibility.

Several approaches have attempted to narrow this gap.
Ontology-based methods aim to provide formal, machine-readable representations of model components.
Polhill and Gotts~\cite{Polhill2009} explored the use of ontologies for integrated human-natural system modeling, and Polhill~\cite{Polhill2015} developed a NetLogo extension for extracting OWL ontologies from agent-based models.
While conceptually promising, these approaches have seen limited adoption due to their technical complexity and the overhead they impose on modelers.

Literate programming~\cite{Knuth1984} offers an alternative approach by interleaving documentation and code.
However, literate programming has not been widely adopted in the ABM community, and its focus on documenting implementation rather than specifying conceptual models limits its applicability for inter-platform reproducibility.

The emergence of large language models (LLMs) has introduced both new opportunities and new urgency for structured model specifications.
Kazieva~\cite{Kazieva2026} investigated LLM-based replication of agent-based models and found that the quality and structure of the model description significantly affect reproduction success rates.
This finding suggests that structured, symbol-based specifications---which reduce ambiguity and provide clear cross-references---could substantially improve the feasibility of automated model reproduction.

\subsection{Summary}

The existing landscape of ABM documentation reveals significant progress alongside persistent gaps.
The ODD protocol and its extensions have established a culture of structured documentation, while infrastructure such as CoMSES has promoted code sharing.
However, three fundamental limitations remain.
First, existing protocols are predominantly text-centric, relying on natural language that introduces interpretation variability.
Second, the proliferation of extensions has created redundancy and overlap, increasing documentation burden without proportional gains in reproducibility.
Third, no existing protocol provides a complete, symbol-based specification that is simultaneously human-readable and machine-parseable, directly bridging conceptual description and computational implementation.
VISA addresses these gaps through an alternative paradigm: a table-based, symbol-driven protocol designed for completeness, minimality, and machine interpretability.

\section{The VISA Protocol}
\label{sec:visa}

VISA organizes the complete specification of an agent-based simulation model into eight interconnected tables across two levels.
The protocol is designed around three principles: \emph{completeness}---every element necessary for unambiguous model reproduction is captured; \emph{minimality}---no redundant or optional elements are retained; and \emph{machine interpretability}---symbol-based notation and explicit cross-references enable automated parsing and code generation.
To illustrate how each table is constructed, we use a \emph{multi-newsvendor inventory system} as a running example throughout this section.
In this model, multiple vendor agents compete by setting retail prices and ordering from a single wholesaler; a customer agent allocates demand via a multinomial logit rule; and an environment agent manages vendor entry and exit.

\subsection{Protocol Overview}
\label{sec:visa-overview}

\begin{figure}[htbp]
  \centering
  \includegraphics[width=\textwidth]{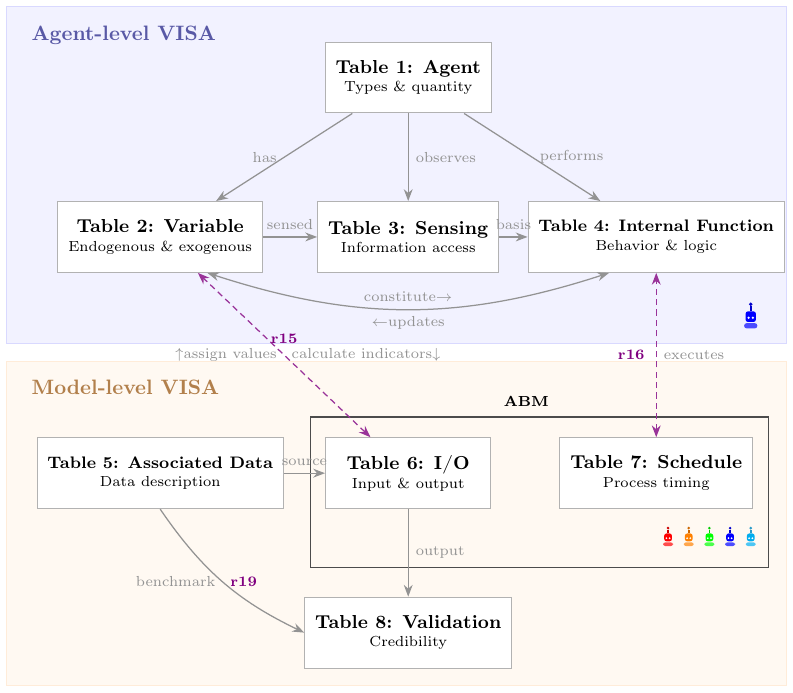}
  \caption{Structure of the VISA protocol. Eight tables are organized into Agent-level VISA (blue background) and Model-level VISA (orange background). Solid gray arrows indicate intra-level relationships; dashed violet arrows indicate cross-level connections, annotated with the consistency-rule IDs they realize (r15 input--output coverage, r16 schedule coverage, r19 data-reference resolution). The ABM box groups the core simulation components. The four within-table rules (r1--r4) and the full formal predicates are in Tables~\ref{tab:visa-rules-within}--\ref{tab:visa-rules-cross}.}
  \label{fig:visa-overview}
\end{figure}

The eight VISA tables are divided into two levels that mirror the fundamental architecture of agent-based models (Figure~\ref{fig:visa-overview}).

The \emph{Agent level} (Tables~\ref{tab:visa-t1}--\ref{tab:visa-t4}) specifies the micro-level structure and behavior of individual agents.
Table~\ref{tab:visa-t1} (Agent) enumerates all agent types and their roles.
Table~\ref{tab:visa-t2} (Variable) catalogs every state variable with its classification and provenance.
Table~\ref{tab:visa-t3} (Sensing) defines the information access structure---which agent observes which variables from which other agents.
Table~\ref{tab:visa-t4} (Internal Function) describes the behavioral rules that process information, make decisions, and update agent states.

The \emph{Model level} (Tables~\ref{tab:visa-t5}--\ref{tab:visa-t8}) describes the macro-level configuration and evaluation of the simulation.
Table~\ref{tab:visa-t5} (Associated Data) documents all external data sources.
Table~\ref{tab:visa-t6} (Input and Output) specifies model-level inputs and output indicators.
Table~\ref{tab:visa-t7} (Schedule) defines the temporal execution sequence.
Table~\ref{tab:visa-t8} (Validation) establishes credibility assessment criteria.

The tables are interconnected through explicit cross-references (Figure~\ref{fig:visa-overview}).
Within the Agent level, variables (Table~\ref{tab:visa-t2}) serve as both inputs to and outputs of functions (Table~\ref{tab:visa-t4}), forming a bidirectional dependency: functions constitute the logic that governs variables, while variables provide the state that functions read and update.
Sensing (Table~\ref{tab:visa-t3}) determines what information from other agents is available as the basis for functions.
Between the two levels, the Input section of Table~\ref{tab:visa-t6} assigns values to exogenous variables in Table~\ref{tab:visa-t2}, while its Output section computes indicators from endogenous variables.
Table~\ref{tab:visa-t7} (Schedule) orchestrates the execution of functions from Table~\ref{tab:visa-t4}.
Within the Model level, Table~\ref{tab:visa-t5} provides data for Table~\ref{tab:visa-t6} inputs and Table~\ref{tab:visa-t8} benchmarks.

\subsection{Notation Conventions}
\label{sec:visa-notation}

A defining feature of VISA is the systematic use of mathematical symbols to describe agents, sets, variables, and functions.
Existing protocols such as ODD rely primarily on natural-language narratives to communicate model structure, which inevitably introduces ambiguity and interpretation variability~\cite{Grimm2020}.
By adopting a formal notation system, VISA ensures that every model element is expressed unambiguously, facilitating both human understanding and machine parsing.
The conventions below draw on standard practices in management science and operations research, and are designed to map naturally to object-oriented programming constructs (e.g., class vs.\ instance).

\begin{itemize}
\item \textbf{Agent-type sets} are denoted by calligraphic uppercase letters (e.g., $\mathcal{V}$ for the set of all vendors, $\mathcal{E}$ for the environment). This corresponds to the concept of a \emph{class} in object-oriented programming.
\item \textbf{Agent instances} are denoted by lowercase italic letters with subscripts (e.g., $v_i$ for the $i$-th vendor instance, $e$ for the single environment instance). This corresponds to an \emph{object} instantiated from a class.
\item \textbf{Exogenous parameters and aggregate quantities} are denoted by uppercase italic letters (e.g., $N$ for population size, $D$ for demand, $T$ for time horizon). These are model-level inputs that remain constant or are drawn from specified distributions.
\item \textbf{Endogenous variables} are denoted by lowercase italic letters with subscripts and carry a time index $t$ (e.g., $q_{i,t}$ for the order quantity of vendor $v_i$ at step $t$, $x_{i,t}$ for an agent's position). These are agent-level state variables whose values are determined within the simulation and therefore evolve over time; the time index is written explicitly in Table~\ref{tab:visa-t2} (where each variable is defined) and may be suppressed in equations and running text for readability.
\item \textbf{Vectors and matrices} are denoted by boldface (e.g., $\mathbf{d}_i$ for the vector of historical demand observations of vendor $v_i$).
\item \textbf{Reserved indices:} The letters $i, j, k$ are reserved for indexing agent instances; $t$ for simulation time steps.
\item \textbf{Cardinality:} $N$ denotes a fixed, predetermined number of instances (e.g., $N_\mathcal{W} = 1$ wholesaler); $n$ denotes a variable number of instances that may change during simulation (e.g., $n_\mathcal{V}$ vendors, where vendors can enter or exit).
\end{itemize}

These conventions are applied consistently across all eight VISA tables and throughout the remainder of this paper.

\subsection{Agent-Level Description}
\label{sec:visa-agent}

\subsubsection{Table 1: Agent}
The Agent table establishes the model's ontology by enumerating all agent types, their functional roles, descriptions, and population sizes (Table~\ref{tab:visa-t1}).
The \emph{Name} column identifies each agent type; the \emph{Set} column assigns the corresponding calligraphic symbol; the \emph{Instances} column shows representative instance symbols.
The \emph{Category} field classifies each agent into one of four system roles:
\begin{description}
  \item[Environment] An agent with two duties: managing the lifecycle of other agent instances (creation, removal, ordering) and computing model-level statistics and output aggregates (e.g., a mean price, a Gini coefficient). Parameters and state that genuinely belong to a specific decision-maker are attributed to that decision-maker, not stored on the Environment.
  \item[Space] An \emph{active} agent that owns the spatial or topological \emph{geometry} of the world (dimensions, boundary conditions such as edge wrapping, a room layout) and maintains the spatial or topological relationships among the currently-alive agent instances---a grid that returns the agents within a vision range, or a network that exposes each agent's neighbors---and typically answers spatial-query functions for the other agents, rather than being a passive container.
  \item[Decision-maker] An agent that executes behavioral rules and makes autonomous choices based on its internal state and sensed information.
  \item[Passive] An agent that serves as a static information carrier without autonomous behavior or decision-making capability.
\end{description}
A defining trait of VISA is that the agent taxonomy reflects the \emph{model's own semantics}, not the host platform's object graph: only entities that behave autonomously become agents. Two things that platform code frequently exposes as separate classes are therefore \emph{not} agents---a simulation \emph{output datum} (e.g., a finalized contact record) is data produced by a function and is modelled as an Environment aggregate or a T6 output, and \emph{reference / lookup data} (a registry, a survey, a template library) is a dataset recorded in T5. Correspondingly, a parameter read by several decision-makers is attributed to each of those types rather than parked on the Environment.
The \emph{Quantity} column specifies the number of instances, using $N$ for fixed counts and $n$ for variable counts (agents that may be created or destroyed during simulation).

\begin{table}[htbp]
\caption{VISA Table 1: Agent types and roles (multi-newsvendor example)}
\label{tab:visa-t1}
\centering\small
\begin{tabular}{@{}llllp{4.5cm}c@{}}
\toprule
Name & Set & Instances & Category & Description & Quantity \\
\midrule
Environment & $\mathcal{E}$ & $e$ & Environment & Manages vendor entry/exit; computes the average price & $N_\mathcal{E} = 1$ \\
SensingNet & $\mathcal{S}$ & $s$ & Space & Maintains the vendor network topology; answers neighbor queries & $N_\mathcal{S} = 1$ \\
Wholesaler & $\mathcal{W}$ & $w$ & Passive & Provides wholesale price & $N_\mathcal{W} = 1$ \\
Vendor & $\mathcal{V}$ & $v_1, v_2, \ldots$ & Decision-maker & Decides order qty.\ and retail price under demand uncertainty & $n_\mathcal{V}$ \\
Customer & $\mathcal{C}$ & $c$ & Decision-maker & Allocates demand to vendors via multinomial logit & $N_\mathcal{C} = 1$ \\
\bottomrule
\end{tabular}
\end{table}

\subsubsection{Table 2: Variable}
\begin{figure}[htbp]
  \centering
  \includegraphics[width=0.75\textwidth]{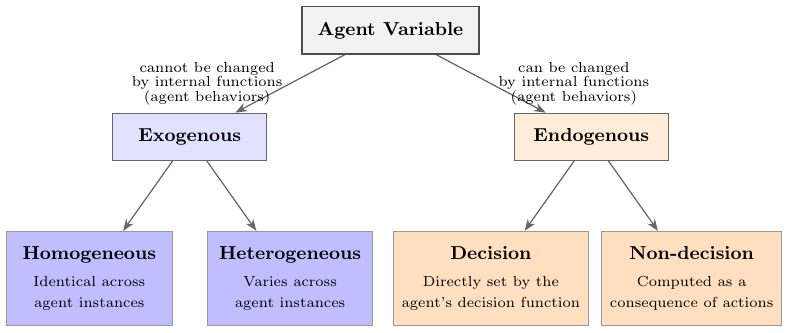}
  \caption{Classification of agent variables in VISA. Variables are first distinguished by whether their values can be changed by internal functions (agent behaviors): Exogenous variables cannot, while Endogenous variables can. Each category is further divided into two subtypes.}
  \label{fig:variable-types}
\end{figure}

The Variable table provides a complete specification of the model's state space by cataloging every state variable for every agent type (Table~\ref{tab:visa-t2}).
Each variable is classified by \emph{Type} along two dimensions (Figure~\ref{fig:variable-types}):
\begin{description}
  \item[Exogenous-homogeneous] A variable whose value is provided from outside the model and is identical across all agent instances.
  \item[Exogenous-heterogeneous] A variable whose value is provided from outside the model but varies across agent instances (e.g., drawn from a distribution).
  \item[Endogenous-decision] A variable whose value is directly set by the agent's internal decision function.
  \item[Endogenous (non-decision)] A variable whose value is determined within the model as a computational consequence of other variables or functions, but is not directly controlled by the agent's decisions.
\end{description}

The \emph{Symbol} column assigns a mathematical identifier using the notation conventions of Section~\ref{sec:visa-notation}; temporal dynamics are captured through subscripts ($t$, $t{-}1$).
The \emph{Data Type} column specifies the variable's data structure using programming-language notation (e.g., Integer, Float, Boolean, List[Integer]) to facilitate implementation.
The \emph{Value} column traces each variable's provenance: exogenous variables are marked as ``Input'' (referencing Table~\ref{tab:visa-t6}), while endogenous variables reference the function ID (f1--f13) from Table~\ref{tab:visa-t4} that computes them.
For an agent type with a single instance ($N{=}1$), the homogeneous/heterogeneous distinction is vacuous, so its exogenous variables may carry either label without affecting the specification; this is a notation convenience rather than a binding constraint.

\begin{table}[htbp]
\caption{VISA Table 2: Agent variables (multi-newsvendor example)}
\label{tab:visa-t2}
\centering\footnotesize
\begin{tabular*}{\textwidth}{@{\extracolsep{\fill}}lclllcp{2cm}@{}}
\toprule
Variable & Symbol & Type & Data Type & Value & Unit & Desc. \\
\midrule
\multicolumn{6}{l}{\textbf{\textit{\underline{Env. $\mathcal{E}$}}}} \\
SensingNet count & $N_\mathcal{S}$ & Exog.-homo. & Integer & Input & -- & Fixed agent count \\
Wholesaler count & $N_\mathcal{W}$ & Exog.-homo. & Integer & Input & -- & Fixed agent count \\
Customer count & $N_\mathcal{C}$ & Exog.-homo. & Integer & Input & -- & Fixed agent count \\
Vendor count & $n_{\mathcal{V},t}$ & Endog. & Integer & f1, f2 & -- & Current number of vendors \\
Vendor list & $\mathcal{L}_{\mathcal{V},t}$ & Endog. & List[Integer] & f1, f2 & -- & List of active vendor IDs \\
Avg.\ retail price & $\bar{p}_t$ & Endog. & Float & f3 & \$/unit & Mean price across vendors \\
\midrule
\multicolumn{6}{l}{\textbf{\textit{\underline{SensingNet $\mathcal{S}$}}}} \\
Connection probability & $\rho$ & Exog.-homo. & Float & Input & -- & Edge formation probability \\
Initial adjacency matrix & $A^{(0)}$ & Exog.-homo. & Matrix[Integer] & Input & -- & Initial vendor network topology \\
Current adjacency matrix & $A_t$ & Endog. & Matrix[Integer] & f4 & -- & Current vendor network topology \\
\midrule
\multicolumn{6}{l}{\textbf{\textit{\underline{Wholesaler $\mathcal{W}$}}}} \\
Wholesale price & $p_w$ & Exog.-homo. & Float & Input & \$/unit & Fixed wholesale cost \\
\midrule
\multicolumn{6}{l}{\textbf{\textit{\underline{Vendor $\mathcal{V}$}}}} \\
Pricing params & $M_a, M_b$ & Exog.-homo. & Float & Input & -- & Multiplier distribution bounds \\
Pricing multiplier & $m_i$ & Exog.-hetero. & Float & Input & -- & $\sim U(M_a, M_b)$ \\
Order quantity & $q_{i,t}$ & Endog.-dec. & Integer & f8 & units & Ordering decision \\
Retail price & $p_{i,t}$ & Endog.-dec. & Float & f7 & \$/unit & Pricing decision \\
Local avg price & $\hat{p}_{i,t}$ & Endog. & Float & f6 & \$/unit & Avg of self \& neighbor prices \\
Current \& prev.\ sales & $s_{i,t}, s_{i,t-1}$ & Endog. & Integer & f13 & units & Period sales \\
Sales history & $\mathbf{s}_{i,t}$ & Endog. & List[Integer] & f5 & -- & Historical sales vector \\
Sales increase flag & $\delta_{i,t}$ & Endog. & Boolean & f11 & -- & Boolean indicator \\
Profit & $\pi_{i,t}$ & Endog. & Float & f9 & \$ & Current-period profit \\
Wealth & $w_{i,t}$ & Endog. & Float & f10 & \$ & Cumulative wealth \\
\midrule
\multicolumn{6}{l}{\textbf{\textit{\underline{Customer $\mathcal{C}$}}}} \\
Base demand & $D$ & Exog.-homo. & Integer & Input & units & Total market demand \\
Demand noise & $\sigma$ & Exog.-homo. & Float & Input & -- & Std.\ dev.\ of demand shock \\
\bottomrule
\end{tabular*}
\end{table}

\subsubsection{Table 3: Sensing}
The Sensing table specifies the information access structure as a quasi-matrix: rows represent observer agents (active agents plus the environment), columns represent observed agents (all agent types), and each cell lists the variables that the row agent can read from the column agent (Table~\ref{tab:visa-t3}).
The table is divided into two sections by a vertical line: active agents (Environment, Space, Decision-makers) appear to the left and can both observe and be observed, while passive agents appear to the right and can only be observed.
The symbol $*$ denotes access to all attributes of the observed agent; $\emptyset$ denotes no access.
On the diagonal, a cell states what an instance may observe of \emph{other} instances of its own type (its peers): the list of accessible peer attributes, or $\emptyset$ if same-type agents observe nothing of one another. A single-instance type ($N{=}1$) has no peers, so its diagonal is $\emptyset$. (Every agent reads its own complete state by default, so self-observation is implicit and is not recorded in the matrix.)
Space agents appear as columns---they mediate information access through topological structure (e.g., a vendor can read the adjacency matrix from the Space agent to identify its neighbors).
Passive agents (e.g., the Wholesaler) also appear only as columns.
By making this structure explicit, VISA ensures that the model's rationality assumptions are unambiguous and that inter-agent information flows can be fully traced.

\begin{table}[htbp]
\caption{VISA Table 3: Agent sensing (multi-newsvendor example)}
\label{tab:visa-t3}
\centering\small
\begin{tabular}{@{}l|cccc|c@{}}
\toprule
\multicolumn{1}{c|}{Observer} & \multicolumn{4}{c}{Observed (active)} & \multicolumn{1}{c}{Observed (passive)} \\
 & $\mathcal{E}$ (Env.) & $\mathcal{S}$ (SensingNet) & $\mathcal{V}$ (Vendor) & $\mathcal{C}$ (Cust.) & $\mathcal{W}$ (Whol.) \\
\midrule
$\mathcal{E}$ (Env.) & $\emptyset$ & $\emptyset$ & $p_i, \pi_i, w_i$ & $\emptyset$ & $\emptyset$ \\
$\mathcal{S}$ (SensingNet) & $\mathcal{L}_\mathcal{V}$ & $\emptyset$ & $w_i$ & $\emptyset$ & $\emptyset$ \\
$\mathcal{V}$ (Vendor) & $\emptyset$ & $A_t$ & $p_j$ & $\emptyset$ & $p_w$ \\
$\mathcal{C}$ (Cust.) & $\emptyset$ & $\emptyset$ & $p_i$ & $\emptyset$ & $\emptyset$ \\
\bottomrule
\end{tabular}
\end{table}

Reading the matrix row by row (an agent's own state is read by default and not shown): the environment observes each vendor's price, profit, and wealth; SensingNet reads the vendor list and wealth (for preferential attachment) and maintains its own adjacency matrix; each vendor reads the adjacency matrix (to identify neighbors), its peer neighbors' prices $p_j$ on the diagonal, and the wholesale price, but not non-neighbors' other attributes; the customer observes all vendor prices. Singleton types (environment, SensingNet, customer) have $\emptyset$ diagonals.

\subsubsection{Table 4: Internal Function}
The Internal Function table specifies the behavioral logic and computational procedures of every agent (Table~\ref{tab:visa-t4}).
Each function is assigned a unique ID (f1--f13) that is referenced by the Value column in Table~\ref{tab:visa-t2}.
Beyond the ID, each function has three input--output specifications: \emph{Decision Basis} (information used, from Tables~\ref{tab:visa-t2}--\ref{tab:visa-t3}); \emph{Self-state Update} (the agent's own endogenous variables modified); and \emph{External Effect} (modifications to other agents' variables). \emph{Ref.} cites the literature motivating the \emph{Method}, making each behavioral choice traceable (blank ``---'' for purely mechanical computations). A core VISA principle follows: every internal function exists to update endogenous variables---its own or another agent's.

\begin{table}[htbp]
\caption{VISA Table 4: Internal functions (multi-newsvendor example)}
\label{tab:visa-t4}
\centering\footnotesize
\resizebox{\textwidth}{!}{%
\begin{tabular}{@{}lllllll@{}}
\toprule
ID & Function & Method & Decision Basis & Self-state Update & External Effect & Ref. \\
\midrule
\multicolumn{7}{l}{\textbf{\textit{\underline{Environment $\mathcal{E}$}}}} \\
f1 & create\_vendor & Instantiation & $\pi_i$ (all $v_i \in \mathcal{V}$) & $n_\mathcal{V} \leftarrow n_\mathcal{V} + 1$ & New $v_i$ & \citep{Nelson1982} \\
f2 & remove\_vendor & Removal & $w_i$ (any $v_i$: $w_i < 0$) & $n_\mathcal{V} \leftarrow n_\mathcal{V} - 1$ & Delete $v_i$ & \citep{Nelson1982} \\
f3 & compute\_avg\_price & Mean & $p_i$ (all $v_i \in \mathcal{V}$) & $\bar{p} \leftarrow \frac{1}{n_\mathcal{V}}\sum p_i$ & --- & --- \\
\midrule
\multicolumn{7}{l}{\textbf{\textit{\underline{SensingNet $\mathcal{S}$}}}} \\
f4 & create\_adjacency & Preferential attachment & $\mathcal{L}_\mathcal{V},\, w_i$ (all $v_i \in \mathcal{V}$) & $A_t \leftarrow$ preferential($w_i$) & --- & \citep{Barabasi1999} \\
\midrule
\multicolumn{7}{l}{\textbf{\textit{\underline{Vendor $\mathcal{V}$}}}} \\
f5 & update\_history & Append & $s_{i,t-1}$ & $\mathbf{s}_i \leftarrow \mathbf{s}_i \cup \{s_{i,t-1}\}$ & --- & --- \\
f6 & compute\_local\_avg & Mean & $p_j\,(j \in \mathcal{N}_i)$ & $\hat{p}_i \leftarrow \frac{1}{|\mathcal{N}_i|+1}(p_i + \sum_{j \in \mathcal{N}_i} p_j)$ & --- & --- \\
f7 & set\_price & Conditional & $\hat{p}_i,\, \delta_i,\, m_i$ & $p_i \leftarrow$ Eq.~\eqref{eq:price} & --- & --- \\
f8 & decide\_order & Optimization & $\mathbf{s}_i,\, p_i,\, p_w$ & $q_i \leftarrow$ Eq.~\eqref{eq:order} & --- & \citep{Arrow1951} \\
f9 & compute\_profit & Arithmetic & $p_i,\, p_w,\, s_{i,t},\, q_i$ & $\pi_i \leftarrow p_i s_{i,t} - p_w q_i$ & --- & --- \\
f10 & update\_wealth & Addition & $\pi_i$ & $w_i \leftarrow w_i + \pi_i$ & --- & --- \\
f11 & update\_sales\_change & Comparison & $s_{i,t},\, s_{i,t-1}$ & $\delta_i \leftarrow (s_{i,t} > s_{i,t-1})$ & --- & --- \\
f12 & record\_sales & Transfer & $s_{i,t}$ & $s_{i,t-1} \leftarrow s_{i,t}$ & --- & --- \\
\midrule
\multicolumn{7}{l}{\textbf{\textit{\underline{Customer $\mathcal{C}$}}}} \\
f13 & allocate\_demand & Multinomial logit & $p_j$ (all $v_j \in \mathcal{V}$), $D, \sigma$ & --- & $s_{i,t}$ via Eq.~\eqref{eq:demand} & \citep{McFadden1974} \\
\bottomrule
\end{tabular}}
\end{table}

Three key functions warrant further detail.
\emph{Local average price computation} aggregates the vendor's own price with its neighbors' prices according to the network topology defined by the adjacency matrix:
\begin{equation}
  \label{eq:local-avg}
  \hat{p}_i = \frac{1}{|\mathcal{N}_i|+1}\!\left(p_i + \sum_{j \in \mathcal{N}_i} p_j\right),
\end{equation}
where $\mathcal{N}_i = \{j : A_{ij} = 1\}$ denotes the neighbor set of vendor $v_i$ derived from the current adjacency matrix $A_t$.
\emph{Vendor pricing} then follows a simple adaptive rule:
\begin{equation}
  \label{eq:price}
  p_i =
  \begin{cases}
    \hat{p}_i\,(1 + m_i) & \text{if } \delta_i = \text{true} \quad\text{(sales increased)}, \\
    \hat{p}_i\,(1 - m_i) & \text{otherwise} \quad\text{(sales decreased)}.
  \end{cases}
\end{equation}
Each vendor raises its price above the local average when its sales increase, and lowers it otherwise, with the heterogeneous multiplier $m_i \sim U(M_a, M_b)$ determining the magnitude of adjustment.

\emph{Vendor ordering} applies the classical newsvendor critical fractile~\citep{Arrow1951}:
\begin{equation}
  \label{eq:order}
  q_i = F_i^{-1}\!\left(\frac{p_i - p_w}{p_i}\right),
\end{equation}
where $F_i^{-1}$ is the inverse CDF computed from vendor $v_i$'s accumulated historical sales $\mathbf{s}_i$.

\emph{Customer demand allocation} uses a multinomial logit model~\citep{McFadden1974}:
\begin{equation}
  \label{eq:demand}
  s_{i,t} = \frac{e^{-p_i}}{1 + \displaystyle \sum_{v_j \in \mathcal{V}} e^{-p_j}} \cdot D + \varepsilon,
  \quad \varepsilon \sim \mathcal{N}(0, \sigma^2),
\end{equation}
ensuring that vendors with higher prices receive less demand.

\subsection{Model-Level Description}
\label{sec:visa-model}

\subsubsection{Table 5: Associated Data}

The Associated Data table (Table~\ref{tab:visa-t5}) documents all external data sources referenced by the model.
Data provenance is a critical but often under-specified aspect of ABM documentation.
\citet{Laatabi2018}, in the \textbf{ODD+2D} extension, identify several shortcomings in current practice: information about data sources, collection methods, preprocessing, and the mapping from data to model entities is typically scattered across the description or omitted entirely.
They emphasize that a model description should make transparent: (i)~\emph{where} data comes from and how it was collected; (ii)~\emph{how} raw data is selected, filtered, and transformed before entering the model; (iii)~\emph{which} data attributes map to which model variables; and (iv)~whether the data is sufficient for model calibration and validation.
VISA's Table~\ref{tab:visa-t5} consolidates these requirements into a single structured table, capturing the complete data provenance chain: source, collection method, preprocessing steps, number of records, and availability.

Several columns use predefined, controlled vocabularies to ensure unambiguous documentation.
We formally define these options below.

\begin{description}

\item[Type] The origin category of the data source:
\begin{itemize}
\item \textbf{Empirical}: Data obtained from direct real-world observations, measurements, surveys, or experiments---reflecting actual phenomena in the target system.
\item \textbf{Literature}: Parameter values or datasets sourced from published academic works (journal articles, books, technical reports) that have undergone peer review or editorial quality control.
\item \textbf{Generated}: Data produced by computational processes, including synthetic population generation, random sampling from specified distributions, or output from auxiliary simulation models.
\end{itemize}

\item[Temporal] The time-varying nature of the data within the simulation:
\begin{itemize}
\item \textbf{Static}: The data remains constant throughout the simulation run and is loaded once (typically at initialization).
\item \textbf{Dynamic}: The data is updated or reloaded at one or more time steps during the simulation, reflecting temporal changes in the external environment.
\end{itemize}

\item[Collection] The method by which the data was originally obtained:
\begin{itemize}
\item \textbf{Survey}: Data collected through structured questionnaires, interviews, or field surveys involving human or organizational respondents.
\item \textbf{Administrative}: Data obtained from institutional records, government databases, census data, or organizational information systems.
\item \textbf{Sensor}: Data collected automatically via sensors, monitoring devices, remote sensing, or other automated measurement instruments.
\item \textbf{Experimental}: Data produced through controlled experiments designed for the modeling project.
\item \textbf{Computational}: Data generated by algorithms, mathematical procedures, or auxiliary simulation models.
\end{itemize}

\item[Pre-processing] The primary transformation applied to raw data before entering the model:
\begin{itemize}
\item \textbf{None}: Raw data is used directly without any modification.
\item \textbf{Selected}: A subset of records or variables is extracted from a larger dataset based on specified criteria (e.g., date range, quality threshold, geographic area).
\item \textbf{Aggregated}: Multiple records are combined into summary statistics (e.g., mean, sum, distribution fitting) or coarser-grained categories, reducing the number of records.
\item \textbf{Transformed}: Data values are converted through mathematical operations, type casting, normalization, discretization, or mapping rules, without changing the number of records.
\end{itemize}

\item[Availability] The access level of the data:
\begin{itemize}
\item \textbf{Open}: The data is publicly accessible under an open license or no license restrictions.
\item \textbf{Restricted}: The data is available under specific conditions (e.g., license agreement, registration, institutional affiliation).
\item \textbf{Private}: The data is not publicly available due to privacy, confidentiality, or proprietary constraints.
\end{itemize}

\end{description}

In the multi-newsvendor example, the model draws on regional dairy market records to parameterize vendor pricing and customer demand.

\begin{table}[htbp]
\caption{VISA Table 5: Associated data (multi-newsvendor example)}
\label{tab:visa-t5}
\centering\scriptsize
\resizebox{\textwidth}{!}{%
\begin{tabular}{@{}clllllccl@{}}
\toprule
ID & Title & Type & Temporal & Source & Collection & Pre-processing & \#Rec. & Avail. \\
\midrule
d1 & Milk vendor prices & Empirical & Static & Source A, 2025 & Administrative & None & 1{,}200 & Open \\
d2 & Milk vendor sales & Empirical & Static & Source A, 2025 & Administrative & None & 1{,}200 & Open \\
\bottomrule
\end{tabular}}
\end{table}

\subsubsection{Table 6: Input and Output}
The Input and Output table bridges external data and simulation results through two sub-tables.
The \emph{Input} sub-table (Table~\ref{tab:visa-t6}a) specifies every exogenous parameter: its value or probability distribution, the data source (referencing Table~\ref{tab:visa-t5}), the derivation method, the algorithm used, and a reference grounding that algorithm.
The \emph{Output} sub-table (Table~\ref{tab:visa-t6}b) defines the simulation's output indicators: computation formula, data type (following the vocabulary of Table~\ref{tab:visa-t2}), unit, sampling frequency, and description.

Several columns of Table~\ref{tab:visa-t6} use predefined values; we specify their meaning below.

\begin{description}

\item[Derivation] (Input sub-table) The method by which an exogenous parameter value is obtained from its data source:
\begin{itemize}
\item \textbf{Direct}: The value is taken verbatim from the referenced data source (Table~\ref{tab:visa-t5}) without any further processing.
\item \textbf{Estimated}: The value---or the parameters of its probability distribution---is statistically inferred from the referenced data (e.g., fitting distribution parameters, estimating a mean or variance from a sample).
\item \textbf{Computed}: The value is produced by a deterministic or stochastic procedure applied to other known quantities (e.g., a formula, or a randomized construction such as the random-connect build of $A^{(0)}$), without statistical inference from data.
\item \textbf{Assumed}: The value is fixed by model design, expert judgment, convention, or the literature, and is not obtained from the referenced data.
\end{itemize}

\item[Algorithm] (Input sub-table) When \emph{Derivation} is \emph{Estimated} or \emph{Computed}, this column names the specific procedure applied---either a standard method (e.g., ``maximum likelihood estimation'') or a model-specific construction (e.g., the random-connect build of $A^{(0)}$).
As with the \emph{Method} column of Table~\ref{tab:visa-t4}, the choice among alternative procedures is modeler-dependent; the adjacent \emph{Ref.} column therefore cites the literature or formula that motivates each choice.

\item[Frequency] (Output sub-table) The sampling interval of an output indicator, recorded as an integer counting the number of time steps between consecutive samples:
\begin{itemize}
\item $k \geq 1$: the indicator is sampled once every $k$ time steps---$k = 1$ records every step (a complete time series), while $k > 1$ downsamples.
\item $-1$: the indicator is recorded only once, at the final time step (the terminal state of the run), following Python's convention that index $-1$ selects the last element.
\end{itemize}

\end{description}

The \emph{Data Type} column of the Output sub-table reuses the type vocabulary of Table~\ref{tab:visa-t2} (e.g., Integer, Float, List[Integer], Matrix[Integer]) and is therefore not redefined here.
In the multi-newsvendor example, the input parameters comprise author-specified structural constants (agent counts and network settings), pricing and wholesale parameters drawn from data record d1, and demand parameters from d2 (Table~\ref{tab:visa-t5}); the initial vendor network is constructed by the random-connect procedure (an Erd\H{o}s--R\'enyi graph with probability $\rho$); and the three output indicators span the \emph{Float} and \emph{List[Integer]} data types, recorded at frequency $1$ (every step) for the two scalar means and $-1$ (terminal step only) for the degree sequence.

\begin{table}[htbp]
\caption{VISA Table 6: Input and Output (multi-newsvendor example)}
\label{tab:visa-t6}
\centering
\textbf{(a) Input}\par\smallskip\scriptsize
\begin{tabular}{@{}clllll@{}}
\toprule
Symbol & Value / Distribution & Data source & Derivation & Algorithm & Ref. \\
\midrule
\multicolumn{6}{l}{\textbf{\textit{\underline{Environment $\mathcal{E}$}}}} \\
$N_\mathcal{S}, N_\mathcal{W}, N_\mathcal{C}$ & 1, 1, 1 & Author & Assumed & --- & --- \\
\midrule
\multicolumn{6}{l}{\textbf{\textit{\underline{SensingNet $\mathcal{S}$}}}} \\
$\rho$ & 0.3 & Author & Assumed & --- & --- \\
$A^{(0)}$ & Random adjacency & Author & Computed & Erd\H{o}s--R\'enyi ($\rho$) & \citep{ErdosRenyi1959} \\
\midrule
\multicolumn{6}{l}{\textbf{\textit{\underline{Wholesaler $\mathcal{W}$}}}} \\
$p_w$ & 100 & d1 & Direct & --- & --- \\
\midrule
\multicolumn{6}{l}{\textbf{\textit{\underline{Vendor $\mathcal{V}$}}}} \\
$M_a, M_b$ & 0.1, 0.5 & d1 & Estimated & Empirical range & --- \\
$m_i$ & $U(M_a, M_b)$ & d1 & Computed & Uniform sampling & --- \\
\midrule
\multicolumn{6}{l}{\textbf{\textit{\underline{Customer $\mathcal{C}$}}}} \\
$D$ & 10000 & d2 & Estimated & Mean & --- \\
$\sigma$ & 500 & d2 & Estimated & Std.\ dev. & --- \\
\bottomrule
\end{tabular}
\par\bigskip
\textbf{(b) Output}\par\smallskip\scriptsize
\begin{tabular*}{\textwidth}{@{\extracolsep{\fill}}llp{1.5cm}lcll@{}}
\toprule
Symbol & Indicator & Formula & Data Type & Unit & Frequency & Desc. \\
\midrule
$\bar{\pi}$ & Avg.\ profit & $\frac{1}{n_\mathcal{V}}\sum_{v_i \in \mathcal{V}} \pi_i$ & Float & \$ & $1$ & Mean vendor profit \\
$\bar{S}r$ & Avg.\ service level & $\frac{1}{n_\mathcal{V}}\sum_i s_{i,t}/q_i$ & Float & --- & $1$ & Fill rate across vendors \\
$\mathbf{d}$ & Degree distribution & $\text{degree}(A_t)$ & List[Integer] & -- & $-1$ & Network degree sequence \\
\bottomrule
\end{tabular*}
\end{table}

The initial adjacency matrix $A^{(0)}$ of the SensingNet space is a symmetric Erd\H{o}s--R\'enyi random graph: each off-diagonal entry $A^{(0)}_{ij}=A^{(0)}_{ji}$ is set to $1$ independently with probability $\rho\in(0,1)$ and to $0$ otherwise~\citep{ErdosRenyi1959}.

\subsubsection{Table 7: Schedule}
The Schedule table defines the temporal execution sequence of the simulation (Table~\ref{tab:visa-t7}).
The execution schedule (Table~\ref{tab:visa-t7}a) lists the per-step loop: each step identifies the acting agent, the function (by its ID and name, referencing Table~\ref{tab:visa-t4}), and the execution mode.
Initial values are not part of this loop; they are supplied as exogenous inputs via Table~\ref{tab:visa-t6}.
The execution mode specifies how the agent instances of a function are scheduled within a step---a crucial but often under-documented aspect of ABM~\cite{Axtell1996}---and is orthogonal to the trigger, which is recorded in the \emph{Condition} column. We define its values below.
\begin{description}
\item[Synchronous] All agent instances observe the state at the start of the step and apply their updates simultaneously; no instance sees another's within-step update (lockstep, parallel update).
\item[Sequential] Instances act one at a time in a fixed, deterministic order; each instance observes the updates of those earlier in the order (serial update with first-mover effects).
\item[Random-order] As Sequential, but the order is uniformly randomized and reshuffled each step, eliminating systematic ordering bias.
\item[Asynchronous] Instances are not synchronized to the global step; each activates at independently drawn times or in response to events (event-driven scheduling).
\end{description}
Where a mode admits mode-specific parameters, they are appended in parentheses after the mode name---for instance, \emph{Sequential (by $w_i$, descending)} records that instances act in descending order of wealth, and \emph{Asynchronous (Poisson, $\lambda$)} records the activation rate---so the schedule remains fully specified without a separate parameter column.
Two further scheduling notions are deliberately not treated as base modes: \emph{conditional} activation (a function firing only when a trigger holds) is recorded in the \emph{Condition} column rather than as an execution mode; and \emph{staged} execution---a step decomposed into synchronized sub-phases such as sense $\to$ decide $\to$ update, in which \emph{all} agents complete one phase before any begins the next---is a composite pattern built from the four modes above. Staging is useful when within-step information contamination must be ruled out (every decision then rests on the same start-of-step snapshot), though it is a local organizing principle rather than a fixed global structure, since a complex model may run several sense--decide--update rounds within a single step for distinct subsystems (e.g., reputation or regulation).
Different execution modes can produce qualitatively different outcomes even with identical behavioral rules---a fact long recognized in work on synchronization in parallel agent-based simulation~\cite{Xu2017}---making this specification essential for reproducibility.
The schedule also specifies termination conditions (Table~\ref{tab:visa-t7}b) as a logical combination of indicator-based criteria referencing Table~\ref{tab:visa-t6}.

\begin{table}[htbp]
\caption{VISA Table 7: Schedule (multi-newsvendor example)}
\label{tab:visa-t7}
\centering
\textbf{(a) Execution schedule}\par\smallskip\scriptsize
\begin{tabular*}{\textwidth}{@{\extracolsep{\fill}}llllll@{}}
\toprule
Step & Agent & ID & Function & Exec.\ mode & Condition \\
\midrule
1 & $\mathcal{E}$ & f1 & create\_vendor & Synchronous & Every step; only if all $\pi_i > 0$ \\
2 & $\mathcal{S}$ & f4 & create\_adjacency & Synchronous & Only when a new vendor is added \\
3 & $\mathcal{V}$ & f5 & update\_history & Synchronous & Every step \\
4 & $\mathcal{V}$ & f6 & compute\_local\_avg & Synchronous & Every step \\
5 & $\mathcal{E}$ & f3 & compute\_avg\_price & Sequential (by vendor ID) & Every step \\
6 & $\mathcal{V}$ & f7 & set\_price & Synchronous & Every step \\
7 & $\mathcal{V}$ & f8 & decide\_order & Synchronous & Every step \\
8 & $\mathcal{C}$ & f13 & allocate\_demand & Sequential (by vendor ID) & Every step \\
9 & $\mathcal{V}$ & f9 & compute\_profit & Synchronous & Every step \\
10 & $\mathcal{V}$ & f10 & update\_wealth & Synchronous & Every step \\
11 & $\mathcal{V}$ & f11 & update\_sales\_change & Synchronous & Every step \\
12 & $\mathcal{V}$ & f12 & record\_sales & Synchronous & Every step \\
13 & $\mathcal{E}$ & f2 & remove\_vendor & Synchronous & Every step; only if any $w_i < 0$ \\
\bottomrule
\end{tabular*}
\par\bigskip
\textbf{(b) Termination conditions}\par\smallskip\small
\begin{tabular}{@{}cllll|l@{}}
\toprule
ID & Indicator & Condition & Description & Value Source/Ref & Termination logic \\
\midrule
c1 & $t$ & $t \geq 1000$ & Max. time steps reached & Author Assumed & \multirow{2}{*}{Stop when $c1 \lor c2$} \\
c2 & $\bar{\pi}$ & $|\bar{\pi}^{(t)} - \bar{\pi}^{(t-1)}| < 0.01$ & Avg.\ profit stable & Author Assumed & \\
\bottomrule
\end{tabular}
\end{table}

\subsubsection{Table 8: Validation}
The Validation table establishes the model's credibility assessment framework (Table~\ref{tab:visa-t8}).
It instantiates the \emph{Hierarchical ABM Validation (HAV)} framework of \citet{he2025hav} which organizes validation across three levels aligned with the agent-based modeling process and distinguishes data-grounded from non-data-grounded methods.
Each entry carries a stable ID (v1, v2, \ldots) so that readers can cite individual validation criteria, and is grouped into one of three levels mirroring the HAV hierarchy: \emph{Agent} (individual behavior), \emph{Model} (structural mechanisms), or \emph{Output} (aggregate results).
The benchmark data, method, indicator, and passing condition together define an unambiguous, testable criterion; the \emph{Ref} column points to prior work that employs the same validation method.
Modelers are encouraged to add further entries---for example, additional agents, mechanisms, or output indicators---to achieve a more comprehensive validation.

\begin{table}[htbp]
\caption{VISA Table 8: Validation (multi-newsvendor example)}
\label{tab:visa-t8}
\centering\footnotesize
\resizebox{\textwidth}{!}{%
\begin{tabular}{@{}lllllll@{}}
\toprule
ID & Validation object & Benchmark data & Method & Indicator & Passing cond. & Ref \\
\midrule
\multicolumn{7}{l}{\textbf{\textit{\underline{Agent level}}}} \\
v1 & Individual vendor sales & d2 (milk vendor sales) & K-S test (two-sample) & $p$-value & $p > 0.05$ & \citep{he2025hav} \\
\midrule
\multicolumn{7}{l}{\textbf{\textit{\underline{Model level}}}} \\
v2 & Price correlation across SensingNet & Positive autocorrelation expected & Moran's $I$ (network) & Moran's $I$ & $I > 0$, $p < 0.05$ & --- \\
\midrule
\multicolumn{7}{l}{\textbf{\textit{\underline{Output level}}}} \\
v3 & Avg.\ profit vs.\ vendor count & Profit decreases with $n_\mathcal{V}$ & Regression $\bar{\pi} \sim n_\mathcal{V}$ & slope $\beta$ & $\beta < 0$, $p < 0.05$ & --- \\
\bottomrule
\end{tabular}}
\end{table}

\section{Operationalization: Consistency Rules and Skills}
\label{sec:guidelines}

Beyond the eight-table format of Section~\ref{sec:visa}, VISA contributes two artifacts that make the format practical at scale: a set of \emph{consistency rules} that turn model validity into a checkable property, and three \emph{skills} that automate authoring, checking, and code generation.

\subsection{Consistency Rules}
\label{sec:guide-modelers}

The eight VISA tables are interrelated and governed by 19 consistency rules that ensure model completeness and logical validity (Tables~\ref{tab:visa-rules-within} and~\ref{tab:visa-rules-cross}).
They fall into two categories: \emph{within-table} rules (r1--r4), which govern consistency within a single table, and \emph{cross-table} rules (r5--r19), which link entries across different tables; when all are satisfied, the specification is guaranteed complete, consistent, and reproducible.
In the \emph{Tables} column, T$k$ denotes Table~$k$ of the VISA protocol (T7b denotes the termination sub-table of T7); an arrow ($\to$) indicates a directional dependency from source to target, a double arrow ($\leftrightarrow$) marks a bidirectional rule that is checked in both directions, and a within-table rule lists a single table.

\begin{table}[htbp]
\caption{VISA within-table consistency rules (r1--r4).}
\label{tab:visa-rules-within}
\centering\scriptsize\renewcommand{\arraystretch}{0.95}
\begin{tabularx}{\textwidth}{@{}llX@{}}
\toprule
ID & Tables & Rule \\
\midrule
r1 & T2 & \textbf{Variable-type coverage and time-indexing.} T2 must classify every variable with one of the four leaf types---Exog.-homo., Exog.-hetero., Endog.-dec., or Endog.\ (non-decision); a bare exogenous/endogenous label or a blank Type leaves provenance ambiguous. In addition, every \emph{endogenous} variable must carry the time index $t$ in its symbol (e.g., $x_{i,t}$, $s_{i,t}$), reflecting that it evolves over the simulation, whereas exogenous variables do not. \\
r2 & T4 & \textbf{Function-ID uniqueness.} Every function in T4 must carry a unique f-ID. \\
r3 & T4 & \textbf{Function productivity.} Every function in T4 must modify at least one endogenous variable, listed in its Self-state Update or External Effect; a function that reads inputs but writes nothing is incomplete. \\
r4 & T7 & \textbf{Step--Exec.\ mode consistency.} Within the execution schedule of T7, all functions assigned the same Step number must share the same Exec.\ mode, since a Step schedules functions that execute together; conversely, functions that must execute simultaneously are assigned a common Step. \\
\bottomrule
\end{tabularx}
\end{table}

\begin{table}[htbp]
\caption{VISA cross-table consistency rules (r5--r19). An arrow ($\to$) indicates a directional dependency from source to target; a double arrow ($\leftrightarrow$) marks a bidirectional rule checked in both directions.}
\label{tab:visa-rules-cross}
\centering\scriptsize\renewcommand{\arraystretch}{0.95}
\begin{tabularx}{\textwidth}{@{}llX@{}}
\toprule
ID & Tables & Rule \\
\midrule
r5 & T1 $\to$ T3 & \textbf{Same-type (peer) sensing.} Every agent type with variable quantity $n{>}1$ must record on the T3 diagonal the attributes its instances observe of one another (the peer set; $\emptyset$ if they observe nothing of each other); single-instance types ($N{=}1$) have no peers and their diagonal is $\emptyset$. (Self-observation is implicit and not recorded.) \\
r6 & T1 $\to$ T2, T3, T4 & \textbf{Passive-agent implications.} Agents categorized as \emph{Passive} in T1: (i)~have no endogenous variables in T2; (ii)~appear only as columns in T3 (right of the vertical line); (iii)~have no internal functions in T4. \\
r7 & T1 $\to$ T3 & \textbf{Observer-row completeness.} Every non-passive agent type in T1 must have exactly one row in T3; passive agents have no rows. \\
r8 & T1 $\to$ T2, T4 & \textbf{Population-dynamics consistency.} An agent type whose T1 Quantity is variable ($n$) must be created or removed by at least one T4 function; a type with a fixed Quantity ($N$) must not. \\
r9 & T1 $\to$ T4 & \textbf{Active-agent function coverage.} Every active (i.e., non-\emph{Passive}) agent type in T1 must own at least one internal function in T4, and is therefore represented in the execution schedule of T7. \\
r10 & T2 $\to$ T3 & \textbf{Variable observability.} Every variable in T2 must be specified as observable in at least one cell of T3 (except variables never accessed by any function). \\
r11 & T2 $\to$ T4 & \textbf{Endogenous-variable completeness.} Every endogenous variable in T2 must be updated by at least one T4 function whose f-ID is listed in the Value column and exists in T4; a variable may be updated by multiple functions across a step. Orphaned variables or dangling f-IDs indicate incomplete specification. \\
r12 & T2 $\to$ T4 & \textbf{Self-state-update validity.} All variables in the \emph{Self-state Update} column of T4 must be endogenous variables of that agent (as defined in T2). Exogenous variables cannot appear in Self-state Update. \\
r13 & T2 $\to$ T4 & \textbf{External-effect validity.} Each entry in the \emph{External Effect} column of T4 is either (i)~an endogenous variable of \emph{another} agent (per T2), or (ii)~the creation or removal of an instance of a variable-quantity type (governed by r8). Modifying another agent's exogenous variables is forbidden. \\
r14 & T3 $\to$ T4 & \textbf{Information-access validation.} For every function in T4, all variables in the \emph{Decision Basis} column must be either: (i)~an attribute of the agent itself; or (ii)~a variable from T3 that the agent is authorized to observe. \\
r15 & T6 $\leftrightarrow$ T2 & \textbf{Input--output coverage.} Every \emph{Input} symbol in T6a must correspond to an exogenous variable in T2, and every T2 variable marked ``Input'' must have a T6a entry; every \emph{Output} indicator in T6b must be computable from variables in T2. \\
r16 & T7 $\leftrightarrow$ T4 & \textbf{Schedule coverage.} Every function in T4 must appear at least once in the \emph{Execution schedule} of T7, and every function ID in T7a must exist in T4; a function may be scheduled multiple times within a step, but none may be left unscheduled or phantom. \\
r17 & T7b $\to$ T2, T6 & \textbf{Termination-indicator source.} Every indicator referenced in the termination conditions of T7b must be either a variable of T2 or an output indicator of T6 (the global time index $t$ is always available and need not be declared). \\
r18 & T8 $\to$ T2, T6 & \textbf{Validation-object coverage.} Every quantity submitted to validation in T8 (the \emph{Validation object}) must be expressed in terms of T2 variables or T6b output indicators; the test statistic in the \emph{Indicator} column is computed during validation and is not itself a model variable. \\
r19 & T8, T6a $\to$ T5 & \textbf{Data-reference resolution.} Every empirical data reference in T8 (\emph{Benchmark data}) and in T6a (\emph{Data source}) must resolve to a T5 record (d-ID); references marked ``Author'' and purely theoretical benchmarks are exempt. \\
\bottomrule
\end{tabularx}
\end{table}

These 19 rules serve two audiences at once: a modeler runs them as a self-check before release, and an evaluator (reviewers, program officers, stakeholders) uses them as a review checklist. Because VISA distributes a model's assumptions across the tables---scale and persistence in T1, provenance in T2, information flow in T3, behavior in T4, data in T5, parameters in T6, timing in T7, and validation in T8---each assumption is locatable and contestable rather than buried in narrative; parameters whose \emph{Derivation} is \emph{Assumed} (T6a) and the thresholds of T7b surface the model's subjective choices directly, and T8 functions as a contract an evaluator may extend with further tests.

\newpage
\subsection{Skills}
\label{sec:guide-evaluators}

Manually maintaining eight interrelated tables is tedious and error-prone---itself a source of the reproducibility failures VISA aims to eliminate.
We therefore provide three reusable skills---self-contained instruction sets that an LLM agent executes---that operationalize the protocol: an \emph{authoring}, a \emph{checking}, and a \emph{code-generation} skill, released as supplementary material and on GitHub (\url{https://github.com/AgentLabCn/visa}).

The \emph{authoring skill} constructs the eight tables in canonical pipeline order (T1--T8), since each table supplies identifiers that later tables reference---agent types from T1 anchor T2's variables, and function IDs from T4 populate T2's Value column---while enforcing the controlled vocabularies and notation of Section~\ref{sec:visa} so the tables are immediately machine-parseable.

The \emph{checking skill} verifies all 19 rules (Tables~\ref{tab:visa-rules-within} and~\ref{tab:visa-rules-cross}) automatically, reporting each rule's pass/fail status, the implicated cells, and a concrete fix---turning the rules into an executable validator.

The \emph{code-generation skill} turns a verified specification into runnable code: for each agent it emits the state variables (Table~\ref{tab:visa-t2}), functions (Table~\ref{tab:visa-t4}) as methods, the sensing neighborhood (Table~\ref{tab:visa-t3}), and the data and parameter wiring (Tables~\ref{tab:visa-t5}--\ref{tab:visa-t6}); the schedule (Table~\ref{tab:visa-t7}a) becomes the step loop and the validation entries (Table~\ref{tab:visa-t8}) become executable checks. The three skills form one loop---author the tables, run the checker until all 19 rules pass, generate code, and re-run the validation---at which point the specification is complete, consistent, and reproduced.


\section{Empirical Evaluation}
\label{sec:evaluation}

To test the protocol's central claim---that a VISA specification is sufficient to reproduce a model---we applied the full author--check--code loop to three external, independently authored agent-based models that the present authors did \emph{not} write. Targeting models we have no stake in defeats the obvious objection that a description protocol can be made to look good on a model designed for it. Table~\ref{tab:eval-summary} summarizes the outcome; the complete eight-table specifications, the 19-rule reports, and the full reproduction results are in the supplementary material, and all runnable code is on GitHub (\url{https://github.com/AgentLabCn/visa}).

\begin{table}[htbp]
  \centering
  \small
  \caption{The VISA loop applied to three external ABMs. \texttt{author}/\texttt{check}/\texttt{code} are the three skills; ``reproduce'' re-executes the T8 validation. Models 1--2 are reproduced cross-language (NetLogo~$\to$~Python); Model~3 is a describe-only expressiveness case.}
  \label{tab:eval-summary}
  \begin{tabular}{@{}llccccc@{}}
    \toprule
    Model & Platform & author & check & code & reproduce & Outcome \\
    \midrule
    Rebellion~\citep{Epstein2002}        & NetLogo$\to$Python & \checkmark & 19/19 & \checkmark & \checkmark & v1--v3 pass \\
    Wolf Sheep Stride~\citep{Novak2006}  & NetLogo$\to$Python & \checkmark & 19/19 & \checkmark & \checkmark & v2--v5 pass; v1 dir. \\
    AgedCareContactModel                 & AnyLogic           & \checkmark & 19/19 & ---        & ---        & blocked \\
    \bottomrule
  \end{tabular}
\end{table}

\paragraph{Setup.}
The two primary models are NetLogo library models; we re-implemented each in Python directly from its VISA specification, making the reproduction \emph{cross-language} (NetLogo~$\to$~Python), a stronger test than re-implementing in the same language. The reproduction target is each model's qualitative signature---a pattern that is structurally determined and not sensitive to the random seed---rather than a bit-exact trajectory. Both pass all 19 consistency rules, generate runnable Python, and reproduce their signature dynamics. The third is an industrial AnyLogic model used as an expressiveness test and an honest limit case.

\begin{figure}[htbp]
  \centering
  \includegraphics[width=0.85\textwidth]{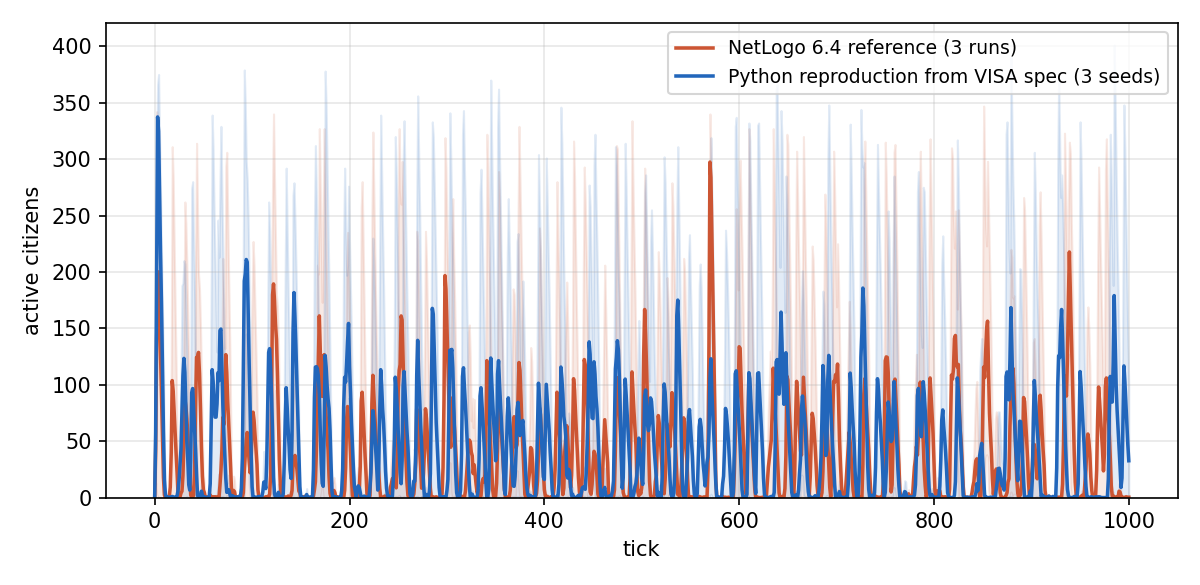}
  \caption{Rebellion reproduction---active-citizen count over time. The NetLogo~6.4 reference (3 runs) and the Python re-implementation from the VISA specification (3 seeds) are min--max bands with solid means. The punctuated-equilibrium signature is reproduced cross-language; the pattern is robust to the random seed.}
  \label{fig:rebellion-overlay}
\end{figure}

\paragraph{Rebellion.}
The civil-violence model of \citet{Epstein2002} reproduces its punctuated-equilibrium signature: the active-citizen count shows periodic rebellion episodes (peaks near 300, roughly 30 per 1000 ticks) separated by long quiescent stretches. Figure~\ref{fig:rebellion-overlay} overlays the NetLogo~6.4 reference against the Python re-implementation generated from the specification; the two reproduce the same qualitative pattern. T8 criteria v1--v3 pass.

\begin{figure}[htbp]
  \centering
  \includegraphics[width=0.7\textwidth]{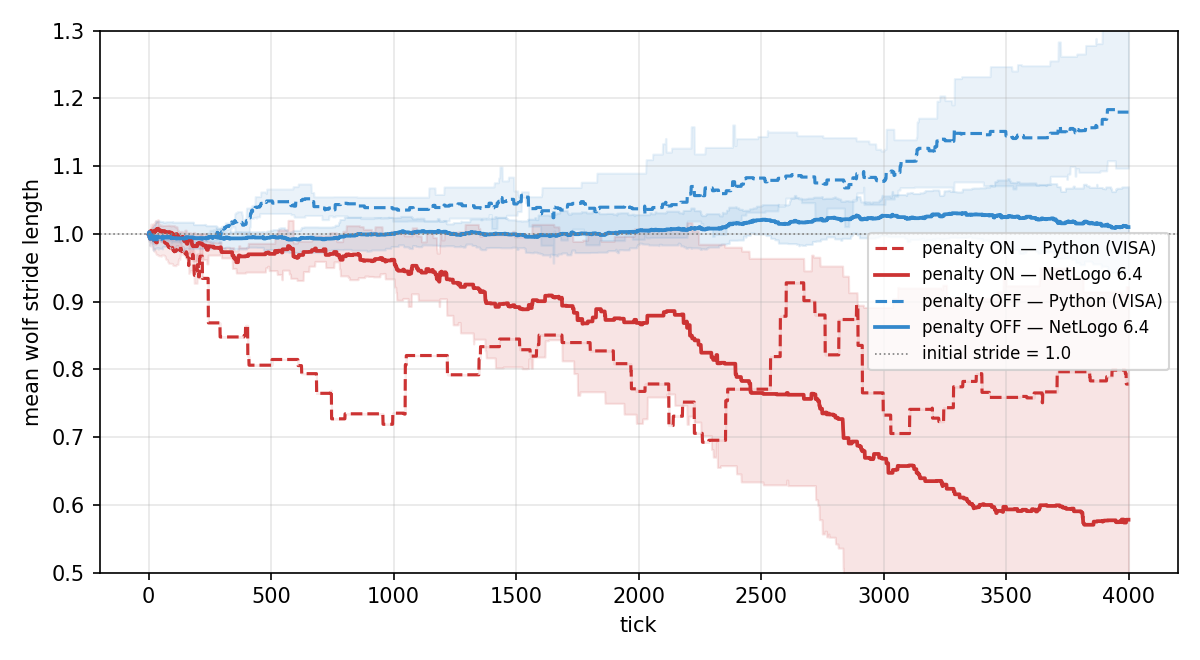}
  \caption{Wolf stride-length selection (default model). The NetLogo~6.4 reference (solid) and the Python re-implementation generated from the VISA specification (dashed)---each a mean over 3 runs/seeds---both reproduce the causal reversal: penalty on~$\to$~shorter stride; off~$\to$~longer.}
  \label{fig:wolf-stride}
\end{figure}

\paragraph{Wolf Sheep Stride Inheritance.}
This model carries an inheritable, mutating \emph{stride-length} trait---the distance an animal moves per step, passed to offspring with random mutation---so natural selection on movement distance can be observed, and the population's mean stride length is the model's headline selection indicator. Its headline result concerns the \emph{stride-length energy penalty}, a switch that, when on, charges each animal energy equal to its stride length per step (moving farther costs more); when off, movement is free. The penalty \emph{causally reverses} the direction of wolf-stride selection: with the penalty on, the cost of moving far outweighs the extra prey caught, so wolves evolve a shorter stride (1.0$\to$0.78); with it off, moving far is free and catches more sheep, so they evolve a longer one (1.0$\to$1.18). Figure~\ref{fig:wolf-stride} overlays the NetLogo~6.4 reference against the VISA-driven Python reproduction, showing that both platforms reproduce the reversal. Parameter-dependent extinction and predator--prey oscillation also reproduce (full overlay figures in the supplementary). T8 v2--v5 pass; v1 (sheep stride converging toward~1) is reproduced only directionally, a shortfall we attribute to the model's grass-abundance structure rather than to any specification error.

\paragraph{AgedCareContactModel.}
This AnyLogic~8.9 model of an aged-care facility produces epidemiological contact matrices, and it exercises VISA's expressiveness on a real, industrially authored model: \texttt{visa-author} captures its 29 object types and 240 functions in eight tables (consolidated to 4 agent sets and 14 behavioural functions), and \texttt{visa-check} passes 19/19 with no structural accommodation. Full reproduction is blocked, and this is itself a finding rather than a VISA failure. The model's behavior splits into (a)~reproducible custom Java logic, fully captured in the tables, and (b)~movement executed by AnyLogic's proprietary Pedestrian Library, which dominates the contact-matrix output and has no open specification. VISA makes~(a) explicit and exposes~(b) as a named, localized dependency---the real barrier---rather than letting it hide in prose; the unavailable backing data is a second, independent blocker. No reproduction figure is shown for this case, by design.

Across the three models, VISA scales from a textbook NetLogo model to an industrial AnyLogic model, reproduces the two whose behavior is entirely in open code, and honestly demarcates its limit on the one that depends on a proprietary library. The reproduction barrier moves from the model---where it is invisible---to a named dependency, where it is actionable.

\section{Discussion}
\label{sec:discussion}

We discuss two aspects of VISA that bear on its broader adoption: its relationship to visual model documentation, and its applicability to agents whose behavior is learned rather than analytically specified.

\subsection{Visualization and Human Readability}
\label{sec:disc-viz}

VISA and visualization are complementary, not competing: the table-based specification is in fact a stronger substrate for producing diagrams than narrative prose. Several tables correspond directly to established diagram types---the sensing matrix (Table~\ref{tab:visa-t3}) is a directed graph, the agent--function ownership of Tables~\ref{tab:visa-t1} and~\ref{tab:visa-t4} maps onto a UML class diagram, and the schedule (Table~\ref{tab:visa-t7}) onto a sequence diagram---and because every table is structured and symbol-based, these visuals can be \emph{generated automatically} from the specification, whereas rendering an ODD narrative requires manual translation. VISA thus provides the machine-interpretable source from which visual layers such as Visual ODD~\citep{Szangolies2024,Lee2026} and UML~\citep{Bersini2012} render the human-readable view; the two are designed to be used together.

\subsection{Complex Agent Behaviors: Reinforcement Learning and LLM-Based Agents}
\label{sec:disc-complex}

VISA can also accommodate agents governed by a reinforcement-learning (RL) policy or a large language model (LLM), without structural change, because it specifies behavior at the level of a function's input--output contract rather than its internal mechanism. Both fit Table~\ref{tab:visa-t4} directly: the \emph{Decision Basis} is the observed state or the variables serialized into a prompt; the \emph{Self-state Update} is the chosen action or parsed decision; the \emph{Method} names the learned policy or language model. The nineteen consistency rules are mechanism-agnostic, governing structural completeness rather than behavioral form, so an RL- or LLM-driven model satisfies exactly the same rules as the multi-newsvendor.

The genuine difficulty is reproducibility, not specification. A learned component is defined by its training configuration---algorithm, hyperparameters, data, and seed for RL; exact model version, prompt, decoding parameters, and seed for an LLM---all of which VISA can capture (data in Table~\ref{tab:visa-t5}, parameters and seeds in Table~\ref{tab:visa-t6}, the model in the \emph{Ref} column). Even so, LLM-based agents carry a residual reproducibility hazard that no description protocol can fully eliminate: model versions drift and inference is stochastic. As with the proprietary-library dependency of Section~\ref{sec:evaluation}, VISA's contribution is to make the configuration explicit and to localize the irreproducibility to a named, versioned component rather than letting it hide in prose; bounding it remains an open problem.

\section{Conclusion}
\label{sec:conclusion}

Agent-based modeling has become an indispensable methodology for complex systems, yet its scientific credibility rests on a reproducibility that the field has struggled to deliver.
The root difficulty is descriptive: existing protocols are predominantly text-centric, not machine-readable, and leave a wide gap between a model's documentation and its executable implementation.
We have argued that the way forward is to first ask what information is truly indispensable for reproducing an agent-based model, and then to fix that information in a structured, symbol-based form.

This paper proposed \textbf{VISA}, a table-based description protocol that answers that question.
VISA organizes a model's complete specification into eight interconnected tables across two levels---four at the agent level (Agent, Variable, Sensing, Internal Function) and four at the model level (Associated Data, Input/Output, Schedule, Validation)---under the principle of minimality with completeness: every element is indispensable, and together they are necessary and sufficient for a model to be unambiguously understood, faithfully reproduced, critically evaluated, and productively extended.
Symbol-based notation removes the ambiguity of natural language, explicit cross-references make every dependency traceable, and nineteen consistency rules turn model validity from a matter of judgment into a checkable property, as illustrated throughout by the multi-newsvendor example.

To move the protocol from specification to practice, we operationalized it with two artifacts: nineteen executable consistency rules and three reusable skills (authoring, checking, code generation). Together they target the three properties the field most lacks---\emph{transparency}, through explicit and locatable tables; \emph{reproducibility}, through automated code generation and pinned execution modes; and \emph{verifiability}, through executable rules and an extensible validation table.

Empirically, the full author--check--code loop closes on three external, independently authored models: the Rebellion and Wolf Sheep Stride Inheritance models are reproduced cross-language (NetLogo to Python) directly from their VISA specifications, each passing all nineteen rules and recovering its qualitative signature, while a large industrial AnyLogic model is captured faithfully in eight tables (again 19/19), with its reproduction barrier---a proprietary movement library and unavailable backing data---made explicit rather than hidden.

The protocol has clear limits, and each points to a direction for improvement. VISA complements rather than replaces visual documentation, so a natural next step is visualization tools that render its tables automatically, bridging to the visual standards modelers already use. Its benefits over narrative description still await quantitative evidence: a systematic comparison with the ODD family would measure the trade-off in information coverage and documentation burden, and a controlled reproduction study---independent teams re-implementing the same model from a VISA specification and from a narrative description---would quantify the effect on faithful replication. VISA accommodates reinforcement-learning and LLM-based agents without structural change, yet the residual hazard of versioned, stochastic black-box components is one no description protocol can fully eliminate; extending the skills to further platforms, and sharpening how such components are specified, would both broaden the protocol's reach and refine this honest limit. Throughout, the aim is constant: to move agent-based models from implicit description toward specification that is explicit, locatable, and executable, bringing simulation-based research closer to the reproducibility standards expected of experimental science.

\section*{Acknowledgments}
This work was supported in part by the National Natural Science Foundation of China (grant number 72271227), the Fundamental Research Funds for the Central Universities (grant numbers E1E40810X2), the Youth Innovation Promotion Association CAS (grant number 110800EAG2), and the MOE Social Science Laboratory of Digital Economic Forecasts and Policy Simulation at UCAS.

\bibliographystyle{plainnat}
\bibliography{RefVisa}

@article{Stone2026,
  title   = {Generating a Contact Matrix for Aged Care Settings in Australia: an agent-based model study},
  author  = {Stone, Haley and MacIntyre, C. Raina and Kunasekaran, Mohana and Poulos, Chris and Heslop, David},
  journal = {Journal of Artificial Societies and Social Simulation},
  year    = {2026},
  volume  = {29},
  number  = {2},
  pages   = {7},
  doi     = {10.18564/jasss.5993},
}

@article{Helbing2000,
  author  = {Helbing, Dirk and Moln{\'a}r, P{\'e}ter and Farkas, Ill{\'e}s J. and Bolay, Kai},
  title   = {Self-organizing pedestrian movement},
  journal = {Environment and Planning B: Planning and Design},
  year    = {2001},
  volume  = {28},
  number  = {3},
  pages   = {361--383},
}

@article{Szangolies2024,
  author  = {Szangolies, Leonna and Rohw{\"a}der, Marie-Sophie and Ahmed, Hazem and Jahanmiri, Fatima and Wagner, Alexander and Souto-Veiga, Rodrigo and Grimm, Volker and Gallagher, Cara A.},
  title   = {Visual {ODD}: A Standardised Visualisation Illustrating the Narrative of Agent-Based Models},
  journal = {Journal of Artificial Societies and Social Simulation},
  year    = {2024},
  volume  = {27},
  number  = {4},
  pages   = {1},
  doi     = {10.18564/jasss.5450},
  url     = {http://jasss.soc.surrey.ac.uk/27/4/1.html},
}

@article{Grimm2020,
  author  = {Grimm, Volker and Railsback, Steven F. and Vincenot, Christian E. and Berger, Uta and Gallagher, Cara and DeAngelis, Donald L. and Edmonds, Bruce and Ge, Jiaqi and Giske, Jarl and Groeneveld, J{\"u}rgen and Johnston, Alice S. A. and Milles, Alexander and Nabe-Nielsen, Jacob and Polhill, J. Gareth and Radchuk, Viktoriia and Rohw{\"a}der, Marie-Sophie and Stillman, Richard A. and Thiele, Jan C. and Ayll{\'o}n, Daniel},
  title   = {The {ODD} Protocol for Describing Agent-Based and Other Simulation Models: A Second Update to Improve Clarity, Replication, and Structural Realism},
  journal = {Journal of Artificial Societies and Social Simulation},
  year    = {2020},
  volume  = {23},
  number  = {2},
  pages   = {7},
  doi     = {10.18564/jasss.4259},
  url     = {http://jasss.soc.surrey.ac.uk/23/2/7.html},
}

@article{Laatabi2018,
  author  = {Laatabi, Ahmed and Marilleau, Nicolas and Nguyen-Huu, Tri and Hbid, Hassan and Babram, Mohamed Ait},
  title   = {{ODD+2D}: An {ODD} Based Protocol for Mapping Data to Empirical {ABM}s},
  journal = {Journal of Artificial Societies and Social Simulation},
  year    = {2018},
  volume  = {21},
  number  = {2},
  pages   = {9},
  doi     = {10.18564/jasss.3646},
  url     = {http://jasss.soc.surrey.ac.uk/21/2/9.html},
}

@article{Grimm2010,
  author  = {Grimm, Volker and Berger, Uta and DeAngelis, Donald L. and Polhill, J. Gary and Giske, Jarl and Railsback, Steven F.},
  title   = {The {ODD} protocol: A review and first update},
  journal = {Ecological Modelling},
  year    = {2010},
  volume  = {221},
  number  = {23},
  pages   = {2760--2768},
  doi     = {10.1016/j.ecolmodel.2010.08.019},
}

@inproceedings{Kazieva2026,
  author    = {Kazieva, Victoria},
  title     = {Towards Model Reproducibility: Insights from Replicating an Agent-Based Model},
  booktitle = {Proceedings of the 2026 Winter Simulation Conference},
  year      = {2026},
  editor    = {Ramamohan, V. and Djanatliev, A. and Fakhimi, M. and Krejci, C. and Ruiz Martin, C. and Onggo, B. S. and Mustafee, N.},
  publisher = {{IEEE}},
}

@inproceedings{Lee2026,
  author    = {Lee, Hui Min and Lazarova-Molnar, Sanja},
  title     = {{vODD-DD}: A Visual Documentation Protocol for Transparent and Reproducible Data-Driven Agent-Based Models},
  booktitle = {Proceedings of the 2026 Winter Simulation Conference},
  year      = {2026},
  editor    = {Ramamohan, V. and Djanatliev, A. and Fakhimi, M. and Krejci, C. and Ruiz Martin, C. and Onggo, B. S. and Mustafee, N.},
  publisher = {{IEEE}},
}

@incollection{Parker2006,
  author    = {Parker, Dawn C. and Brown, Daniel G. and Polhill, J. Gary and Deadman, Peter J. and Manson, Steven M.},
  title     = {Illustrating a new `conceptual design pattern' for agent-based models of land use via five case studies---the {MR POTATOHEAD} framework},
  booktitle = {Agent-Based Modelling in Natural Resource Management},
  year      = {2006},
  editor    = {L{\'o}pez Paredes, Adolfo and Hern{\'a}ndez Iglesias, Ces{\'a}reo},
  publisher = {Valladolid: INSISOC},
}

@article{Grimm2006,
  author  = {Grimm, Volker and Berger, Uta and Bastiansen, Finn and Eliassen, Sigmund and Ginot, Vincent and Giske, Jarl and Goss-Custard, John and Grand, Tanya and Heinz, Stephan K. and Huse, Geir and Huth, Andreas and Jepsen, Jes\'{u}s U. and J\o{}rgensen, Christoper and Mooij, Wolf M. and M\"{u}ller, Birgit and Pe'er, Guy and Piou, Cyril and Railsback, Steven F. and Robbins, Andrew M. and Robbins, Martha M. and Rossmanith, Erika and R\"{u}ger, Nadja and Strand, Espen and Souissi, Sami and Stillman, Richard A. and Vab\o{}, Rune and Visser, Ute and DeAngelis, Donald L.},
  title   = {A standard protocol for describing individual-based and agent-based models},
  journal = {Ecological Modelling},
  year    = {2006},
  volume  = {198},
  number  = {1-2},
  pages   = {115--126},
  doi     = {10.1016/j.ecolmodel.2006.04.023},
}

@article{Mueller2013,
  author  = {M\"{u}ller, Birgit and Bohn, Franziska and Dre\ss{}ler, Gunnar and Groeneveld, J\"{u}rgen and Klassert, Christian and Martin, Romina and Schl\"{u}ter, Maja and Schulze, J\"{o}rg and Weise, Henriette and Schwarz, Nina},
  title   = {Describing human decisions in agent-based models -- {ODD} + {D}, an extension of the {ODD} protocol},
  journal = {Environmental Modelling \& Software},
  year    = {2013},
  volume  = {48},
  pages   = {37--48},
  doi     = {10.1016/j.envsoft.2013.06.003},
}

@article{Vincenot2018,
  author  = {Vincenot, Christian E.},
  title   = {How new concepts become universal scientific approaches: Insights from citation network analysis of agent-based complex systems science},
  journal = {Proceedings of the Royal Society B: Biological Sciences},
  year    = {2018},
  volume  = {285},
  number  = {1874},
  pages   = {20172360},
  doi     = {10.1098/rspb.2017.2360},
}

@article{Bersini2012,
  author  = {Bersini, Hugues},
  title   = {{UML} for {ABM}},
  journal = {Journal of Artificial Societies and Social Simulation},
  year    = {2012},
  volume  = {15},
  number  = {1},
  pages   = {9},
}

@article{Polhill2008,
  author  = {Polhill, J. Gary and Parker, Dawn and Brown, Daniel and Grimm, Volker},
  title   = {Using the {ODD} protocol for describing three agent-based social simulation models of land-use change},
  journal = {Journal of Artificial Societies and Social Simulation},
  year    = {2008},
  volume  = {11},
  number  = {2},
  pages   = {3},
}

@article{Polhill2010,
  author  = {Polhill, J. Gary},
  title   = {{ODD} updated},
  journal = {Journal of Artificial Societies and Social Simulation},
  year    = {2010},
  volume  = {13},
  number  = {4},
  pages   = {9},
}

@article{Innocenti2020,
  author  = {Innocenti, Elena and Detotto, Claudio and Idda, Carlo and Parker, Dawn C. and Prunetti, Donatella},
  title   = {An iterative process to construct an interdisciplinary {ABM} using {MR POTATOHEAD}: An application to housing market models in touristic areas},
  journal = {Ecological Complexity},
  year    = {2020},
  volume  = {44},
  pages   = {100882},
  doi     = {10.1016/j.ecocom.2020.100882},
}

@article{Daly2022,
  author  = {Daly, Andrew J. and De Visscher, Laurens and Baetens, Jan M. and De Baets, Bernard},
  title   = {Quo vadis, agent-based modelling tools?},
  journal = {Environmental Modelling \& Software},
  year    = {2022},
  volume  = {157},
  pages   = {105514},
  doi     = {10.1016/j.envsoft.2022.105514},
}

@article{Squazzoni2020,
  author  = {Squazzoni, Flaminio and Polhill, J. Gareth and Edmonds, Bruce and Ahrweiler, Petra and Antosz, Patrycja and Scholz, Gerd and Chappin, Emile and Borit, Melania and Verhagen, Harko and Giardini, Francesca and Gilbert, Nigel},
  title   = {Computational Models that Matter During a Global Pandemic Outbreak: A Call to Action},
  journal = {Journal of Artificial Societies and Social Simulation},
  year    = {2020},
  volume  = {23},
  number  = {2},
  pages   = {10},
  doi     = {10.18564/jasss.4296},
}

@article{Smaldino2020,
  author  = {Smaldino, Paul E.},
  title   = {How to Translate a Verbal Theory Into a Formal Model},
  journal = {Social Psychology},
  year    = {2020},
  volume  = {51},
  number  = {4},
  pages   = {207--218},
  doi     = {10.1027/1864-9335/a000425},
}

@article{Augusiak2014,
  author  = {Augusiak, Jaideep and Van den Brink, Paul J. and Grimm, Volker},
  title   = {Merging validation and evaluation of ecological models to `evaludation': A review of terminology and a practical approach},
  journal = {Ecological Modelling},
  year    = {2014},
  volume  = {280},
  pages   = {117--128},
  doi     = {10.1016/j.ecolmodel.2013.11.009},
}

@article{Grimm2014,
  author  = {Grimm, Volker and Augusiak, Jaideep and Focks, Andre and Frank, Bj\"{o}rn M. and Gabsi, Faten and Johnston, Alice S. A. and Liu, Canran and Martin, Benjamin T. and Meli, Marco and Radchuk, Viktoriia and Thorbek, Pernille and Railsback, Steven F.},
  title   = {Towards better modelling and decision support: Documenting model development, testing, and analysis using {TRACE}},
  journal = {Ecological Modelling},
  year    = {2014},
  volume  = {280},
  pages   = {129--139},
  doi     = {10.1016/j.ecolmodel.2013.11.018},
}

@article{Grimm2005,
  author  = {Grimm, Volker and Revilla, Eric and Berger, Uta and Jeltsch, Florian and Mooij, Wolf M. and Railsback, Steven F. and Thulke, Hans-Hermann and Weiner, Jacob and Wiegand, Thorsten and DeAngelis, Donald L.},
  title   = {Pattern-oriented modeling of agent-based complex systems: Lessons from ecology},
  journal = {Science},
  year    = {2005},
  volume  = {310},
  number  = {5750},
  pages   = {987--991},
  doi     = {10.1126/science.1116681},
}

@article{Railsback2001,
  author  = {Railsback, Steven F.},
  title   = {Concepts from complex adaptive systems as a framework for individual-based modelling},
  journal = {Ecological Modelling},
  year    = {2001},
  volume  = {139},
  number  = {1},
  pages   = {47--62},
  doi     = {10.1016/S0304-3800(01)00228-6},
}

@article{Waltemath2011,
  author  = {Waltemath, Dagmar and Adams, Richard and Beard, Daniel A. and Bergmann, Frank T. and Bhalla, Upinder S. and Britten, Randall and Chelliah, Vijayalakshmi and Cooling, Michael T. and Cooper, Jonathan and Crampin, Edmund J.},
  title   = {Minimum information about a simulation experiment ({MIASE})},
  journal = {PLoS Computational Biology},
  year    = {2011},
  volume  = {7},
  number  = {4},
  pages   = {e1001122},
  doi     = {10.1371/journal.pcbi.1001122},
}

@article{Wilkinson2016,
  author  = {Wilkinson, Mark D. and Dumontier, Michel and Aalbersberg, Isabel J. and Appleton, Gabrielle and Axton, Myles and Baak, Arie and Blomberg, Niklas and Boiten, Jan-Willem and da Silva Santos, Luiz Bonino and Bourne, Philip E. and others},
  title   = {The {FAIR} guiding principles for scientific data management and stewardship},
  journal = {Scientific Data},
  year    = {2016},
  volume  = {3},
  pages   = {160018},
  doi     = {10.1038/sdata.2016.18},
}

@article{Fanelli2018,
  author  = {Fanelli, Daniele},
  title   = {Opinion: Is science really facing a reproducibility crisis, and do we need it to?},
  journal = {Proceedings of the National Academy of Sciences},
  year    = {2018},
  volume  = {115},
  number  = {11},
  pages   = {2628--2631},
  doi     = {10.1073/pnas.1708272114},
}

@article{Monks2019,
  author  = {Monks, Thomas and Currie, Christine S. M. and Onggo, Bhakti S. and Robinson, Stewart and Kunc, Martin and Taylor, Simon J. E.},
  title   = {Strengthening the reporting of empirical simulation studies: Introducing the {STRESS} guidelines},
  journal = {Journal of Simulation},
  year    = {2019},
  volume  = {13},
  number  = {1},
  pages   = {55--67},
  doi     = {10.1080/17477778.2018.1536465},
}

@article{Knuth1984,
  author  = {Knuth, Donald E.},
  title   = {Literate programming},
  journal = {The Computer Journal},
  year    = {1984},
  volume  = {27},
  number  = {2},
  pages   = {97--111},
  doi     = {10.1093/comjnl/27.2.97},
}

@misc{Wilensky1999,
  author       = {Wilensky, Uri},
  title        = {{NetLogo}},
  year         = {1999},
  howpublished = {Center for Connected Learning and Computer-Based Modeling, Northwestern University, Evanston, IL},
  url          = {http://ccl.northwestern.edu/netlogo/},
}

@article{Polhill2015,
  author  = {Polhill, J. Gary},
  title   = {Extracting {OWL} ontologies from agent-based models: A {NetLogo} extension},
  journal = {Journal of Artificial Societies and Social Simulation},
  year    = {2015},
  volume  = {18},
  number  = {2},
  pages   = {15},
  doi     = {10.18564/jasss.2710},
}

@article{Polhill2009,
  author  = {Polhill, J. Gary and Gotts, Nicholas M.},
  title   = {Ontologies for transparent integrated human-natural system modelling},
  journal = {Landscape Ecology},
  year    = {2009},
  volume  = {24},
  number  = {9},
  pages   = {1255},
  doi     = {10.1007/s10980-009-9391-7},
}

@article{OpenScience2015,
  author  = {{Open Science Collaboration}},
  title   = {Estimating the reproducibility of psychological science},
  journal = {Science},
  year    = {2015},
  volume  = {349},
  number  = {6251},
  pages   = {aac4716},
  doi     = {10.1126/science.aac4716},
}

@article{Mueller2014,
  author  = {M\"{u}ller, Birgit and Balbi, Stefano and Buchmann, C. M. and De Sousa, L. and Dressler, Gunnar and Groeneveld, J\"{u}rgen and Klassert, C. J. and Le, Q. B. and Millington, J. D. and Nolzen, H. and others},
  title   = {Standardised and transparent model descriptions for agent-based models: Current status and prospects},
  journal = {Environmental Modelling \& Software},
  year    = {2014},
  volume  = {55},
  pages   = {156--163},
  doi     = {10.1016/j.envsoft.2014.01.029},
}

@article{Richiardi2006,
  author  = {Richiardi, Matteo and Leombruni, Roberto and Saam, Nicole and Sonnessa, Michele},
  title   = {A common protocol for agent-based social simulation},
  journal = {Journal of Artificial Societies and Social Simulation},
  year    = {2006},
  volume  = {9},
  number  = {1},
  pages   = {15},
}

@article{Janssen2017,
  author  = {Janssen, Marco A.},
  title   = {The practice of archiving model code of agent-based models},
  journal = {Journal of Artificial Societies and Social Simulation},
  year    = {2017},
  volume  = {20},
  number  = {1},
  pages   = {2},
  doi     = {10.18564/jasss.3317},
}

@article{Donkin2017,
  author  = {Donkin, E. and Dennis, P. and Ustalakov, A. and Warren, J. and Clare, A.},
  title   = {Replicating complex agent based models, a formidable task},
  journal = {Environmental Modelling \& Software},
  year    = {2017},
  volume  = {92},
  pages   = {142--151},
  doi     = {10.1016/j.envsoft.2017.01.014},
}

@article{Filatova2015,
  author  = {Filatova, Tatiana},
  title   = {Empirical agent-based land market: Integrating adaptive economic behavior in urban land-use models},
  journal = {Computers, Environment and Urban Systems},
  year    = {2015},
  volume  = {54},
  pages   = {397--413},
  doi     = {10.1016/j.compenvurbsys.2014.12.005},
}

@article{Groeneveld2017,
  author  = {Groeneveld, J\"{u}rgen and Klabunde, Anna and O'Brien, Michelle L. and Grow, Andr\'{e}s},
  title   = {How to describe agent-based models in population studies?},
  journal = {Agent-Based Modelling in Population Studies},
  year    = {2017},
  pages   = {237--254},
  publisher = {Springer},
}

@article{Wolf2013,
  author  = {Wolf, Sebastian and Bouchaud, Jean-Philippe and Cecconi, Furio and Cincotti, Silvano and Dawid, Herbert and Gintis, Herbert and van der Hoog, Sander and Jaeger, Carlo C. and Kovalevsky, Dimitri V. and Mandel, Antoine and others},
  title   = {Describing economic agent-based models -- {Dahlem ABM} documentation guidelines},
  journal = {Complexity Economics},
  year    = {2013},
  volume  = {2},
  number  = {1},
}

@incollection{Parker2008,
  author    = {Parker, Dawn C. and Brown, Daniel and Polhill, J. Gary and Manson, Steven and Deadman, Peter},
  title     = {Illustrating a new conceptual design pattern for agent-based models and land use via five case studies: The {MR POTATOHEAD} framework},
  booktitle = {Agent-Based Modeling in Natural Resource Management},
  year      = {2008},
  editor    = {L{\'o}pez Paredes, A. and Hern{\'a}ndez Iglesias, C.},
  publisher = {Valladolid: INSISOC},
}

@article{Janssen2008,
  author  = {Janssen, Marco A. and Alessa, Lilian N. and Barton, Michael and Bergin, Sean and Lee, Allen},
  title   = {Towards a community framework for agent-based modelling},
  journal = {Journal of Artificial Societies and Social Simulation},
  year    = {2008},
  volume  = {11},
  number  = {2},
  pages   = {6},
}

@article{Axtell1996,
  author  = {Axtell, Robert and Axelrod, Robert and Epstein, Joshua M. and Cohen, Michael D.},
  title   = {Aligning simulation models: A case study and results},
  journal = {Computational and Mathematical Organization Theory},
  year    = {1996},
  volume  = {1},
  number  = {2},
  pages   = {123--141},
  doi     = {10.1007/BF01299065},
}

@article{Edmonds2003,
  author  = {Edmonds, Bruce and Hales, David},
  title   = {Replication, replication and replication: Some hard lessons from model alignment},
  journal = {Journal of Artificial Societies and Social Simulation},
  year    = {2003},
  volume  = {6},
  number  = {4},
  pages   = {11},
}

@article{Angus2015,
  author  = {Angus, Simon D. and Hassani-Mahmooei, Behrooz},
  title   = {{`Anarchy'} Reigns: A Quantitative Analysis of Agent-Based Modelling Publication Practices in {JASSS}, 2001--2012},
  journal = {Journal of Artificial Societies and Social Simulation},
  year    = {2015},
  volume  = {18},
  number  = {4},
  pages   = {16},
  doi     = {10.18564/jasss.2862},
}

@article{Zhang2021replication,
  author  = {Zhang, Jiaying and Robinson, Derek T.},
  title   = {Replication of an agent-based model using the {Replication Standard}},
  journal = {Environmental Modelling \& Software},
  year    = {2021},
  volume  = {139},
  pages   = {105016},
  doi     = {10.1016/j.envsoft.2021.105016},
}

@article{Epstein2009,
  author  = {Epstein, Joshua M.},
  title   = {Modelling to contain pandemics},
  journal = {Nature},
  year    = {2009},
  volume  = {460},
  number  = {7256},
  pages   = {687--687},
  doi     = {10.1038/460687a},
}

@inproceedings{Macal2014,
  author    = {Macal, Charles and North, Michael},
  title     = {Introductory tutorial: Agent-based modeling and simulation},
  booktitle = {2014 Winter Simulation Conference (WSC)},
  year      = {2014},
  pages     = {6--20},
  doi       = {10.1109/WSC.2014.7019874},
  publisher = {{IEEE}},
}

@book{North2007,
  author    = {North, Michael J. and Macal, Charles M.},
  title     = {Managing Business Complexity: Discovering Strategic Solutions with Agent-Based Modeling and Simulation},
  year      = {2007},
  publisher = {Oxford University Press},
  address   = {New York},
}

@book{Batty2007,
  author    = {Batty, Michael},
  title     = {Cities and Complexity: Understanding Cities with Cellular Automata, Agent-Based Models, and Fractals},
  year      = {2007},
  publisher = {The MIT Press},
  address   = {Cambridge, Massachusetts},
}

@Article{Caiani2016,
 author  = {Alessandro, Caiani and Antoine, Godin and Eugenio, Caverzasi and Mauro, Gallegati and Stephen, Kinsella and Joseph E., Stiglitz},
 journal  = {Journal of Economic Dynamics and Control},
 title   = {Agent based-stock flow consistent macroeconomics: Towards a benchmark model},
 year   = {2016},
 month   = aug,
 pages   = {375--408},
 volume  = {69},
 doi    = {10.1016/j.jedc.2016.06.001},
 publisher = {Elsevier {BV}},
}

@Article{Mahmood2020,
 author  = {Imran Mahmood and Hamid Arabnejad and Diana Suleimenova and Isabel Sassoon and Alaa Marshan and Alan Serrano-Rico and Panos Louvieris and Anastasia Anagnostou and Simon J E Taylor and David Bell and Derek Groen},
 journal  = {Journal of Simulation},
 title   = {{FACS}: A geospatial agent-based simulator for analysing {COVID}-19 spread and public health measures on local regions},
 year   = {2020},
 month   = aug,
 number  = {4},
 pages   = {355--373},
 volume  = {16},
 doi    = {10.1080/17477778.2020.1800422},
 publisher = {Informa {UK} Limited},
}

@Article{Siebers2010,
 author  = {P O Siebers and C M Macal and J Garnett and D Buxton and M Pidd},
 journal  = {Journal of Simulation},
 title   = {Discrete-event simulation is dead, long live agent-based simulation!},
 year   = {2010},
 month   = sep,
 number  = {3},
 pages   = {204--210},
 volume  = {4},
 doi    = {10.1057/jos.2010.14},
 publisher = {Informa {UK} Limited},
}

@Article{Bonzelet2022,
 author  = {Sabrina Bonzelet},
 journal  = {International Journal of Production Economics},
 title   = {How increasing relative risk aversion affects retailer orders under coordinating contracts},
 year   = {2022},
 month   = sep,
 pages   = {108500},
 volume  = {251},
 doi    = {10.1016/j.ijpe.2022.108500},
 publisher = {Elsevier {BV}},
}

@Article{Guan2018,
 author  = {Peiqiu Guan and Jing Zhang and Vineet M. Payyappalli and Jun Zhuang},
 journal  = {Decision Analysis},
 title   = {Modeling and validating public{\textendash}private partnerships in disaster management},
 year   = {2018},
 month   = jun,
 number  = {2},
 pages   = {55--71},
 volume  = {15},
 doi    = {10.1287/deca.2017.0361},
 publisher = {Institute for Operations Research and the Management Sciences ({INFORMS})},
}

@Article{Tian2022,
 author  = {Chen Tian and Tiaojun Xiao and Jennifer Shang},
 journal  = {European Journal of Operational Research},
 title   = {Channel differentiation strategy in a dual-channel supply chain considering free riding behavior},
 year   = {2022},
 month   = sep,
 number  = {2},
 pages   = {473--485},
 volume  = {301},
 doi    = {10.1016/j.ejor.2021.10.034},
 publisher = {Elsevier {BV}},
}

@Article{Dosi2006,
 author  = {Giovanni, Dosi and Giorgio, Fagiolo and Andrea, Roventini},
 journal  = {Computational Economics},
 title   = {An evolutionary model of endogenous business cycles},
 year   = {2006},
 month   = may,
 number  = {1},
 pages   = {3--34},
 volume  = {27},
 doi    = {10.1007/s10614-005-9014-2},
 publisher = {Springer Science and Business Media {LLC}},
}

@article{he2025hav,
  author = {He, Zhou and Song, Qi and Lian, Junchao and Liu, Yiming},
  title = {Towards Standardizing Validation Practices in Agent-Based Modeling: A Hierarchical ABM Validation Framework},
  journal = {ACM Transactions on Modeling and Computer Simulation},
  year = {2026},
  volume = {36},
  number = {1},
  pages = {1--25},
  publisher = {ACM},
  doi = {10.1145/3769857}
}

@Article{Tracy2018,
 author  = {Melissa Tracy and Magdalena Cerd{\'{a}} and Katherine M. Keyes},
 journal  = {Annual Review of Public Health},
 title   = {Agent-based modeling in public health: Current applications and future directions},
 year   = {2018},
 month   = apr,
 number  = {1},
 pages   = {77--94},
 volume  = {39},
 doi    = {10.1146/annurev-publhealth-040617-014317},
 publisher = {Annual Reviews},
}

@article{Arrow1951,
  author  = {Arrow, Kenneth J. and Harris, Theodore and Marschak, Jacob},
  title   = {Optimal Inventory Policy},
  journal = {Econometrica},
  year    = {1951},
  volume  = {19},
  number  = {3},
  pages   = {250--272},
  doi     = {10.2307/1906813},
}

@incollection{McFadden1974,
  author    = {McFadden, Daniel},
  title     = {Conditional Logit Analysis of Qualitative Choice Behavior},
  booktitle = {Frontiers in Econometrics},
  editor    = {Zarembka, Paul},
  publisher = {Academic Press},
  address   = {New York},
  year      = {1974},
  pages     = {105--142},
}

@article{Barabasi1999,
  author  = {Barab{\'a}si, Albert-L{\'a}szl{\'o} and Albert, R{\'e}ka},
  title   = {Emergence of Scaling in Random Networks},
  journal = {Science},
  year    = {1999},
  volume  = {286},
  number  = {5439},
  pages   = {509--512},
  doi     = {10.1126/science.286.5439.509},
}

@article{ErdosRenyi1959,
  author  = {Erd{\H{o}}s, Paul and R{\'e}nyi, Alfr{\'e}d},
  title   = {On Random Graphs~I},
  journal = {Publicationes Mathematicae Debrecen},
  year    = {1959},
  volume  = {6},
  pages   = {290--297},
}

@book{Nelson1982,
  author    = {Nelson, Richard R. and Winter, Sidney G.},
  title     = {An Evolutionary Theory of Economic Change},
  publisher = {Belknap Press of Harvard University Press},
  address   = {Cambridge, MA},
  year      = {1982},
}

@article{Epstein2002,
  author  = {Epstein, Joshua M.},
  title   = {Modeling Civil Violence: An Agent-Based Computational Approach},
  journal = {Proceedings of the National Academy of Sciences},
  volume  = {99},
  number  = {Suppl. 3},
  pages   = {7243--7250},
  year    = {2002},
  doi     = {10.1073/pnas.092080199}
}

@misc{Novak2006,
  author       = {Novak, Michael and Wilensky, Uri},
  title        = {NetLogo {Wolf Sheep Stride Inheritance} model},
  howpublished = {Center for Connected Learning and Computer-Based Modeling,
                  Northwestern University, Evanston, IL},
  year         = {2006},
  note         = {NetLogo 6.x; CC BY-NC-SA 3.0}
}

@article{Abar2017,
  author  = {Abar, Sameera and Theodoropoulos, Georgios K. and Lemarinier, Pierre and O'Hare, Gregory M. P.},
  title   = {Agent Based Modelling and Simulation tools: A review of the state-of-art software},
  journal = {Computer Science Review},
  year    = {2017},
  volume  = {24},
  pages   = {13--33},
  doi     = {10.1016/j.cosrev.2017.03.001},
}

@article{Xu2017,
  author  = {Xu, Yadong and Cai, Wentong and Aydt, Heiko and Lees, Michael and Zehe, Daniel},
  title   = {Relaxing Synchronization in Parallel Agent-Based Road Traffic Simulation},
  journal = {ACM Transactions on Modeling and Computer Simulation},
  year    = {2017},
  volume  = {27},
  number  = {2},
  pages   = {14:1--14:24},
  doi     = {10.1145/2994143},
}

@article{Li2017,
  author  = {Li, Xiaosong and Cai, Wentong and Turner, Stephen J.},
  title   = {Cloning Agent-Based Simulation},
  journal = {ACM Transactions on Modeling and Computer Simulation},
  year    = {2017},
  volume  = {27},
  number  = {2},
  pages   = {15:1--15:24},
  doi     = {10.1145/3013529},
}

\appendix
\onecolumn
\newgeometry{margin=1.8cm}  

\section*{Supplementary Material}
\label{sec:supplementary}

\noindent This appendix contains the full eight-table VISA specifications, the
19-rule consistency reports, and the reproduction figures for the three
external, independently authored agent-based models studied in the paper
(Rebellion; Wolf Sheep Stride Inheritance; AgedCareContactModel), together with
the three LLM-executable skills that operationalize the protocol. Each
specification below is the exact artifact on which the \texttt{visa-check} skill
reported 19/19~PASS; it is reproduced here verbatim. Runnable Python
reproductions, raw CSV outputs, the NetLogo reference trajectory, and the
AnyLogic extractor are on the companion GitHub repository
(\url{https://github.com/AgentLabCn/visa}).

\section{Model 1 --- Rebellion (reproduced)}
\label{sec:rebellion}

\begin{figure}[H]
  \centering
  \includegraphics[width=\linewidth]{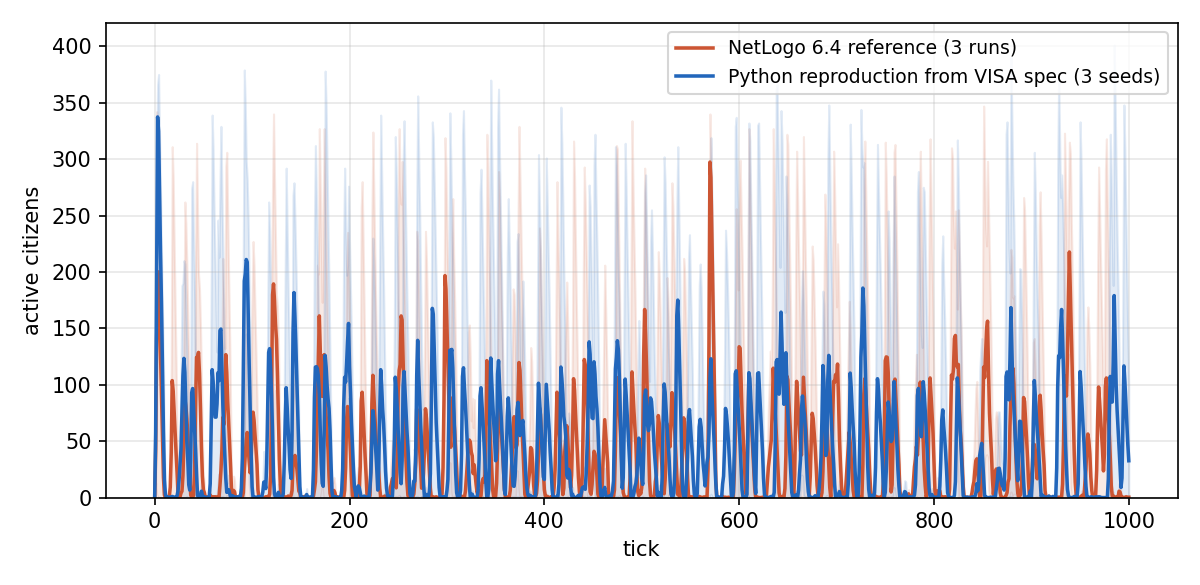}
  \caption{Rebellion reproduction --- active-citizen count over time. The
  NetLogo~6.4 reference (3 runs) and the Python re-implementation generated from
  the VISA specification (3 seeds) are shown as min--max bands with solid means.
  The two use different RNGs, so trajectories are not bit-identical, but the
  punctuated-equilibrium signature is reproduced cross-language. T8 v1--v3:
  PASS.}
  \label{fig:reb}
\end{figure}

Epstein (2002) civil-violence model, as implemented in the NetLogo Rebellion
model (Wilensky 2004). Reproduced cross-language (NetLogo $\to$ Python). The
specification below passes 19/19 rules; the reproduction reproduces the
punctuated-equilibrium signature (Fig.~\ref{fig:reb}).

{\small 
\subsection{VISA Specification --- Rebellion Model (Exp1)}

\textbf{Model.} Civil-violence rebellion model of Epstein (2002), \emph{Modeling civil
violence: An agent-based computational approach}, PNAS 99(suppl. 3), as implemented in
Wilensky (2004) as the \textbf{NetLogo Rebellion model}, shipped in the
\textbf{NetLogo Models Library} (Sample Models \ensuremath{\rightarrow} Social Science \ensuremath{\rightarrow} Rebellion; NetLogo 6.x,
CC BY-NC-SA 3.0).

\textbf{Reproduction goal.} Reproduce the \emph{punctuated-equilibrium /
periodic-rebellion} signature in the active-count time series (Epstein 2002, Fig. 1) by
re-implementing the NetLogo model in Python directly from this VISA specification. The
target is the qualitative signature --- structurally determined and robust to the random
seed --- not a bit-exact trajectory.

\textbf{Authoring note (space).} The model runs on a 40\ensuremath{\times}40 \emph{toroidal} grid with
radius-\(V\) metric sensing. Following VISA's principle that a spatial structure mediating
agent interaction should be an \textbf{explicit active agent}, the grid is modelled as a
\textbf{Space agent} \(\mathcal{G}\) that owns the world geometry (width, height, and the
two edge-wrap flags) and whose function (f5) answers vision-range queries: given an
agent's position and vision radius it returns the list of all agents within toroidal
distance \(V\). This list is an endogenous variable of each citizen and cop
(\(\mathbf{n}_{i,t}\)), refreshed every tick by f5. The vision radius \(V\) itself belongs
to the \emph{sensing} agents (citizen/cop), not to the Grid --- the Grid only reads it as
a query argument. This mirrors the running example, where SensingNet is an explicit Space
agent that maintains the network topology; here the Space agent maintains the
\emph{metric} topology. (VISA reserves implicit space only for models with no spatial
interaction; Rebellion is not such a model.)

\textbf{Authoring note (what the Environment is --- and is not).} In VISA the Environment
agent has two duties: (i) manage other agent instances (here: spawn the fixed citizen/cop
populations at setup, governed by the initial densities); and (ii) compute model-level
statistics (here: the active / jailed / quiet counts). Every parameter that conceptually
belongs to a \emph{decision-maker} is attributed to that decision-maker, not dumped into
the Environment: the citizen's grievance/arrest parameters (\(L,\tau,k,\mu\)) and its
sensing radius (\(V\)) sit on the Citizen, the sentence length (\(T_{\max}\)) and sensing
radius (\(V\)) sit on the Cop, and the world geometry (\(W_x,W_y,w_x,w_y\)) sits on the
Grid. This keeps each agent's T2 block a faithful list of \emph{its own} state.

\textbf{Notation.} Sets calligraphic
(\(\mathcal{E},\mathcal{G},\mathcal{A},\mathcal{P}\)); instances lowercase + subscript;
fixed counts \(N\); exogenous params uppercase (\(L, V, T_{\max}, k, \tau, W_x, W_y\));
endogenous lowercase + subscript with a time index
(\(a_{i,t}, j_{i,t}, x_{i,t}, y_{i,t}\)); the time index \(t\) marks a variable as
endogenous and may be suppressed in prose for readability.

\begin{center}\rule{0.5\linewidth}{0.5pt}\end{center}

\subsubsection{T1 --- Agent}

\bgroup
\setlength{\tabcolsep}{4pt}
\renewcommand{\arraystretch}{1.12}
\begin{longtable}{@{}>{\raggedright\arraybackslash}p{1.9cm}
                    >{\raggedright\arraybackslash}p{1.0cm}
                    >{\raggedright\arraybackslash}p{1.7cm}
                    >{\raggedright\arraybackslash}p{2.4cm}
                    >{\raggedright\arraybackslash}p{6.5cm}
                    >{\raggedright\arraybackslash}p{1.8cm}@{}}
\toprule
\textbf{Name} & \textbf{Set} & \textbf{Instances} & \textbf{Category} & \textbf{Description} & \textbf{Quantity} \\
\midrule
Environment & $\mathcal{E}$ & $e$ & Environment & Spawns the fixed populations at setup; computes output aggregates & $N_{\mathcal{E}}=1$ \\
Grid & $\mathcal{G}$ & $g$ & Space & Toroidal grid; owns world geometry; answers vision-range spatial queries & $N_{\mathcal{G}}=1$ \\
Citizen & $\mathcal{A}$ & $a_1,a_2,\ldots$ & Decision-maker & Decides to rebel or stay quiet based on grievance and arrest risk & $N_{\mathcal{A}}$ (fixed; set by init-agent-density, see T6a) \\
Cop & $\mathcal{P}$ & $p_1,p_2,\ldots$ & Decision-maker & Patrols; arrests a random active citizen within vision & $N_{\mathcal{P}}$ (fixed; set by init-cop-density, see T6a) \\
\bottomrule
\end{longtable}
\egroup

Populations are \textbf{fixed} (jail is a state, not removal) \ensuremath{\rightarrow} all decision-maker
quantities use \(N\), satisfying the population-dynamics rule trivially (no create/remove
functions). The Environment and Grid are singletons (\(=1\)); the concrete citizen/cop
counts and the grid dimensions are exogenous inputs (T6a).

\begin{center}\rule{0.5\linewidth}{0.5pt}\end{center}

\subsubsection{T2 --- Variable}

\bgroup
\setlength{\tabcolsep}{4pt}
\renewcommand{\arraystretch}{1.12}
\begin{longtable}{@{}>{\raggedright\arraybackslash}p{2.7cm}
                    >{\raggedright\arraybackslash}p{1.4cm}
                    >{\raggedright\arraybackslash}p{1.5cm}
                    >{\raggedright\arraybackslash}p{1.7cm}
                    >{\raggedright\arraybackslash}p{1.1cm}
                    >{\raggedright\arraybackslash}p{0.8cm}
                    >{\raggedright\arraybackslash}p{6.3cm}@{}}
\toprule
\textbf{Variable} & \textbf{Symbol} & \textbf{Type} & \textbf{Data type} & \textbf{Value} & \textbf{Unit} & \textbf{Description} \\
\midrule
\multicolumn{7}{@{}l}{\textbf{\underline{Environment $\mathcal{E}$}}} \\
Init agent density & --- & Exog.-homo. & Float & Input & \% & used only at setup to spawn citizens \\
Init cop density & --- & Exog.-homo. & Float & Input & \% & used only at setup to spawn cops \\
Active count & $A_t$ & Endog. & Integer & f4 & agents & \# active citizens \\
Jailed count & $J_t$ & Endog. & Integer & f4 & agents & \# jailed citizens \\
Quiet count & $Q_t$ & Endog. & Integer & f4 & agents & \# quiet citizens \\
\midrule
\multicolumn{7}{@{}l}{\textbf{\underline{Grid $\mathcal{G}$ (Space)}}} \\
Width & $W_x$ & Exog.-homo. & Integer & Input & patches & grid columns; toroidal extent \\
Height & $W_y$ & Exog.-homo. & Integer & Input & patches & grid rows; toroidal extent \\
Wrap left--right & $w_x$ & Exog.-homo. & Boolean & Input & --- & horizontal edges connected \\
Wrap top--bottom & $w_y$ & Exog.-homo. & Boolean & Input & --- & vertical edges connected \\
\midrule
\multicolumn{7}{@{}l}{\textbf{\underline{Citizen $\mathcal{A}$}}} \\
Risk aversion & $R_i$ & Exog.-hetero. & Float & Input & --- & drawn from a uniform distribution; fixed at birth \\
Perceived hardship & $H_i$ & Exog.-hetero. & Float & Input & --- & drawn from a uniform distribution; fixed at birth \\
Vision radius & $V$ & Exog.-homo. & Integer & Input & patches & sensing radius; passed to f5 as a query argument \\
Government legitimacy & $L$ & Exog.-homo. & Float & Input & --- & enters grievance $g_{i,t}=H_i(1-L)$ \\
Rebellion threshold & $\tau$ & Exog.-homo. & Float & Input & --- & $g{-}RP$ threshold for rebelling \\
Arrest constant & $k$ & Exog.-homo. & Float & Input & --- & scale in the arrest-probability formula \\
Movement? & $\mu$ & Exog.-homo. & Boolean & Input & --- & whether free citizens move \\
x-coordinate & $x_{i,t}$ & Endog. & Integer & f1 & patch & updated by f1 \\
y-coordinate & $y_{i,t}$ & Endog. & Integer & f1 & patch & updated by f1 \\
Active flag & $a_{i,t}$ & Endog.-dec. & Boolean & f1 & --- & true if openly rebelling \\
Jail term & $j_{i,t}$ & Endog. & Integer & f2; f3 (ext) & ticks & set by arresting cop, decremented each tick \\
Agents in vision & $\mathbf{n}_{i,t}$ & Endog. & List[$\mathcal{A}{\cup}\mathcal{P}$] & f5 (ext) & agents & all agents within toroidal distance $V$ \\
\midrule
\multicolumn{7}{@{}l}{\textbf{\underline{Cop $\mathcal{P}$}}} \\
Max jail term & $T_{\max}$ & Exog.-homo. & Integer & Input & ticks & upper bound on a sentence \\
Vision radius & $V$ & Exog.-homo. & Integer & Input & patches & sensing radius; shared with Citizen (declared on both) \\
x-coordinate & $x_{i,t}$ & Endog. & Integer & f3 & patch & updated by f3 \\
y-coordinate & $y_{i,t}$ & Endog. & Integer & f3 & patch & updated by f3 \\
Agents in vision & $\mathbf{n}_{i,t}$ & Endog. & List[$\mathcal{A}{\cup}\mathcal{P}$] & f5 (ext) & agents & all agents within toroidal distance $V$ \\
\bottomrule
\end{longtable}
\egroup

\emph{Transient quantities (not stored state):} grievance \(g_{i,t}=H_i(1-L)\) and
estimated arrest probability \(P_{i,t}=1-\exp(-k\lfloor c_{i,t}/(1+\alpha_{i,t})\rfloor)\)
are recomputed inside f1 each tick from \(\mathbf{n}_{i,t}\), where \(c_{i,t}\) = cops in
vision and \(\alpha_{i,t}\) = active citizens in vision.

\begin{center}\rule{0.5\linewidth}{0.5pt}\end{center}

\subsubsection{T3 --- Sensing}

The Grid (f5) realises the spatial modality: an agent ``senses'' another iff the
\textbf{toroidal Euclidean distance} between their positions is \(\le V\) (vision). Rows =
observers (all non-Passive agents); columns = observed.

\bgroup
\setlength{\tabcolsep}{5pt}
\renewcommand{\arraystretch}{1.15}
\begin{longtable}{@{}>{\raggedright\arraybackslash}p{2.7cm}
                    >{\centering\arraybackslash}p{1.6cm}
                    >{\centering\arraybackslash}p{1.6cm}
                    >{\centering\arraybackslash}p{3.0cm}
                    >{\centering\arraybackslash}p{2.6cm}@{}}
\toprule
\textbf{Observer} & $\mathcal{E}$ & $\mathcal{G}$ & $\mathcal{A}$ & $\mathcal{P}$ \\
\midrule
$\mathcal{E}$ (Env) & $\emptyset$ & $\emptyset$ & $a_{i,t}, j_{i,t}$ & $\emptyset$ \\
$\mathcal{G}$ (Grid) & $\emptyset$ & $\emptyset$ & $x_{i,t}, y_{i,t}$ & $x_{i,t}, y_{i,t}$ \\
$\mathcal{A}$ (Citizen) & $\emptyset$ & $\emptyset$ & $a_{j,t}, j_{j,t}, x_{j,t}, y_{j,t}$ & $x_{j,t}, y_{j,t}$ \\
$\mathcal{P}$ (Cop) & $\emptyset$ & $\emptyset$ & $a_{j,t}, j_{j,t}, x_{j,t}, y_{j,t}$ & $x_{j,t}, y_{j,t}$ \\
\bottomrule
\end{longtable}
\egroup

Reading the matrix: the Environment reads every citizen's active/jail flags to form the
aggregates; the Grid reads every citizen and cop position to compute the vision lists
\(\mathbf{n}_{i,t}\) that it pushes to each agent; a citizen reads the attributes of the
peers and cops that f5 has placed in its \(\mathbf{n}_{i,t}\) (active flag for the
cop-to-active ratio in \(P_{i,t}\); jail-term and position to pick a move target free of
free agents; cop positions to avoid cop-occupied patches); a cop reads citizens' active
flag and position (to find and move to a suspect) and peer cop positions (move validity).
All citizen/cop parameters (\(L,\tau,k,\mu,T_{\max},V\)) are self-attributes and so do not
appear as cross-agent sensing.

\begin{center}\rule{0.5\linewidth}{0.5pt}\end{center}

\subsubsection{T4 --- Internal Function}

Complex updates are given as numbered equations below the table and referenced by their
equation number; the Self-state Update and External Effect columns record \emph{which
endogenous variable} is written and the equation that governs it.

\bgroup
\setlength{\tabcolsep}{4pt}
\renewcommand{\arraystretch}{1.12}
\begin{longtable}{@{}>{\raggedright\arraybackslash}p{0.7cm}
                    >{\raggedright\arraybackslash}p{2.2cm}
                    >{\raggedright\arraybackslash}p{2.2cm}
                    >{\raggedright\arraybackslash}p{3.3cm}
                    >{\raggedright\arraybackslash}p{3.7cm}
                    >{\raggedright\arraybackslash}p{2.8cm}
                    >{\raggedright\arraybackslash}p{1.0cm}@{}}
\toprule
\textbf{ID} & \textbf{Function} & \textbf{Method} & \textbf{Decision basis} & \textbf{Self-state update} & \textbf{External effect} & \textbf{Ref.} \\
\midrule
\multicolumn{7}{@{}l}{\textbf{\underline{Grid $\mathcal{G}$}}} \\
f5 & query\_vision & Spatial query & $(x_{i,t},y_{i,t})\,\forall i$ (T3); $V$ & --- & $\mathbf{n}_{i,t}\leftarrow$ \eqref{eq:reb-vision}, $\forall i\in\mathcal{A}\cup\mathcal{P}$ & --- \\
\midrule
\multicolumn{7}{@{}l}{\textbf{\underline{Citizen $\mathcal{A}$}}} \\
f1 & rebel\_or\_move & Conditional + Random & $\mathbf{n}_{i,t}$; $H_i,R_i,L,\tau,k,\mu$ (self) & $x_{i,t},y_{i,t},a_{i,t}$ via \eqref{eq:reb-active} & --- & \citep{Epstein2002} \\
f2 & decrement\_jail & Decrement & $j_{i,t}$ (self) & $j_{i,t}\leftarrow j_{i,t}-1$ & --- & --- \\
\midrule
\multicolumn{7}{@{}l}{\textbf{\underline{Cop $\mathcal{P}$}}} \\
f3 & patrol\_and\_arrest & Random + Random selection & $\mathbf{n}_{i,t}$; $T_{\max}$ (self) & $x_{i,t},y_{i,t}$ via \eqref{eq:reb-copmove} & $a_{s,t},j_{s,t}$ via \eqref{eq:reb-enforce} & \citep{Epstein2002} \\
\midrule
\multicolumn{7}{@{}l}{\textbf{\underline{Environment $\mathcal{E}$}}} \\
f4 & compute\_counts & Count & $a_{i,t}, j_{i,t}\,\forall i\in\mathcal{A}$ (T3) & $A_t,J_t,Q_t$ via \eqref{eq:reb-counts} & --- & --- \\
\bottomrule
\end{longtable}
\egroup

\textbf{Behavioural equations.} The Grid's vision query is

\begin{equation}
\mathbf{n}_{i,t} = \big\{\,a'\in\mathcal{A}\cup\mathcal{P}\;:\;
d_{\mathrm{tor}}\big((x_{i,t},y_{i,t}),\,(x_{a',t},y_{a',t})\big)\le V\,\big\}.
\label{eq:reb-vision}
\end{equation}

A free citizen (\(j_{i,t}=0\)) evaluates grievance and arrest probability,

\begin{equation}
g_{i,t} = H_i(1-L),\qquad
P_{i,t}=1-\exp\!\Big(\!-k\Big\lfloor\tfrac{c_{i,t}}{1+\alpha_{i,t}}\Big\rfloor\Big),
\label{eq:reb-gp}
\end{equation}

with \(c_{i,t}=|\mathbf{n}_{i,t}\cap\mathcal{P}|\) (cops in vision) and
\(\alpha_{i,t}=|\{a'\in\mathbf{n}_{i,t}\cap\mathcal{A}:a_{a',t}\}|\) (active citizens in
vision), and then rebels and/or moves:

\begin{equation}
a_{i,t} \leftarrow \mathbf{1}\big[\,g_{i,t}-R_i\,P_{i,t}>\tau\,\big],
\qquad
(x_{i,t},y_{i,t})\leftarrow \mathrm{Unif}\,\mathcal{M}_i\ \ (\text{if }\mu\wedge j_{i,t}{=}0).
\label{eq:reb-active}
\end{equation}

Here
\(\mathcal{M}_i=\{(x,y):d_{\mathrm{tor}}((x,y),(x_{i,t},y_{i,t}))\le V,\ \text{no cop and no free citizen at }(x,y)\}\)
is the set of admissible move patches in vision. The cop moves likewise and arrests one
active citizen:

\begin{equation}
(x_{i,t},y_{i,t})\leftarrow \mathrm{Unif}\,\mathcal{M}_i,\qquad
s\sim\mathrm{Unif}\{a'\in\mathbf{n}_{i,t}\cap\mathcal{A}:a_{a',t}\}.
\label{eq:reb-copmove}
\end{equation}

\begin{equation}
a_{s,t}\leftarrow\text{false},\qquad
j_{s,t}\sim U\{0,\ldots,T_{\max}{-}1\}.
\label{eq:reb-enforce}
\end{equation}

The Environment's aggregates are

\begin{equation}
A_t=\big|\{a'\in\mathcal{A}:a_{a',t}\}\big|,\;\;
J_t=\big|\{a'\in\mathcal{A}:j_{a',t}{>}0\}\big|,\;\;
Q_t=N_{\mathcal{A}}-A_t-J_t .
\label{eq:reb-counts}
\end{equation}

The \textbf{floor} in \(P_{i,t}\) \eqref{eq:reb-gp} is the well-known deviation from
Epstein's original formula: without it, the model does \emph{not} exhibit punctuated
equilibrium. Its effect is \(P_{i,t}\approx 0\) when cops are outnumbered by active
citizens, \(\approx 0.99\) otherwise.

\begin{center}\rule{0.5\linewidth}{0.5pt}\end{center}

\subsubsection{T5 --- Associated Data}

{\def\LTcaptype{none} 
\begin{longtable}[]{@{}lllllllll@{}}
\toprule\noalign{}
ID & Title & Type & Temporal & Source & Collection & Pre-processing & \#Rec. & Avail. \\
\midrule\noalign{}
\endhead
\bottomrule\noalign{}
\endlastfoot
\end{longtable}
}

\emph{(empty --- the model has no stand-alone dataset)}

The model has \textbf{no stand-alone dataset}: it is neither backed by an empirical
dataset nor by a synthetic one. \(R_i,H_i\) are drawn from a uniform distribution in the
sampling step itself, and every structural parameter is an author-assumed value (the
canonical settings reported in Epstein 2002). With no separable data record backing any
parameter, T5 is empty; accordingly every T6a entry has Data source \(=\) Author (no
\(d\)-ID to cite).

\begin{center}\rule{0.5\linewidth}{0.5pt}\end{center}

\subsubsection{T6 --- Input and Output}

\paragraph{(a) Input}

\bgroup
\setlength{\tabcolsep}{4pt}
\renewcommand{\arraystretch}{1.12}
\begin{longtable}{@{}>{\raggedright\arraybackslash}p{1.8cm}
                    >{\raggedright\arraybackslash}p{4.3cm}
                    >{\raggedright\arraybackslash}p{2.0cm}
                    >{\raggedright\arraybackslash}p{1.9cm}
                    >{\raggedright\arraybackslash}p{3.5cm}
                    >{\raggedright\arraybackslash}p{2.0cm}@{}}
\toprule
\textbf{Symbol} & \textbf{Value / Distribution} & \textbf{Data source} & \textbf{Derivation} & \textbf{Algorithm} & \textbf{Ref.} \\
\midrule
\multicolumn{6}{@{}l}{\textbf{\underline{Environment $\mathcal{E}$}}} \\
init-agent-density & 70 (\% of patches) & Author & Assumed & --- & --- \\
init-cop-density & 4 (\% of patches) & Author & Assumed & --- & --- \\
\midrule
\multicolumn{6}{@{}l}{\textbf{\underline{Citizen $\mathcal{A}$}}} \\
$L$ & 0.82 & Author & Assumed & --- & --- \\
$\tau$ & 0.1 & Author & Assumed & --- & --- \\
$k$ & 2.3 & Author & Assumed & --- & --- \\
$\mu$ & true & Author & Assumed & --- & --- \\
$V$ & 7 & Author & Assumed & --- & --- \\
$R_i$ & $U(0,1)$ & Author & Computed & uniform sampling & --- \\
$H_i$ & $U(0,1)$ & Author & Computed & uniform sampling & --- \\
\midrule
\multicolumn{6}{@{}l}{\textbf{\underline{Cop $\mathcal{P}$}}} \\
$T_{\max}$ & 30 & Author & Assumed & --- & --- \\
\midrule
\multicolumn{6}{@{}l}{\textbf{\underline{Grid $\mathcal{G}$}}} \\
$W_x, W_y$ & 40, 40 & Author & Assumed & --- & --- \\
$w_x, w_y$ & true, true & Author & Assumed & --- & --- \\
\bottomrule
\end{longtable}
\egroup

\paragraph{(b) Output}

\bgroup
\setlength{\tabcolsep}{4pt}
\renewcommand{\arraystretch}{1.12}
\begin{longtable}{@{}>{\raggedright\arraybackslash}p{1.2cm}
                    >{\raggedright\arraybackslash}p{2.6cm}
                    >{\raggedright\arraybackslash}p{4.3cm}
                    >{\raggedright\arraybackslash}p{1.7cm}
                    >{\raggedright\arraybackslash}p{1.3cm}
                    >{\centering\arraybackslash}p{1.0cm}
                    >{\raggedright\arraybackslash}p{3.3cm}@{}}
\toprule
\textbf{Symbol} & \textbf{Indicator} & \textbf{Formula} & \textbf{Data type} & \textbf{Unit} & \textbf{Freq.} & \textbf{Desc.} \\
\midrule
$A_t$ & Active citizens & $\sum_{a_i\in\mathcal{A}}\mathbf{1}[a_{i,t}]$ & Integer & agents & 1 & openly rebelling \\
$J_t$ & Jailed citizens & $\sum_{a_i\in\mathcal{A}}\mathbf{1}[j_{i,t}>0]$ & Integer & agents & 1 & in jail \\
$Q_t$ & Quiet citizens & $N_{\mathcal{A}}-A_t-J_t$ & Integer & agents & 1 & inactive \& free \\
\bottomrule
\end{longtable}
\egroup

\begin{center}\rule{0.5\linewidth}{0.5pt}\end{center}

\subsubsection{T7 --- Schedule}

\paragraph{(a) Execution}

\bgroup
\setlength{\tabcolsep}{4pt}
\renewcommand{\arraystretch}{1.12}
\begin{longtable}{@{}>{\centering\arraybackslash}p{0.9cm}
                    >{\centering\arraybackslash}p{1.2cm}
                    >{\centering\arraybackslash}p{0.9cm}
                    >{\raggedright\arraybackslash}p{3.5cm}
                    >{\raggedright\arraybackslash}p{3.0cm}
                    >{\raggedright\arraybackslash}p{4.0cm}@{}}
\toprule
\textbf{Step} & \textbf{Agent} & \textbf{ID} & \textbf{Function} & \textbf{Exec. mode} & \textbf{Condition} \\
\midrule
1 & $\mathcal{G}$ & f5 & query\_vision & Synchronous & every tick \\
2 & $\mathcal{A}$ & f1 & rebel\_or\_move & Random-order & free citizen \\
2 & $\mathcal{P}$ & f3 & patrol\_and\_arrest & Random-order & --- \\
3 & $\mathcal{A}$ & f2 & decrement\_jail & Random-order & $j_{i,t}>0$ \\
4 & $\mathcal{E}$ & f4 & compute\_counts & Synchronous & every tick \\
\bottomrule
\end{longtable}
\egroup

Step 2's two rows are a \textbf{single interleaved} random-order activation (citizens and
cops shuffled together): each agent atomically performs its move then its rebel/arrest
decision before the next agent acts (first-mover faithful). Splitting them into two rows
only separates the two functions; the \textbf{Random-order} mode eliminates systematic
ordering bias. f5 (Step 1) refreshes every \(\mathbf{n}_{i,t}\) before any agent acts, so
all decision-makers see the start-of-tick configuration.

\paragraph{(b) Termination}

\bgroup
\setlength{\tabcolsep}{4pt}
\renewcommand{\arraystretch}{1.12}
\begin{longtable}{@{}>{\centering\arraybackslash}p{0.9cm}
                    >{\raggedright\arraybackslash}p{2.2cm}
                    >{\raggedright\arraybackslash}p{2.4cm}
                    >{\raggedright\arraybackslash}p{3.6cm}
                    >{\raggedright\arraybackslash}p{2.8cm}
                    >{\raggedright\arraybackslash}p{3.6cm}@{}}
\toprule
\textbf{ID} & \textbf{Indicator} & \textbf{Condition} & \textbf{Description} & \textbf{Value source/ref} & \textbf{Termination logic} \\
\midrule
c1 & $t$ & $t\ge 1000$ & Reproduction horizon reached & Author assumed & \multirow{1}{*}{Stop when $c_1$} \\
\bottomrule
\end{longtable}
\egroup

The model runs indefinitely; c1 fixes the reproduction run length.

\begin{center}\rule{0.5\linewidth}{0.5pt}\end{center}

\subsubsection{T8 --- Validation}

\bgroup
\setlength{\tabcolsep}{4pt}
\renewcommand{\arraystretch}{1.12}
\begin{longtable}{@{}>{\centering\arraybackslash}p{0.8cm}
                    >{\raggedright\arraybackslash}p{3.6cm}
                    >{\raggedright\arraybackslash}p{2.7cm}
                    >{\raggedright\arraybackslash}p{3.0cm}
                    >{\raggedright\arraybackslash}p{2.0cm}
                    >{\raggedright\arraybackslash}p{3.6cm}
                    >{\raggedright\arraybackslash}p{1.0cm}@{}}
\toprule
\textbf{ID} & \textbf{Validation object} & \textbf{Benchmark data} & \textbf{Method} & \textbf{Indicator} & \textbf{Passing cond.} & \textbf{Ref.} \\
\midrule
\multicolumn{7}{@{}l}{\textbf{\underline{Output level}}} \\
v1 & Active-count time series $A_t$ & Epstein 2002 Fig. 1 (qualitative) & Peak/episode detection + visual inspection & $A_t$ over $t$ & periodic rebellion episodes (peaks $\gg 0$) separated by quiescent troughs & \citep{Epstein2002} \\
v2 & Arrest-probability floor effect & the formula itself & unit check of $P_{i,t}$ & $P_{i,t}$ & $\approx 0$ when $c_{i,t}<\alpha_{i,t}$; $\approx 0.99$ when $c_{i,t}\ge\alpha_{i,t}$ & --- \\
v3 & Population conservation & identity $A_t{+}J_t{+}Q_t{=}N_{\mathcal{A}}$ & invariant check & $A_t{+}J_t{+}Q_t$ & $=N_{\mathcal{A}}$ at every $t$ & --- \\
\bottomrule
\end{longtable}
\egroup

v1 is the headline reproduction criterion (punctuated equilibrium).

\textbf{Author's note.} Epstein (2002) reports the punctuated-equilibrium result
(Fig.\textasciitilde1) but states no formal validation criteria. The three entries above
(v1--v3) are authored for this case study to illustrate the T8 validation table; they are
not transcribed from the source paper.

\begin{center}\rule{0.5\linewidth}{0.5pt}\end{center}

\subsubsection{Consistency check (19 rules)}

Within-table (r1--r4) and cross-table (r5--r19), per the revised VISA rule set.

{\def\LTcaptype{none} 
\begin{longtable}[]{@{}
  >{\raggedright\arraybackslash}p{(\linewidth - 4\tabcolsep) * \real{0.2842}}
  >{\centering\arraybackslash}p{(\linewidth - 4\tabcolsep) * \real{0.0632}}
  >{\raggedright\arraybackslash}p{(\linewidth - 4\tabcolsep) * \real{0.6526}}@{}}
\toprule\noalign{}
\begin{minipage}[b]{\linewidth}\raggedright
Rule
\end{minipage} & \begin{minipage}[b]{\linewidth}\centering
Res.
\end{minipage} & \begin{minipage}[b]{\linewidth}\raggedright
Note
\end{minipage} \\
\midrule\noalign{}
\endhead
\bottomrule\noalign{}
\endlastfoot
r1 Variable-type coverage \& time-indexing & PASS & all T2 rows carry a leaf type; every
endogenous variable carries the time index \(t\)
(\(x_{i,t},y_{i,t},a_{i,t},j_{i,t},\mathbf{n}_{i,t},A_t,J_t,Q_t\)) \\
r2 Function-ID uniqueness & PASS & f1--f5 unique \\
r3 Function productivity & PASS & f1 writes \(x_{i,t},y_{i,t},a_{i,t}\); f2 writes
\(j_{i,t}\); f3 writes pos + ext \(a_{s,t},j_{s,t}\); f4 writes \(A_t,J_t,Q_t\); f5 writes
ext \(\mathbf{n}_{i,t}\) \\
r4 Step--Exec.mode consistency & PASS & Step 1 = Synchronous (f5); Step 2 = Random-order
(f1,f3); Step 3 = Random-order (f2); Step 4 = Synchronous (f4) \\
r5 Same-type (peer) sensing & PASS & \(\mathcal{E},\mathcal{G}\) \(N{=}1\)\ensuremath{\rightarrow}\(\emptyset\)
(no peers); \(\mathcal{A},\mathcal{P}\) \(n{>}1\)\ensuremath{\rightarrow} peer sets
\(a_{j,t},j_{j,t},x_{j,t},y_{j,t}\) and \(x_{j,t},y_{j,t}\) (self implicit) \\
r6 Passive-agent implications & PASS & no Passive agents \\
r7 Observer-row completeness & PASS & \(\mathcal{E},\mathcal{G},\mathcal{A},\mathcal{P}\)
each have exactly one row \\
r8 Population-dynamics consistency & PASS & all quantities fixed (\(N\)); no create/remove
functions \\
r9 Active-agent function coverage & PASS & \(\mathcal{E}\)\ensuremath{\rightarrow}f4; \(\mathcal{G}\)\ensuremath{\rightarrow}f5;
\(\mathcal{A}\)\ensuremath{\rightarrow}f1,f2; \(\mathcal{P}\)\ensuremath{\rightarrow}f3 \\
r10 Variable observability & PASS & all vars sensed (spatially bounded by \(V\) via
\(\mathbf{n}_{i,t}\)) or self-read \\
r11 Endogenous-variable completeness & PASS & \(A_t,J_t,Q_t\)\ensuremath{\leftarrow}f4; citizen
\(x_{i,t},y_{i,t},a_{i,t}\)\ensuremath{\leftarrow}f1, \(j_{i,t}\)\ensuremath{\leftarrow}f2/f3(ext), \(\mathbf{n}_{i,t}\)\ensuremath{\leftarrow}f5(ext); cop
pos\ensuremath{\leftarrow}f3, \(\mathbf{n}_{i,t}\)\ensuremath{\leftarrow}f5(ext); every referenced f-ID exists \\
r12 Self-state-update validity & PASS & all Self-state vars are endogenous of the owning
agent \\
r13 External-effect validity & PASS & f3 ext writes citizen \(a_{s,t},j_{s,t}\); f5 ext
writes \(\mathbf{n}_{i,t}\) (endogenous of citizen/cop) \\
r14 Information-access validation & PASS & every Decision-basis var is a self-attribute or
T3-authorized (within vision \(V\), realised by \(\mathbf{n}_{i,t}\)) \\
r15 Input--output coverage (bidirectional) & PASS & every Input var has a T6a entry; every
T2 ``Input'' var has a value; \(A_t,J_t,Q_t\) computable from T2 \\
r16 Schedule coverage (bidirectional) & PASS & f1,f2,f3,f4,f5 all appear in T7a; every T7a
ID exists in T4 \\
r17 Termination-indicator source & PASS & c1 uses \(t\) (global time index, exempt) \\
r18 Validation-object coverage & PASS & v1\ensuremath{\rightarrow}\(A_t\); v2\ensuremath{\rightarrow}\(P_{i,t}\) (computed in f1);
v3\ensuremath{\rightarrow}\(A_t{+}J_t{+}Q_t\); all trace to T2/T6b \\
r19 Data-reference resolution & PASS & T5 is empty (no stand-alone dataset); all T6a
sources are Author; no empirical data references to resolve \\
\end{longtable}
}

\textbf{Summary: 19/19 PASS.}
}

\section{Model 2 --- Wolf Sheep Stride Inheritance (reproduced)}
\label{sec:wolfsheep}

\begin{figure}[H]
  \centering
  \includegraphics[width=\linewidth]{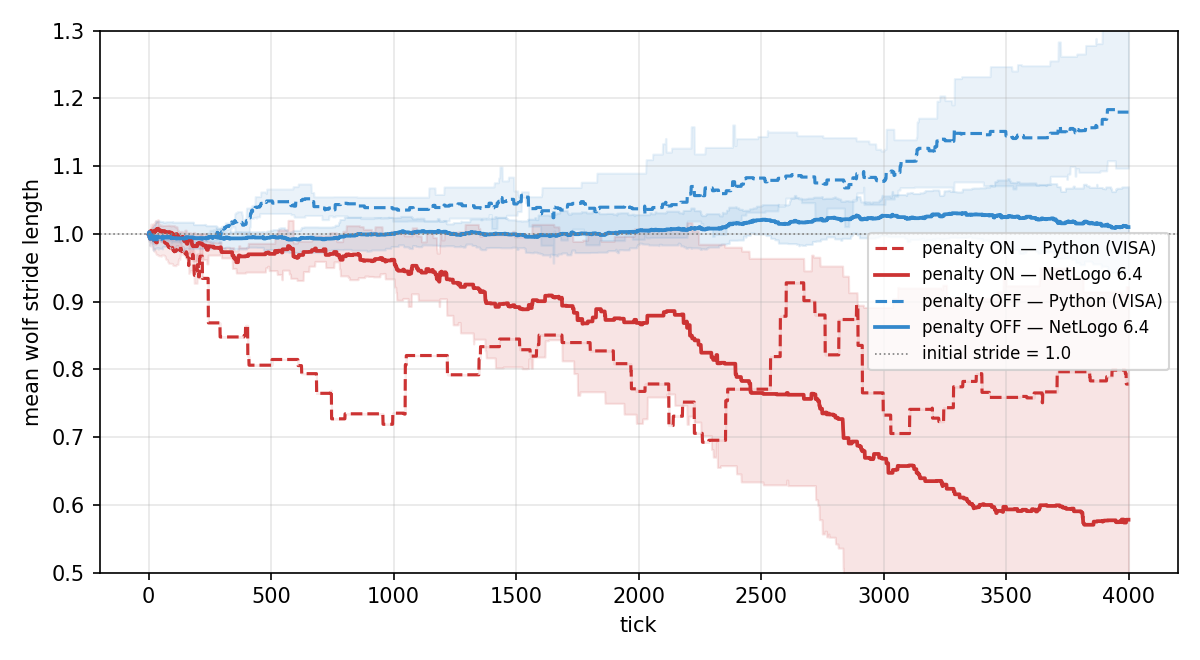}
  \caption{Wolf stride-length selection (default model) --- VISA reproduction
  evidence. NetLogo~6.4 reference (solid) vs.\ the Python re-implementation
  generated from the VISA specification (dashed), each a mean over 3 runs/seeds.
  Both platforms reproduce the causal reversal: penalty ON $\to$ wolves evolve a
  shorter stride; penalty OFF $\to$ a longer one. (T8 v1/v2.)}
  \label{fig:ws1}
\end{figure}

\begin{figure}[H]
  \centering
  \includegraphics[width=\linewidth]{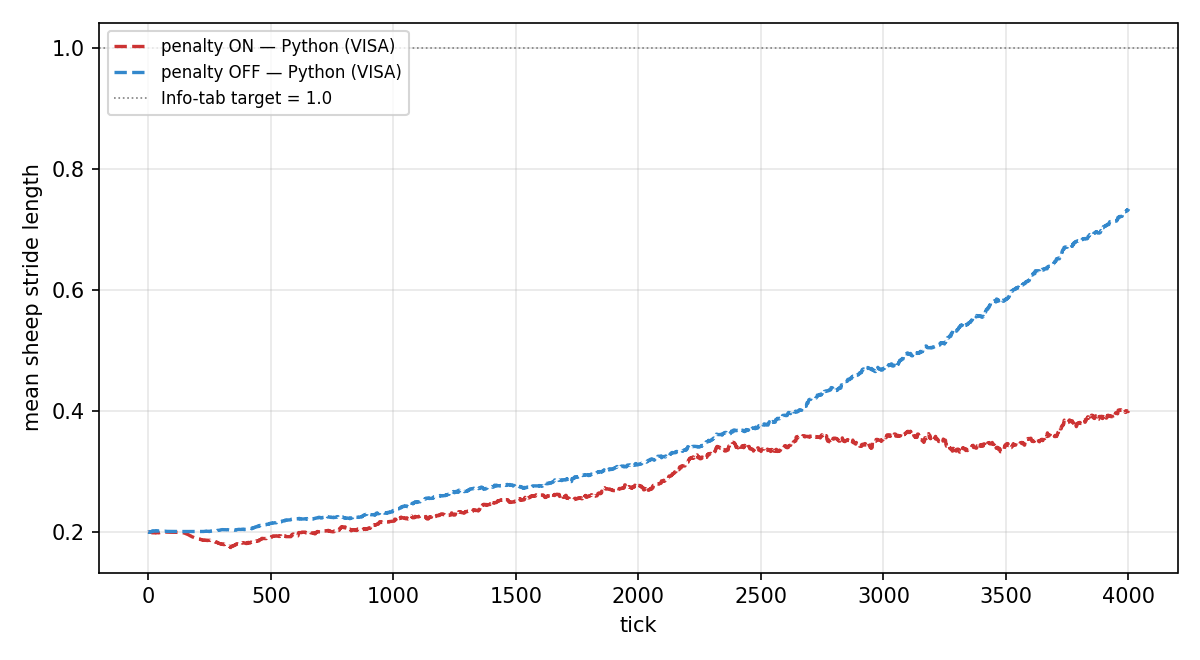}
  \caption{Sheep stride-length evolution (high-density sheep--grass subsystem)
  --- VISA reproduction evidence. NetLogo~6.4 (solid) vs.\ VISA-Python (dashed).
  Both platforms reproduce the divergence: without the penalty, sheep stride
  climbs faster; with it, the cost caps the climb. (T8 v1.)}
  \label{fig:ws2}
\end{figure}

\begin{figure}[H]
  \centering
  \includegraphics[width=\linewidth]{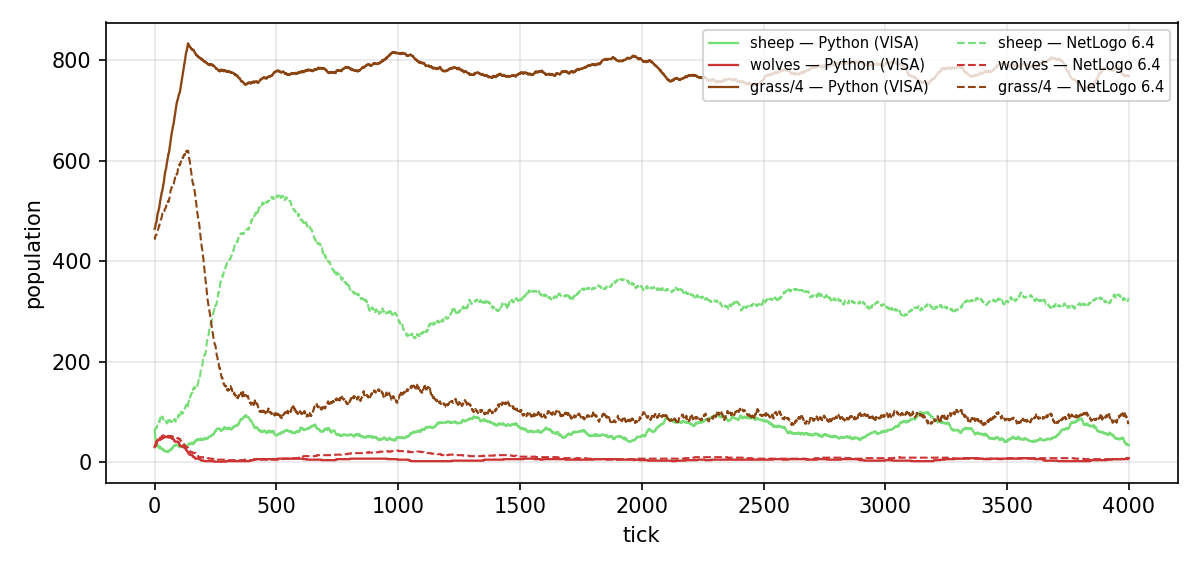}
  \caption{Predator--prey oscillation (default model, penalty on, coexistence)
  --- VISA reproduction evidence. The three populations---sheep (green), wolves
  (red), and grass/4 (brown)---are each plotted twice: the VISA-Python
  reproduction (solid) overlaid on the NetLogo~6.4 reference (dashed). Both
  platforms reproduce the same emergent dynamics---out-of-phase
  sheep--wolf--grass cycles with matching period and amplitude. (T8 v4.)}
  \label{fig:ws3}
\end{figure}

Novak \& Wilensky (2006); reproduced cross-language (NetLogo $\to$ Python). The
specification passes 19/19 rules. Every curve in
Figs.~\ref{fig:ws1}--\ref{fig:ws3} was produced by Python code generated
directly from the eight-table VISA specification via the \texttt{code} skill;
the reproduction target is the model's published Info-tab predictions, not a
bit-exact trajectory. T8 v2--v5 PASS, v1 directional.

{\small 
\subsection{VISA Specification --- Wolf Sheep Stride Inheritance Model (Exp2)}

\textbf{Model.} \emph{Wolf Sheep Stride Inheritance} --- Novak, M. \& Wilensky, U. (2006),
NetLogo Wolf Sheep Stride Inheritance model, shipped in the \textbf{NetLogo Models
Library} (Sample Models \ensuremath{\rightarrow} Biology \ensuremath{\rightarrow} Wolf Sheep Stride Inheritance; CCL, Northwestern
Univ.; NetLogo 6.x, CC BY-NC-SA 3.0) --- a variation on Wolf-Sheep Predation in which
stride length is an \textbf{inheritable, mutating} trait, so natural selection on movement
distance can be observed.

\textbf{Reproduction goal.} Reproduce the model's two signature results directly from this
specification (cross-language, NetLogo \ensuremath{\rightarrow} Python, qualitative --- not bit-exact):

\begin{enumerate}
\def\labelenumi{\arabic{enumi}.}
\tightlist
\item
  \textbf{Sheep mean stride converges to $\approx 1$} when the stride-length energy
  penalty is \textbf{on} (the Info tab's headline result: ``sheep typically converge on an
  average stride length close to 1'').
\item
  With the penalty \textbf{off}, sheep stride grows \textbf{without bound} (\(\gg 1\)) ---
  the counterfactual that isolates \emph{why} \(\approx 1\) is the equilibrium.
\end{enumerate}

Plus the generic predator-prey behaviour inherited from Wolf-Sheep Predation:
parameter-dependent extinction (which species survives) and Lotka--Volterra-style
population oscillations.

\textbf{Authoring note (space and grass).} The world is a 61\ensuremath{\times}61
\textbf{toroidal continuous} space (\texttt{min/max-pxcor\ =\ \ensuremath{\pm}30}, wrapping in both axes;
patch size 6). Two spatial structures that the original code folds into ``patches'' are
made into explicit VISA agents:

\begin{itemize}
\tightlist
\item
  a \textbf{Grid} (Space) agent \(\mathcal{G}\) that owns the world geometry (width,
  height, and the two edge-wrap flags), maintains the position relationships among the
  currently-alive animals, and answers \textbf{patch-co-location} queries --- the spatial
  primitive the model actually uses (an animal interacts with whatever is on \emph{its}
  patch after a forward move of length \(\ell\)). f10 writes, into each animal, the
  co-location list it needs (\(\gamma_{i,t}\) = grass-at-my-patch for a sheep;
  \(\mathbf{h}_{j,t}\) = sheep-at-my-patch for a wolf).
\item
  a \textbf{GrassFieldCell} agent \(\mathcal{F}\) --- one per patch --- whose only
  behaviour is to age its own regrowth clock and turn green (f9). This makes the grass
  field an active agent set rather than an Environment-owned matrix.
\end{itemize}

As in Exp1, this follows VISA's principle that a spatial structure mediating interaction
is an explicit Space agent (cf.~the running example's SensingNet).

\textbf{Notation.} Sets calligraphic (\(\mathcal{E},\mathcal{G},\mathcal{F},\mathcal{S},
\mathcal{W}\)); instances lowercase + subscript; variable quantities \(n\) (birth/death);
fixed \(N\); exogenous params/constants uppercase (\(M_{\max}, S_{\max},
G_f, r, \delta, W_x, W_y, \dots\)); endogenous lowercase + subscript with a time index
(\(e_{i,t}, x_{i,t}, y_{i,t}, \theta_{i,t}, g_{f,t}\)); the time index \(t\) marks a
variable as endogenous and may be suppressed in prose for readability.

\begin{center}\rule{0.5\linewidth}{0.5pt}\end{center}

\subsubsection{T1 --- Agent}

\bgroup
\setlength{\tabcolsep}{4pt}
\renewcommand{\arraystretch}{1.12}
\begin{longtable}{@{}>{\raggedright\arraybackslash}p{1.9cm}
                    >{\raggedright\arraybackslash}p{1.0cm}
                    >{\raggedright\arraybackslash}p{1.7cm}
                    >{\raggedright\arraybackslash}p{2.4cm}
                    >{\raggedright\arraybackslash}p{6.5cm}
                    >{\raggedright\arraybackslash}p{1.8cm}@{}}
\toprule
\textbf{Name} & \textbf{Set} & \textbf{Instances} & \textbf{Category} & \textbf{Description} & \textbf{Quantity} \\
\midrule
Environment & $\mathcal{E}$ & $e$ & Environment & Spawns initial populations; computes output aggregates & $N_{\mathcal{E}}=1$ \\
Grid & $\mathcal{G}$ & $g$ & Space & Toroidal continuous grid; owns world geometry; answers patch-co-location queries & $N_{\mathcal{G}}=1$ \\
GrassFieldCell & $\mathcal{F}$ & $f_1,f_2,\ldots$ & Decision-maker & One patch cell; ages its regrowth clock and turns green & $N_{\mathcal{F}}$ (fixed; one per patch, see T6a) \\
Sheep & $\mathcal{S}$ & $s_1,s_2,\ldots$ & Decision-maker & Eats grass, metabolises, reproduces (inheritable stride), dies & $n_{\mathcal{S}}$ (variable; birth/death) \\
Wolf & $\mathcal{W}$ & $w_1,w_2,\ldots$ & Decision-maker & Catches sheep, metabolises, reproduces (inheritable stride), dies & $n_{\mathcal{W}}$ (variable; birth/death) \\
\bottomrule
\end{longtable}
\egroup

Sheep and Wolf have \textbf{variable} populations (\(n\), not \(N\)) \ensuremath{\rightarrow} r8 requires each to
own a creation function (f3/f7, reproduce) and a removal function (f4/f8, die). The Grid,
GrassFieldCell, and Environment are fixed-quantity singletons/multitons (\(N\)).
GrassFieldCell is categorised Decision-maker because it owns an endogenous state and a
function (the regrowth rule); its ``decision'' is the deterministic countdown trigger, the
only available fit in VISA's four categories.

\begin{center}\rule{0.5\linewidth}{0.5pt}\end{center}

\subsubsection{T2 --- Variable}

\bgroup
\setlength{\tabcolsep}{4pt}
\renewcommand{\arraystretch}{1.12}
\begin{longtable}{@{}>{\raggedright\arraybackslash}p{2.7cm}
                    >{\raggedright\arraybackslash}p{1.4cm}
                    >{\raggedright\arraybackslash}p{1.5cm}
                    >{\raggedright\arraybackslash}p{1.7cm}
                    >{\raggedright\arraybackslash}p{1.1cm}
                    >{\raggedright\arraybackslash}p{0.8cm}
                    >{\raggedright\arraybackslash}p{6.3cm}@{}}
\toprule
\textbf{Variable} & \textbf{Symbol} & \textbf{Type} & \textbf{Data type} & \textbf{Value} & \textbf{Unit} & \textbf{Description} \\
\midrule
\multicolumn{7}{@{}l}{\textbf{\underline{Environment $\mathcal{E}$}}} \\
Initial sheep count & $n_{\mathcal{S}}^{0}$ & Exog.-homo. & Integer & Input & sheep & initial population; used at setup \\
Initial wolf count & $n_{\mathcal{W}}^{0}$ & Exog.-homo. & Integer & Input & wolves & initial population; used at setup \\
Sheep cap & $N_{\max}$ & Exog.-homo. & Integer & Input & sheep & overpopulation termination guard \\
Sheep count & $S_t$ & Endog. & Integer & f11 & sheep & $|\mathcal{S}|$ \\
Wolf count & $W_t$ & Endog. & Integer & f11 & wolves & $|\mathcal{W}|$ \\
Green-patch count & $R_t$ & Endog. & Integer & f11 & patches & $\sum_f g_{f,t}$ \\
Mean sheep stride & $\bar{\ell}^{s}_t$ & Endog. & Float & f11 & patches & $\mathrm{mean}_{i}\,\ell_i$ \\
Mean wolf stride & $\bar{\ell}^{w}_t$ & Endog. & Float & f11 & patches & $\mathrm{mean}_{j}\,\ell_j$ \\
\midrule
\multicolumn{7}{@{}l}{\textbf{\underline{Grid $\mathcal{G}$ (Space)}}} \\
Width & $W_x$ & Exog.-homo. & Integer & Input & patches & grid columns; toroidal extent \\
Height & $W_y$ & Exog.-homo. & Integer & Input & patches & grid rows; toroidal extent \\
Wrap left--right & $w_x$ & Exog.-homo. & Boolean & Input & --- & horizontal edges connected \\
Wrap top--bottom & $w_y$ & Exog.-homo. & Boolean & Input & --- & vertical edges connected \\
\midrule
\multicolumn{7}{@{}l}{\textbf{\underline{GrassFieldCell $\mathcal{F}$}}} \\
Grass regrowth time & $T_g$ & Exog.-homo. & Integer & Input & ticks & brown\ensuremath{\rightarrow}green delay \\
Green flag & $g_{f,t}$ & Endog. & Boolean & f9; f2 (ext) & --- & true = green, false = brown \\
Regrowth countdown & $c_{f,t}$ & Endog. & Integer & f9 & ticks & brown-patch clock \\
\midrule
\multicolumn{7}{@{}l}{\textbf{\underline{Sheep $\mathcal{S}$}}} \\
Energy & $e_{i,t}$ & Endog. & Float & f1; f2; f3 & energy & drives death/reproduction \\
Stride length & $\ell_i$ & Exog.-hetero. & Float & Input (birth) & patches & inherited + mutated at birth; constant per agent \\
Initial sheep stride & $\ell^{0}_{s}$ & Exog.-homo. & Float & Input & patches & stride assigned to a hatched sheep at setup \\
Sheep stride drift & $\delta_s$ & Exog.-homo. & Float & Input & patches & half-width of the per-birth mutation \\
Sheep gain from food & $G_f^{s}$ & Exog.-homo. & Float & Input & energy & energy gained per grass eaten \\
Sheep reproduce chance & $r_s$ & Exog.-homo. & Float & Input & \% & per-tick reproduction probability \\
Max energy & $M_{\max}$ & Exog.-homo. & Float & Input & energy & energy cap; declared on both Sheep \& Wolf \\
Reproduction threshold & $M_{\min}$ & Exog.-homo. & Float & Input & energy & min energy to reproduce; declared on both \\
Max stride & $S_{\max}$ & Exog.-homo. & Float & Input & patches & stride-length upper clamp; declared on both \\
Stride-energy penalty? & $\pi$ & Exog.-homo. & Boolean & Input & --- & whether stride costs energy; declared on both \\
x-coordinate & $x_{i,t}$ & Endog. & Float & f1 & patch & wrapped to torus \\
y-coordinate & $y_{i,t}$ & Endog. & Float & f1 & patch & wrapped to torus \\
Heading & $\theta_{i,t}$ & Endog. & Float & f1 & deg & $[0,360)$; $0=$ +y (north) \\
Grass at my patch & $\gamma_{i,t}$ & Endog. & Boolean & f10 (ext) & --- & green flag of the co-located GrassFieldCell \\
\midrule
\multicolumn{7}{@{}l}{\textbf{\underline{Wolf $\mathcal{W}$}}} \\
Energy & $e_{j,t}$ & Endog. & Float & f5; f6; f7 & energy & drives death/reproduction \\
Stride length & $\ell_j$ & Exog.-hetero. & Float & Input (birth) & patches & inherited + mutated at birth; constant per agent \\
Initial wolf stride & $\ell^{0}_{w}$ & Exog.-homo. & Float & Input & patches & stride assigned to a hatched wolf at setup \\
Wolf stride drift & $\delta_w$ & Exog.-homo. & Float & Input & patches & half-width of the per-birth mutation \\
Wolf gain from food & $G_f^{w}$ & Exog.-homo. & Float & Input & energy & energy gained per sheep caught \\
Wolf reproduce chance & $r_w$ & Exog.-homo. & Float & Input & \% & per-tick reproduction probability \\
Max energy & $M_{\max}$ & Exog.-homo. & Float & Input & energy & energy cap; declared on both Sheep \& Wolf \\
Reproduction threshold & $M_{\min}$ & Exog.-homo. & Float & Input & energy & min energy to reproduce; declared on both \\
Max stride & $S_{\max}$ & Exog.-homo. & Float & Input & patches & stride-length upper clamp; declared on both \\
Stride-energy penalty? & $\pi$ & Exog.-homo. & Boolean & Input & --- & whether stride costs energy; declared on both \\
x-coordinate & $x_{j,t}$ & Endog. & Float & f5 & patch & wrapped to torus \\
y-coordinate & $y_{j,t}$ & Endog. & Float & f5 & patch & wrapped to torus \\
Heading & $\theta_{j,t}$ & Endog. & Float & f5 & deg & $[0,360)$; $0=$ +y (north) \\
Sheep at my patch & $\mathbf{h}_{j,t}$ & Endog. & List[$\mathcal{S}$] & f10 (ext) & sheep & co-located sheep (for catching) \\
\bottomrule
\end{longtable}
\egroup

\textbf{Why stride is Exog.-hetero., not Endogenous.} A given animal's stride is set
\textbf{once,
at birth} (initial slider value, or parent-stride + mutation for offspring) and is never
updated by any function acting on that agent. It is therefore heterogeneous across agents
yet not an endogenous state --- exactly parallel to Rebellion's risk-aversion
\(R_i\sim U(0,1)\) (Exp1). The mutation is part of the \emph{offspring-creation} step
(f3/f7 external effect, \eqref{eq:ws-repro}), not an update to an existing agent.

\emph{Transient quantities (not stored):} the turn deltas and mutation noise are drawn
fresh inside f1/f3/f5/f7 each call.

\begin{center}\rule{0.5\linewidth}{0.5pt}\end{center}

\subsubsection{T3 --- Sensing}

The Grid (f10) realises the spatial modality: animals interact with what is on
\textbf{their own patch} (\(\mathrm{round}(x),\mathrm{round}(y)\)). Rows = observers;
columns = observed.

\bgroup
\setlength{\tabcolsep}{5pt}
\renewcommand{\arraystretch}{1.15}
\begin{longtable}{@{}>{\raggedright\arraybackslash}p{2.4cm}
                    >{\centering\arraybackslash}p{1.5cm}
                    >{\centering\arraybackslash}p{1.5cm}
                    >{\centering\arraybackslash}p{1.5cm}
                    >{\centering\arraybackslash}p{3.2cm}
                    >{\centering\arraybackslash}p{3.2cm}@{}}
\toprule
\textbf{Observer} & $\mathcal{E}$ & $\mathcal{G}$ & $\mathcal{F}$ & $\mathcal{S}$ & $\mathcal{W}$ \\
\midrule
$\mathcal{E}$ (Env) & $\emptyset$ & $\emptyset$ & $g_{f,t}$ & $e_{i,t}, \ell_i, x_{i,t}, y_{i,t}$ & $e_{j,t}, \ell_j, x_{j,t}, y_{j,t}$ \\
$\mathcal{G}$ (Grid) & $\emptyset$ & $\emptyset$ & $g_{f,t}$ & $x_{i,t}, y_{i,t}$ & $x_{j,t}, y_{j,t}$ \\
$\mathcal{F}$ (Grass) & $\emptyset$ & $\emptyset$ & $\emptyset$ & $\emptyset$ & $\emptyset$ \\
$\mathcal{S}$ (Sheep) & $\emptyset$ & $\emptyset$ & $g_{f,t}$ & $\emptyset$ & $\emptyset$ \\
$\mathcal{W}$ (Wolf) & $\emptyset$ & $\emptyset$ & $\emptyset$ & $x_{i,t}, y_{i,t}$ & $\emptyset$ \\
\bottomrule
\end{longtable}
\egroup

Reading the matrix: the Environment reads every animal's energy/stride/position and every
cell's green flag to form the aggregates (f11). The Grid reads animal positions and grass
flags to compute the co-location lists it pushes to each animal. A sheep reads the
\textbf{green flag of the grass cell on its own patch} (\(\gamma_{i,t}\), to eat) and its
own state --- but \textbf{no peer sheep and no wolves} (movement is an undirected random
walk, not chase/flee). A wolf reads the \textbf{positions of co-located sheep} (to pick a
uniformly random co-located prey); it does not sense other wolves. Every same-type
diagonal is \(\emptyset\): neither breed has any same-type sensing, and the singleton
types have no peers (valid per r5; self-observation is implicit).

\begin{center}\rule{0.5\linewidth}{0.5pt}\end{center}

\subsubsection{T4 --- Internal Function}

Complex updates are given as numbered equations below the table; the Self-state Update and
External Effect columns record \emph{which endogenous variable} is written and the
equation that governs it.

\bgroup
\setlength{\tabcolsep}{4pt}
\renewcommand{\arraystretch}{1.12}
\begin{longtable}{@{}>{\raggedright\arraybackslash}p{0.7cm}
                    >{\raggedright\arraybackslash}p{1.7cm}
                    >{\raggedright\arraybackslash}p{2.3cm}
                    >{\raggedright\arraybackslash}p{2.9cm}
                    >{\raggedright\arraybackslash}p{3.6cm}
                    >{\raggedright\arraybackslash}p{3.3cm}
                    >{\raggedright\arraybackslash}p{1.0cm}@{}}
\toprule
\textbf{ID} & \textbf{Function} & \textbf{Method} & \textbf{Decision basis} & \textbf{Self-state update} & \textbf{External effect} & \textbf{Ref.} \\
\midrule
\multicolumn{7}{@{}l}{\textbf{\underline{Grid $\mathcal{G}$}}} \\
f10 & query\_colocation & Spatial query & animal positions + $g_{f,t}$ (T3) & --- & $\gamma_{i,t}\leftarrow g_{f^*(i),t}$ for each sheep; $\mathbf{h}_{j,t}\leftarrow$ sheep on $j$'s patch for each wolf & --- \\
\midrule
\multicolumn{7}{@{}l}{\textbf{\underline{Sheep $\mathcal{S}$}}} \\
f1 & move\_sheep & Random walk + metabolism & $\theta_{i,t}, \ell_i, e_{i,t}, \pi$ (self) & $\theta_{i,t},x_{i,t},y_{i,t},e_{i,t}$ via \eqref{eq:ws-move} & --- & --- \\
f2 & eat\_grass & Conditional & $\gamma_{i,t}$ (self) & $e_{i,t}$ via \eqref{eq:ws-eat} & if $\gamma_{i,t}$: $g_{f^*(i),t}\leftarrow$ False & --- \\
f3 & reproduce\_sheep & Probabilistic + create & $e_{i,t}, \ell_i, \theta_{i,t}, x_{i,t}, y_{i,t}$; $r_s, M_{\min}, \delta_s, S_{\max}$ (self) & if condition: $e_{i,t}\leftarrow e_{i,t}/2$ & \eqref{eq:ws-repro} creates offspring $s_{\text{new}}\in\mathcal{S}$ & \citep{Novak2006} \\
f4 & die\_sheep & Conditional removal & $e_{i,t}$ (self) & if $e_{i,t}<0$: remove self from $\mathcal{S}$ & --- & --- \\
\midrule
\multicolumn{7}{@{}l}{\textbf{\underline{Wolf $\mathcal{W}$}}} \\
f5 & move\_wolf & Random walk + metabolism & $\theta_{j,t}, \ell_j, e_{j,t}, \pi$ (self) & $\theta_{j,t},x_{j,t},y_{j,t},e_{j,t}$ via \eqref{eq:ws-move} & --- & --- \\
f6 & catch\_sheep & Uniform selection & $\mathbf{h}_{j,t}$ (self); $G_f^{w}, M_{\max}$ (self) & $e_{j,t}$ via \eqref{eq:ws-catch} & remove chosen $s_k$ from $\mathcal{S}$ & \citep{Novak2006} \\
f7 & reproduce\_wolf & Probabilistic + create & $e_{j,t}, \ell_j, \theta_{j,t}, x_{j,t}, y_{j,t}$; $r_w, M_{\min}, \delta_w, S_{\max}$ (self) & if condition: $e_{j,t}\leftarrow e_{j,t}/2$ & \eqref{eq:ws-repro} creates offspring $w_{\text{new}}\in\mathcal{W}$ & \citep{Novak2006} \\
f8 & die\_wolf & Conditional removal & $e_{j,t}$ (self) & if $e_{j,t}<0$: remove self from $\mathcal{W}$ & --- & --- \\
\midrule
\multicolumn{7}{@{}l}{\textbf{\underline{GrassFieldCell $\mathcal{F}$}}} \\
f9 & grow\_grass & Field countdown & $c_{f,t}, g_{f,t}$ (self) & $c_{f,t}, g_{f,t}$ via \eqref{eq:ws-grow} & --- & --- \\
\midrule
\multicolumn{7}{@{}l}{\textbf{\underline{Environment $\mathcal{E}$}}} \\
f11 & compute\_counts & Aggregation & all $s_i\in\mathcal{S}$, $w_j\in\mathcal{W}$, $f\in\mathcal{F}$ (T3) & $S_t,W_t,R_t,\bar{\ell}^{s}_t,\bar{\ell}^{w}_t$ via \eqref{eq:ws-counts} & --- & --- \\
\bottomrule
\end{longtable}
\egroup

Here \(f^*(i)\) denotes the GrassFieldCell at sheep \(i\)'s patch, i.e.
\(\mathrm{round}(x_{i,t}){=}\mathrm{round}(x_{f^*})\) and likewise for \(y\). f1/f5 (move)
and f3/f7 (reproduce) share identical logic across breeds; they are listed separately
because they belong to different agent types with different sensing.

\textbf{Behavioural equations.} The move + metabolism step (f1 sheep, f5 wolf) is

\begin{equation}
\theta\leftarrow\theta+U(0,50)-U(0,50),\quad
(x,y)\leftarrow\mathrm{wrap}\big((x,y)+\mathrm{fwd}(\theta,\ell)\big),\quad
e\leftarrow e-0.5-\mathbf{1}[\pi]\,\ell ,
\label{eq:ws-move}
\end{equation}

with the breed's own subscripts (\(i\)/\(j\)) and stride (\(\ell_i\)/\(\ell_j\)). Eating,
catching, growing, and reproducing are

\begin{equation}
\text{if }\gamma_{i,t}:\; e_{i,t}\leftarrow\min(M_{\max},\,e_{i,t}+G_f^{s})
\label{eq:ws-eat}
\end{equation}

\begin{equation}
\text{if }\mathbf{h}_{j,t}\ne\emptyset:\;\text{pick }s_k\sim\mathrm{Unif}(\mathbf{h}_{j,t}),\;
e_{j,t}\leftarrow\min(M_{\max},\,e_{j,t}+G_f^{w})
\label{eq:ws-catch}
\end{equation}

\begin{equation}
\text{if }\neg g_{f,t}:\;c_{f,t}\leftarrow c_{f,t}-1;\quad
\text{if }c_{f,t}\le 0:\;g_{f,t}\leftarrow\text{True},\;c_{f,t}\leftarrow T_g
\label{eq:ws-grow}
\end{equation}

Reproduction (f3 sheep, f7 wolf) fires when a uniform draw beats the reproduce chance and
energy exceeds the threshold; the parent halves its energy and creates one offspring:

\begin{equation}
\text{if }U(0,100)<r\wedge e>M_{\min}:\quad e\leftarrow e/2,\quad
\text{create offspring }(\ell',e',\theta',x',y')
\label{eq:ws-repro}
\end{equation}

with the offspring's traits

\begin{equation}
\ell'=\mathrm{mutate}(\ell,\delta,S_{\max}),\quad
e'=e,\quad
\theta'=\theta+U(0,360),\quad
(x',y')=\mathrm{wrap}\big((x,y)+\mathrm{fwd}(\theta',1)\big),
\label{eq:ws-repro-traits}
\end{equation}

where \(r,\delta\) are the breed's reproduce-chance / drift (\(r_s,\delta_s\) for sheep;
\(r_w,\delta_w\) for wolf). The mutation operator \(\mathrm{mutate}\) adds symmetric
uniform noise and clamps the result:

\begin{equation}
\ell' = \ell + U(0,\delta) - U(0,\delta),\qquad
\mathrm{mutate}=\begin{cases}0 & \ell'<0\\S_{\max}&\ell>S_{\max}\\ \ell'&\text{otherwise.}\end{cases}
\label{eq:ws-mutate}
\end{equation}

The upper branch tests the \textbf{parent's} stride \(\ell\), not the noisy \(\ell'\).
Given the invariant \(\ell\le S_{\max}\) at birth, that branch fires only one generation
after an overshoot, so the effective bound is the reflecting lower clamp at 0; this is
part of the model's selection dynamics, not a defect. Finally, the aggregates (f11) are

\begin{equation}
S_t=|\mathcal{S}|,\;\;W_t=|\mathcal{W}|,\;\;R_t=\sum_f g_{f,t},\;\;
\bar{\ell}^{s}_t=\mathrm{mean}_{i}\ell_i,\;\;\bar{\ell}^{w}_t=\mathrm{mean}_{j}\ell_j .
\label{eq:ws-counts}
\end{equation}

\begin{center}\rule{0.5\linewidth}{0.5pt}\end{center}

\subsubsection{T5 --- Associated Data}

{\def\LTcaptype{none} 
\begin{longtable}[]{@{}lllllllll@{}}
\toprule\noalign{}
ID & Title & Type & Temporal & Source & Collection & Pre-processing & \#Rec. & Avail. \\
\midrule\noalign{}
\endhead
\bottomrule\noalign{}
\endlastfoot
\end{longtable}
}

\emph{(empty --- the model has no stand-alone dataset)}

As in Exp1, the model has \textbf{no stand-alone dataset}: it is neither backed by an
empirical dataset nor by a synthetic one. All structural parameters are author-assumed
values (the canonical settings reported in Novak \& Wilensky 2006), and initial
positions/energies/grass are drawn from uniform distributions in the sampling step itself.
With no separable data record backing any parameter, T5 is empty; accordingly every T6a
entry has Data source \(=\) Author (no \(d\)-ID to cite).

\begin{center}\rule{0.5\linewidth}{0.5pt}\end{center}

\subsubsection{T6 --- Input and Output}

\paragraph{(a) Input}

\bgroup
\setlength{\tabcolsep}{4pt}
\renewcommand{\arraystretch}{1.12}
\begin{longtable}{@{}>{\raggedright\arraybackslash}p{1.8cm}
                    >{\raggedright\arraybackslash}p{4.3cm}
                    >{\raggedright\arraybackslash}p{2.0cm}
                    >{\raggedright\arraybackslash}p{1.9cm}
                    >{\raggedright\arraybackslash}p{3.5cm}
                    >{\raggedright\arraybackslash}p{2.0cm}@{}}
\toprule
\textbf{Symbol} & \textbf{Value / Distribution} & \textbf{Data source} & \textbf{Derivation} & \textbf{Algorithm} & \textbf{Ref.} \\
\midrule
\multicolumn{6}{@{}l}{\textbf{\underline{Environment $\mathcal{E}$}}} \\
$n_{\mathcal{S}}^{0}$ & 64 & Author & Assumed & --- & --- \\
$n_{\mathcal{W}}^{0}$ & 30 & Author & Assumed & --- & --- \\
$N_{\max}$ & 30000 & Author & Assumed & --- & --- \\
\midrule
\multicolumn{6}{@{}l}{\textbf{\underline{Grid $\mathcal{G}$}}} \\
$W_x, W_y$ & 61, 61 & Author & Assumed & --- & --- \\
$w_x, w_y$ & true, true & Author & Assumed & --- & --- \\
\midrule
\multicolumn{6}{@{}l}{\textbf{\underline{GrassFieldCell $\mathcal{F}$}}} \\
$T_g$ & 138 & Author & Assumed & --- & --- \\
grass countdown (per cell) & $U\{0,\ldots,137\}$ & Author & Computed & integer uniform & --- \\
grass flag (per cell) & Bernoulli(0.5) & Author & Computed & integer uniform & --- \\
\midrule
\multicolumn{6}{@{}l}{\textbf{\underline{Sheep $\mathcal{S}$}}} \\
$\ell^{0}_{s}$ & 0.2 & Author & Assumed & --- & --- \\
$\delta_s$ & 0.2 & Author & Assumed & --- & --- \\
$G_f^{s}$ & 20 & Author & Assumed & --- & --- \\
$r_s$ & 5 & Author & Assumed & --- & --- \\
\midrule
\multicolumn{6}{@{}l}{\textbf{\underline{Wolf $\mathcal{W}$}}} \\
$\ell^{0}_{w}$ & 1.0 & Author & Assumed & --- & --- \\
$\delta_w$ & 0.24 & Author & Assumed & --- & --- \\
$G_f^{w}$ & 20 & Author & Assumed & --- & --- \\
$r_w$ & 6 & Author & Assumed & --- & --- \\
\midrule
\multicolumn{6}{@{}l}{\textbf{\underline{Both animals $\mathcal{S}\cup\mathcal{W}$ (shared)}}} \\
$M_{\max}, M_{\min}, S_{\max}$ & 500, 200, 3 & Author & Assumed & --- & --- \\
$\pi$ & true / false & Author & Assumed & --- & --- \\
init energy (per animal) & $U\{0,\ldots,499\}$ & Author & Computed & integer uniform & --- \\
init position (per animal) & $U(-30.5,30.5)^2$ & Author & Computed & continuous uniform & --- \\
\bottomrule
\end{longtable}
\egroup

\paragraph{(b) Output}

\bgroup
\setlength{\tabcolsep}{4pt}
\renewcommand{\arraystretch}{1.12}
\begin{longtable}{@{}>{\raggedright\arraybackslash}p{1.4cm}
                    >{\raggedright\arraybackslash}p{2.8cm}
                    >{\raggedright\arraybackslash}p{3.6cm}
                    >{\raggedright\arraybackslash}p{1.7cm}
                    >{\raggedright\arraybackslash}p{1.3cm}
                    >{\centering\arraybackslash}p{1.0cm}
                    >{\raggedright\arraybackslash}p{4.1cm}@{}}
\toprule
\textbf{Symbol} & \textbf{Indicator} & \textbf{Formula} & \textbf{Data type} & \textbf{Unit} & \textbf{Freq.} & \textbf{Desc.} \\
\midrule
$S_t$ & Sheep population & $|\mathcal{S}|$ & Integer & sheep & 1 & live sheep \\
$W_t$ & Wolf population & $|\mathcal{W}|$ & Integer & wolves & 1 & live wolves \\
$R_t/4$ & Grass (scaled) & $\sum_f g_{f,t}/4$ & Float & patches & 1 & green patches ($\div$4 for plot scale) \\
$\bar{\ell}^{s}_t$ & Mean sheep stride & $\mathrm{mean}_{i}\ell_i$ & Float & patches & 1 & \textbf{headline selection indicator} \\
$\bar{\ell}^{w}_t$ & Mean wolf stride & $\mathrm{mean}_{j}\ell_j$ & Float & patches & 1 & wolf selection indicator \\
\bottomrule
\end{longtable}
\egroup

\begin{center}\rule{0.5\linewidth}{0.5pt}\end{center}

\subsubsection{T7 --- Schedule}

\paragraph{(a) Execution}

\bgroup
\setlength{\tabcolsep}{4pt}
\renewcommand{\arraystretch}{1.12}
\begin{longtable}{@{}>{\centering\arraybackslash}p{0.9cm}
                    >{\centering\arraybackslash}p{1.2cm}
                    >{\centering\arraybackslash}p{0.9cm}
                    >{\raggedright\arraybackslash}p{3.5cm}
                    >{\raggedright\arraybackslash}p{3.0cm}
                    >{\raggedright\arraybackslash}p{5.0cm}@{}}
\toprule
\textbf{Step} & \textbf{Agent} & \textbf{ID} & \textbf{Function} & \textbf{Exec. mode} & \textbf{Condition} \\
\midrule
1 & $\mathcal{G}$ & f10 & query\_colocation & Synchronous & on demand (per animal as it acts) \\
2 & $\mathcal{S}$ & f1 & move\_sheep & Random-order & live sheep \\
2 & $\mathcal{S}$ & f2 & eat\_grass & Random-order & live sheep \\
2 & $\mathcal{S}$ & f4 & die\_sheep & Random-order & live sheep \\
2 & $\mathcal{S}$ & f3 & reproduce\_sheep & Random-order & survived \\
3 & $\mathcal{W}$ & f5 & move\_wolf & Random-order & live wolf \\
3 & $\mathcal{W}$ & f6 & catch\_sheep & Random-order & live wolf \\
3 & $\mathcal{W}$ & f8 & die\_wolf & Random-order & live wolf \\
3 & $\mathcal{W}$ & f7 & reproduce\_wolf & Random-order & survived \\
4 & $\mathcal{F}$ & f9 & grow\_grass & Synchronous & every brown cell \\
5 & $\mathcal{E}$ & f11 & compute\_counts & Synchronous & every tick \\
\bottomrule
\end{longtable}
\egroup

\textbf{Composite schedule.} The tick is a \textbf{breed-sequential composite}: the sheep
block (Step 2) runs to completion before the wolf block (Step 3), which runs to completion
before the grass block (Step 4). Within each block, agents are iterated in
\textbf{random order} (Random-order). The Grid (Step 1) is a live spatial index: each
co-location query is evaluated for the animal that is \emph{currently} acting, so wolves
(Step 3) see the sheep distribution left by Step 2. The sheep-before-wolves ordering is
load-bearing: a sheep born in Step 2 is \textbf{not} processed again in Step 2 (population
snapshot) and is a valid target for wolves in Step 3; a wolf's victim (removed in Step 3)
will not act again. (Staged execution is a composite mode in VISA, built from Random-order
base blocks.)

\textbf{Per-animal sub-ordering within a block} (Step 2, sheep): move+metabolise \ensuremath{\rightarrow}
eat-grass \ensuremath{\rightarrow} maybe-die \ensuremath{\rightarrow} reproduce. Death is checked \textbf{before} reproduction, so a
sheep that starves does not also reproduce that tick.

\paragraph{(b) Termination}

\bgroup
\setlength{\tabcolsep}{4pt}
\renewcommand{\arraystretch}{1.12}
\begin{longtable}{@{}>{\centering\arraybackslash}p{0.9cm}
                    >{\raggedright\arraybackslash}p{2.2cm}
                    >{\raggedright\arraybackslash}p{3.0cm}
                    >{\raggedright\arraybackslash}p{3.1cm}
                    >{\raggedright\arraybackslash}p{2.5cm}
                    >{\raggedright\arraybackslash}p{3.2cm}@{}}
\toprule
\textbf{ID} & \textbf{Indicator} & \textbf{Condition} & \textbf{Description} & \textbf{Value source/ref} & \textbf{Termination logic} \\
\midrule
c1 & $S_t+W_t$ & $=0$ & Extinction of all animals & Model stopping rule & \multirow{3}{*}{Stop when $c_1\lor c_2\lor c_3$} \\
c2 & $W_t,\,S_t$ & $W_t=0\wedge S_t>N_{\max}$ & Wolves extinct; sheep overpopulate & Model stopping rule & \\
c3 & $t$ & $t\ge T$ & Reproduction horizon reached & Author assumed & \\
\bottomrule
\end{longtable}
\egroup

c1/c2 are the model's own stopping rules (total extinction; wolves extinct with sheep over
the cap); c3 bounds the reproduction run length (e.g., \(T=1500\)). The simulation stops
as soon as \textbf{any} of the three fires.

\begin{center}\rule{0.5\linewidth}{0.5pt}\end{center}

\subsubsection{T8 --- Validation}

\bgroup
\setlength{\tabcolsep}{4pt}
\renewcommand{\arraystretch}{1.12}
\begin{longtable}{@{}>{\centering\arraybackslash}p{0.8cm}
                    >{\raggedright\arraybackslash}p{3.6cm}
                    >{\raggedright\arraybackslash}p{3.4cm}
                    >{\raggedright\arraybackslash}p{2.6cm}
                    >{\raggedright\arraybackslash}p{2.0cm}
                    >{\raggedright\arraybackslash}p{3.6cm}
                    >{\raggedright\arraybackslash}p{1.0cm}@{}}
\toprule
\textbf{ID} & \textbf{Validation object} & \textbf{Benchmark data} & \textbf{Method} & \textbf{Indicator} & \textbf{Passing cond.} & \textbf{Ref.} \\
\midrule
\multicolumn{7}{@{}l}{\textbf{\underline{Output level}}} \\
v1 & Sheep stride trajectory $\bar{\ell}^{s}_t$ (penalty \textbf{on}) & Novak \& Wilensky 2006: "sheep typically converge on an average stride length close to 1" & time-series + end-state mean & $\bar{\ell}^{s}_t$ & converges toward $\approx 1$ (end-state mean $\in [0.5, 1.5]$) & \citep{Novak2006} \\
v2 & Sheep stride trajectory (penalty \textbf{off}) & Novak \& Wilensky 2006: "they will become faster and faster" & time-series trend & $\bar{\ell}^{s}_t$ & grows monotonically away from 1 (end-state $\gg 1$, e.g. $>1.5$ and rising) & \citep{Novak2006} \\
v3 & Extinction outcome & Wolf-Sheep Predation family behaviour & per-seed end-state classification & $(S_T, W_T)$ & reproduces parameter-dependent outcome (coexistence / wolf extinction / total extinction) across seeds & \citep{Novak2006} \\
v4 & Predator-prey oscillation & Lotka--Volterra signature & peak/trough detection in $S_t,W_t$ & $S_t, W_t$ & out-of-phase sheep--wolf oscillations while both persist & --- \\
v5 & Energy bound & identity $0\le e_{j,t}\le M_{\max}$ & invariant check & $\max_i e_{i,t}$ & $\le M_{\max}$ at every $t$ & --- \\
\bottomrule
\end{longtable}
\egroup

\textbf{v1 and v2 are the headline reproduction criteria} --- together they show the
emergent selection equilibrium (\(\approx 1\)) and isolate the energy penalty as its
cause. v3 reproduces the inherited Wolf-Sheep Predation extinction dynamics; v4 the
predator-prey cycles; v5 is a structural invariant.

\textbf{Author's note.} Novak \& Wilensky (2006) state the headline behaviors
qualitatively but give no formal validation criteria. The five entries above (v1--v5) are
authored for this case study to illustrate the T8 validation table, not transcribed from
the source paper.

\begin{center}\rule{0.5\linewidth}{0.5pt}\end{center}

\subsubsection{Consistency check (19 rules)}

Within-table (r1--r4) and cross-table (r5--r19), per the revised VISA rule set.

{\def\LTcaptype{none} 
\begin{longtable}[]{@{}
  >{\raggedright\arraybackslash}p{(\linewidth - 4\tabcolsep) * \real{0.2842}}
  >{\centering\arraybackslash}p{(\linewidth - 4\tabcolsep) * \real{0.0632}}
  >{\raggedright\arraybackslash}p{(\linewidth - 4\tabcolsep) * \real{0.6526}}@{}}
\toprule\noalign{}
\begin{minipage}[b]{\linewidth}\raggedright
Rule
\end{minipage} & \begin{minipage}[b]{\linewidth}\centering
Res.
\end{minipage} & \begin{minipage}[b]{\linewidth}\raggedright
Note
\end{minipage} \\
\midrule\noalign{}
\endhead
\bottomrule\noalign{}
\endlastfoot
r1 Variable-type coverage \& time-indexing & PASS & all T2 rows carry a leaf type; every
endogenous variable carries the time index \(t\)
(\(e_{i,t},x_{i,t},y_{i,t},\theta_{i,t},\gamma_{i,t},g_{f,t},c_{f,t},S_t,W_t,R_t,\bar\ell^{s}_t,\bar\ell^{w}_t,\mathbf{h}_{j,t}\)) \\
r2 Function-ID uniqueness & PASS & f1--f11 unique; every f-ID in T2's Value column exists
in T4 \\
r3 Function productivity & PASS & f1 writes \(\theta,x,y,e\); f2 writes \(e\)+ext
\(g_{f,t}\); f3 writes \(e\)+creates instance; f4 removes instance (\ensuremath{\rightarrow}\(n_\mathcal{S}\));
f5--f8 likewise; f9 writes \(g_{f,t},c_{f,t}\); f10 writes ext
\(\gamma_{i,t},\mathbf{h}_{j,t}\); f11 writes 5 aggregates. f4/f8 are
\textbf{removal functions} (r8 pattern): their productivity is the population change
\(n_\mathcal{S}/n_\mathcal{W}\). \\
r4 Step--Exec.mode consistency & PASS & Step 1 = Synchronous (f10); Steps 2--3 =
Random-order (sheep/wolf blocks); Step 4 = Synchronous (f9); Step 5 = Synchronous (f11);
no conflicting modes within a step \\
r5 Same-type (peer) sensing & PASS & \(\mathcal{E},\mathcal{G}\) \(N{=}1\)\ensuremath{\rightarrow}\(\emptyset\);
\(\mathcal{F}\) fixed but no same-type sensing\ensuremath{\rightarrow}\(\emptyset\); \(\mathcal{S},\mathcal{W}\)
\(n{>}1\)\ensuremath{\rightarrow}\(\emptyset\) peer set (no same-type sensing; self implicit) \\
r6 Passive-agent implications & PASS & no Passive agents \\
r7 Observer-row completeness & PASS &
\(\mathcal{E},\mathcal{G},\mathcal{F},\mathcal{S},\mathcal{W}\) each have exactly one
row \\
r8 Population-dynamics consistency & PASS & \(n_\mathcal{S}\)\ensuremath{\rightarrow}create f3 + remove f4;
\(n_\mathcal{W}\)\ensuremath{\rightarrow}create f7 + remove f8;
\(N_\mathcal{E},N_\mathcal{G},N_\mathcal{F}\)\ensuremath{\rightarrow}neither \\
r9 Active-agent function coverage & PASS & \(\mathcal{E}\)\ensuremath{\rightarrow}f11; \(\mathcal{G}\)\ensuremath{\rightarrow}f10;
\(\mathcal{F}\)\ensuremath{\rightarrow}f9; \(\mathcal{S}\)\ensuremath{\rightarrow}f1,f2,f3,f4; \(\mathcal{W}\)\ensuremath{\rightarrow}f5,f6,f7,f8 \\
r10 Variable observability & PASS & all vars self-read or T3-authorized (grass green read
via \(\gamma_{i,t}\)/\(g_{f,t}\); wolf reads co-located sheep positions) \\
r11 Endogenous-variable completeness & PASS & every endog var written by \ensuremath{\geq}1 existing f-ID:
sheep \(e,\theta,x,y\)\ensuremath{\leftarrow}f1, \(e\)\ensuremath{\leftarrow}f2,f3, \(\gamma\)\ensuremath{\leftarrow}f10(ext); wolf \ensuremath{\leftarrow}f5,f6,f7,
\(\mathbf{h}\)\ensuremath{\leftarrow}f10(ext); grass \(g_{f,t},c_{f,t}\)\ensuremath{\leftarrow}f9 (+f2 ext); aggregates\ensuremath{\leftarrow}f11 \\
r12 Self-state-update validity & PASS & all Self-state vars are endogenous of the owning
agent \\
r13 External-effect validity & PASS & f2 ext writes \(g_{f,t}\) (GrassFieldCell endog);
f3/f7 create instances; f4/f6/f8 remove instances; f10 ext writes
\(\gamma_{i,t},\mathbf{h}_{j,t}\) (sheep/wolf endog) --- all valid cross-agent effects \\
r14 Information-access validation & PASS & every Decision-basis var is self-attribute or
T3-authorized (grass at my patch; co-located sheep positions) \\
r15 Input--output coverage (bidirectional) & PASS & every T2 ``Input'' var has a T6a
entry; every T6a symbol is an exog T2 var; all outputs computable from T2 \\
r16 Schedule coverage (bidirectional) & PASS & f1--f11 all appear in T7a; every T7a ID
exists in T4 \\
r17 Termination-indicator source & PASS & c1 uses \(S_t,W_t\) (T6b); c2 uses
\(W_t,S_t,N_{\max}\) (T2/T6b); c3 uses \(t\) (exempt) \\
r18 Validation-object coverage & PASS & v1/v2\ensuremath{\rightarrow}\(\bar{\ell}^{s}_t\); v3\ensuremath{\rightarrow}\((S_T,W_T)\);
v4\ensuremath{\rightarrow}\(S_t,W_t\); v5\ensuremath{\rightarrow}\(e_{i,t}\); all trace to T2/T6b \\
r19 Data-reference resolution & PASS & T5 is empty (no stand-alone dataset); all T6a
sources are Author; no empirical data references to resolve \\
\end{longtable}
}

\textbf{Summary: 19/19 PASS.}
}

\section{Model 3 --- AgedCareContactModel (described; reproduction blocked)}
\label{sec:agedcare}

AnyLogic 8.9 model of an aged-care facility (contact-matrix generation). This is
the expressiveness case: the specification captures a 29-type / 240-function
industrial model in eight tables and passes 19/19 rules with no structural
accommodation. Full reproduction is blocked by (1) the proprietary AnyLogic
Pedestrian Library, which executes all movement and hence dominates the
contact-matrix output, and (2) unavailable backing data; VISA makes both
barriers explicit and localized rather than hidden. No figure (by design; see
the paper).

{\small 
\subsection{VISA Specification --- AgedCareContactModel (Exp3)}

\textbf{Model.} \emph{Generating a Contact Matrix for Aged Care Settings in Australia: An
Agent-Based Model Study} \citep{Stone2026} --- an AnyLogic 8 agent-based model of an
aged-care (nursing-home) facility that produces epidemiological contact matrices from
fine-grained staff--resident and resident--resident proximity. The model and its backing
data are released by the authors at
\texttt{https://github.com/BREATHE-UNSW/AgedCareContactMatrix}. It was developed and
documented following the ODD (+ODD+D) protocol; movement uses the Helbing--Moln\'ar
social-force pedestrian model \citep{Helbing2000}.

\textbf{Role in the experiment.} This is the \textbf{describe-only / expressiveness} case
(see the experiment README, Model 3). The goal is \emph{not} to reproduce the model --- it
is to show that \texttt{visa-author} produces \textbf{eight faithful tables} capturing a
large, real-world, industrially-authored ABM and that \texttt{visa-check} passes
\textbf{19/19}, and then to \textbf{honestly document why full reproduction is blocked}.
That limit is itself a finding for the paper (see the Reproduction blocker section below).

\textbf{Modelling logic (what is --- and is not --- an agent).} The agent taxonomy follows the
\emph{model's own semantics} (per its paper), not the host platform's object graph. The
paper defines \textbf{two agent populations} --- \emph{residents} and \emph{staff} ---
acting inside a \emph{physical facility}. VISA therefore has four agent types:

\begin{itemize}
\tightlist
\item
  \textbf{Environment} (Main) \(\mathcal{M}\) --- the clock, shift roster, person/location
  registries, and the contact-matrix computation. It also \textbf{owns the contact record
  store}: a finalized proximity contact is \emph{data produced by the simulation}, so it
  is an endogenous aggregate of the Environment, not an agent in its own right. The action
  of ``recording a contact'' is an Environment function, not a ContactRecord agent.
\item
  \textbf{Floor Plan} (Location) \(\mathcal{L}\) --- the \textbf{Space agent}: a
  heterogeneous, functionally-typed room hierarchy that manages the spatial relationships
  among residents and staff and is the co-location substrate determining who can contact
  whom.
\item
  \textbf{Resident} \(\mathcal{R}\) and \textbf{Staff} \(\mathcal{S}\) --- the two
  decision-makers.
\end{itemize}

Two things that the AnyLogic object graph exposes as separate classes are \textbf{not}
agents: (i) \emph{Contact records} are an \textbf{action/output} of contact detection,
hence absorbed into the Environment (the matrix \(C_t\) and the record log); (ii)
\emph{reference data} (the floor-plan registry, the staff activity-diary survey, the task
templates, the shift roster) are \textbf{datasets}, hence described by T5 (Associated
Data), not by a T1 agent row. This is the modelling-logic test of VISA: it asks \emph{what
behaves}, and only those things become agents.

\textbf{Authoring note (space).} Like Exp1 (Rebellion) and Exp2 (Wolf Sheep), this model
uses an \textbf{explicit Space agent} --- but where theirs is a \emph{homogeneous}
toroidal grid that answers metric / patch-co-location queries, here space is a
\textbf{heterogeneous,
functionally-typed floor plan} (rooms have different purposes, and \emph{which room a
person is in} determines whom they can contact). Co-location in typed rooms is
load-bearing for the output, so the \texttt{Location} hierarchy is the Space agent set
\(\mathcal{L}\).

\textbf{Authoring note (abstraction).} The model's raw source contains roughly 240
functions across roughly 29 object types. These are collapsed to the \textbf{14
behavioural-level functions} that carry model semantics. Platform plumbing (source / go-to
/ wait / sink navigation blocks, process-flow primitives, process tokens, and the
per-care-level activity sub-controllers) are absorbed into their owners' schedule
functions (they are embedded sub-objects, not independent decision-makers). Every retained
function is traceable to a named source routine.

\textbf{Notation.} Sets calligraphic: \(\mathcal{M}\) (Main/Environment), \(\mathcal{L}\)
(Floor Plan/Space), \(\mathcal{R}\) (Resident), \(\mathcal{S}\) (Staff). Instances
lowercase + subscript (\(r_i, s_j, \ell_k\)). Exogenous params uppercase (\(I_R, \rho,
\delta, \Delta t, T_1,T_2,T_3, M_1,M_2,M_3\)); endogenous lowercase + subscript or symbol
with a time index (\(\tau_t\) time-of-day, \(\sigma_t\) current shift, \(C_t\) contact
matrix, positions \((x^r_{i,t},y^r_{i,t})\)). The time index \(t\) (in seconds) marks a
variable as endogenous and may be suppressed in prose for readability.

\begin{center}\rule{0.5\linewidth}{0.5pt}\end{center}

\subsubsection{T1 --- Agent}

\bgroup
\setlength{\tabcolsep}{4pt}
\renewcommand{\arraystretch}{1.12}
\begin{longtable}{@{}>{\raggedright\arraybackslash}p{2.3cm}
                    >{\raggedright\arraybackslash}p{1.0cm}
                    >{\raggedright\arraybackslash}p{1.7cm}
                    >{\raggedright\arraybackslash}p{2.2cm}
                    >{\raggedright\arraybackslash}p{6.2cm}
                    >{\raggedright\arraybackslash}p{2.0cm}@{}}
\toprule
\textbf{Name} & \textbf{Set} & \textbf{Instances} & \textbf{Category} & \textbf{Description} & \textbf{Quantity} \\
\midrule
Environment (Main) & $\mathcal{M}$ & $m$ & Environment & Clock, shift roster, location/person registries; contact detection policy; contact-matrix computation and record store; output emission & $N_{\mathcal{M}}=1$ \\
Floor Plan (Location) & $\mathcal{L}$ & $\ell_1,\ell_2,\ldots$ & Space & Heterogeneous, typed room hierarchy bound to the floor plan; maintains person--room membership and is the co-location substrate determining who can contact whom & $N_{\mathcal{L}}$ (fixed; loaded from d1) \\
Resident & $\mathcal{R}$ & $r_1,r_2,\ldots$ & Decision-maker & Aged-care resident; care-level-stratified daily activity schedule; proximity contacts tracked & $n_{\mathcal{R}}$ (variable; set by $I_R$, see T6a) \\
Staff & $\mathcal{S}$ & $s_1,s_2,\ldots$ & Decision-maker & Care staff; shift-based, priority-driven work schedule over the daily task set; proximity contacts tracked & $n_{\mathcal{S}}$ (variable; injected per shift, departs at shift end) \\
\bottomrule
\end{longtable}
\egroup

\textbf{Notes.}

\begin{itemize}
\tightlist
\item
  \textbf{Why ContactRecord and ReferenceData are not agents.} A finalized proximity
  contact is a \emph{datum the simulation emits} --- it is produced by an Environment
  function (f4) and stored as the matrix \(C_t\) plus a record log (T6b), not inhabited by
  an agent. The backing lookups (floor-plan registry, staff activity-diary survey, task
  templates, shift roster) are \emph{datasets} and are recorded in T5. This keeps T1 to
  the four things that actually \emph{behave}.
\item
  \textbf{Consolidation of the source types into 4 VISA agent sets.} The 8
  \texttt{Location} subtypes
  (Wing/Bedroom/Bathroom/Bedspace/Dining/Communal/Staff\_Room/Transit) are one Space set
  \(\mathcal{L}\) (distinguished by location type). The activity sub-controllers and
  process tokens are embedded sub-objects of \(\mathcal{R}\)/\(\mathcal{S}\), not
  independent agent sets.
\item
  \textbf{Population dynamics (\ensuremath{\rightarrow} r8).} Variable (\(n\)): \(\mathcal{R}\) created by f5
  (init) and \(\mathcal{S}\) created by f3 (inject at each shift boundary) and removed by
  f13 (depart at shift end); each owns a creation/removal function. Fixed (\(N\)):
  \(\mathcal{M}\) (\(=1\)) and \(\mathcal{L}\) (loaded from d1) have no create/remove
  functions.
\end{itemize}

\begin{center}\rule{0.5\linewidth}{0.5pt}\end{center}

\subsubsection{T2 --- Variable}

\bgroup
\setlength{\tabcolsep}{4pt}
\renewcommand{\arraystretch}{1.12}
\begin{longtable}{@{}>{\raggedright\arraybackslash}p{2.7cm}
                    >{\raggedright\arraybackslash}p{1.4cm}
                    >{\raggedright\arraybackslash}p{1.5cm}
                    >{\raggedright\arraybackslash}p{1.7cm}
                    >{\raggedright\arraybackslash}p{1.1cm}
                    >{\raggedright\arraybackslash}p{0.8cm}
                    >{\raggedright\arraybackslash}p{6.3cm}@{}}
\toprule
\textbf{Variable} & \textbf{Symbol} & \textbf{Type} & \textbf{Data type} & \textbf{Value} & \textbf{Unit} & \textbf{Description} \\
\midrule
\multicolumn{7}{@{}l}{\textbf{\underline{Environment $\mathcal{M}$}}} \\
Initial \# residents & $I_R$ & Exog.-homo. & Integer & Input & --- & facility population \\
Contact-policy range & $\rho$ & Exog.-homo. & Float & Input & m & proximity threshold for a contact \\
Contact duration threshold & $\delta$ & Exog.-homo. & Float & Input & s & min co-presence for a contact \\
Time step & $\Delta t$ & Exog.-homo. & Float & Input & s & integration step \\
Simulation start time & --- & Exog.-homo. & Float & Input & min & clock value at $t=0$ \\
Shift start times & $T_1,T_2,T_3$ & Exog.-homo. & Float & Input & min & start of each of the three daily shifts \\
Staff per shift & $M_1,M_2,M_3$ & Exog.-homo. & Integer & Input & --- & staff injected at each shift boundary \\
Staff per wing & --- & Exog.-homo. & Integer & Input & --- & staff allocated per wing \\
Queue timeout & --- & Exog.-homo. & Float & Input & min & stale-task escalation limit \\
Second of day & $\tau_t$ & Endog. & Float & f2 & s & model clock \\
Current shift & $\sigma_t$ & Endog. & Integer & f3 & --- & $\in\{1,2,3\}$ \\
Contact matrix & $C_t$ & Endog. & Matrix[Integer] & f4 & contacts & role$\times$role aggregated contact counts \\
\midrule
\multicolumn{7}{@{}l}{\textbf{\underline{Floor Plan $\mathcal{L}$ (Space)}}} \\
Location name / node id & --- & Exog.-hetero. & String & f5 & --- & floor-plan node identifier \\
Location type & --- & Exog.-hetero. & Enum & f5 & --- & bedroom / bathroom / dining / communal / staff room / \ldots \\
Geometry / polygon & --- & Exog.-hetero. & Polygon & f5 & m & room footprint (from d1) \\
Associated wing / bed & --- & Exog.-hetero. & Location ref & f5 & --- & structural links \\
Occupancy & --- & Endog. & List[$\mathcal{R}{\cup}\mathcal{S}$] & f6; f10 & --- & persons currently in this room \\
\midrule
\multicolumn{7}{@{}l}{\textbf{\underline{Resident $\mathcal{R}$}}} \\
Care level & $c_i$ & Exog.-hetero. & Enum\{high,\allowbreak med,\allowbreak low\} & f5 & --- & drives the care-level activity schedule \\
Mobility & --- & Exog.-hetero. & Enum & f5 & --- & tied to care level (high\ensuremath{\rightarrow}low mobility) \\
Allocated bed / bath / wing & --- & Exog.-hetero. & Location ref & f5 & --- & set at init \\
Position & $(x^r_{i,t},y^r_{i,t})$ & Endog. & Float$\times$2 & f6 & m & continuous-space pedestrian position \\
Current activity / command & --- & Endog.-dec. & Enum\{move,\allowbreak wait\}\allowbreak+\allowbreak ref & f6 & --- & selected from the care-level schedule (d2) \\
Sleep / assisted / med. flags & --- & Endog. & Boolean & f6 & --- & activity-driven state \\
Persons in range & --- & Endog. & List[$\mathcal{R}{\cup}\mathcal{S}$] & f7 & --- & peers within $\rho$ this step \\
Open contact sessions & --- & Endog. & List[(peer, dur)] & f7; f8 & --- & transient sessions being aged \\
\midrule
\multicolumn{7}{@{}l}{\textbf{\underline{Staff $\mathcal{S}$}}} \\
Working / depart flags & --- & Endog. & Boolean & f9; f13 & --- & on-shift / leaving \\
Current working wing & --- & Endog. & Wing ref & f10 & --- & area of responsibility \\
Task / priority state & --- & Endog.-dec. & Enum+priority & f10 & --- & current task from the priority queue (d2) \\
Residents responsible & --- & Endog. & List[$\mathcal{R}$] & f9 & --- & residents in the assigned wing \\
Position & $(x^s_{j,t},y^s_{j,t})$ & Endog. & Float$\times$2 & f10 & m & continuous-space pedestrian position \\
Persons in range & --- & Endog. & List[$\mathcal{R}{\cup}\mathcal{S}$] & f11 & --- & peers within $\rho$ this step \\
Open contact sessions & --- & Endog. & List[(peer, dur)] & f11; f12 & --- & transient sessions being aged \\
\bottomrule
\end{longtable}
\egroup

\emph{Transient quantities (not stored):} the contact-verdict flag
\(\mathrm{isContact}:=\mathrm{duration}\ge\delta\) is computed inside the aging step
(f8/f12) when a session closes, then handed to the Environment (f4). The activity / task
templates themselves are datasets (d2), read by the decision-makers, not stored as agent
state.

\begin{center}\rule{0.5\linewidth}{0.5pt}\end{center}

\subsubsection{T3 --- Sensing}

\textbf{Spatial modality.} A person ``senses'' another iff the \textbf{continuous-space
Euclidean distance} between their positions is \(\le\rho\) (contact-policy range).
Co-location in a typed room is derived from position. Rows = observers; columns =
observed.

\bgroup
\setlength{\tabcolsep}{5pt}
\renewcommand{\arraystretch}{1.15}
\begin{longtable}{@{}>{\raggedright\arraybackslash}p{3.0cm}
                    >{\centering\arraybackslash}p{1.7cm}
                    >{\centering\arraybackslash}p{1.7cm}
                    >{\centering\arraybackslash}p{2.4cm}
                    >{\centering\arraybackslash}p{2.4cm}@{}}
\toprule
\textbf{Observer} & $\mathcal{M}$ & $\mathcal{L}$ & $\mathcal{R}$ & $\mathcal{S}$ \\
\midrule
$\mathcal{M}$ (Env) & $\emptyset$ & occupancy & $*$ & $*$ \\
$\mathcal{L}$ (Floor) & $\emptyset$ & $\emptyset$ & $\emptyset$ & $\emptyset$ \\
$\mathcal{R}$ (Resident) & $\tau_t,\sigma_t$ & $\emptyset$ & $\mathrm{pos}$ & $\mathrm{pos}$ \\
$\mathcal{S}$ (Staff) & $\tau_t,\sigma_t$ & $\emptyset$ & $\mathrm{pos}$ & $\mathrm{pos}$ \\
\bottomrule
\end{longtable}
\egroup

\textbf{Reading the matrix.} \(\mathrm{pos}\) denotes ``observes positions within radius
\(\rho\)'' --- the proximity query that drives contact detection; it is symmetric, so
\(\mathcal{R}\) senses \(\mathcal{S}\) and vice versa. The \(\mathcal{R}\) and
\(\mathcal{S}\) diagonals carry \(\mathrm{pos}\): a resident (resp. staff) observes the
positions of \emph{other} residents (resp. staff) within \(\rho\); self-state is implicit.
The Environment observes everything (it owns the global registries and computes the
matrix). \(\mathcal{L}\) is a spatial carrier with no sensing of its own; room membership
is \emph{derived} from person positions, not sensed by the room. The activity/task
templates (d2) are read once at init by \(\mathcal{R}\)/\(\mathcal{S}\).

\begin{center}\rule{0.5\linewidth}{0.5pt}\end{center}

\subsubsection{T4 --- Internal Function}

Complex updates are given as numbered equations below the table. The ``External effect:
movement delegated to Pedestrian Library'' flag marks the proprietary movement that is the
reproduction blocker (see the Reproduction blocker section below).

\bgroup
\setlength{\tabcolsep}{4pt}
\renewcommand{\arraystretch}{1.12}
\begin{longtable}{@{}>{\raggedright\arraybackslash}p{0.7cm}
                    >{\raggedright\arraybackslash}p{2.4cm}
                    >{\raggedright\arraybackslash}p{2.2cm}
                    >{\raggedright\arraybackslash}p{3.1cm}
                    >{\raggedright\arraybackslash}p{3.6cm}
                    >{\raggedright\arraybackslash}p{3.2cm}
                    >{\raggedright\arraybackslash}p{1.0cm}@{}}
\toprule
\textbf{ID} & \textbf{Function} & \textbf{Method} & \textbf{Decision basis} & \textbf{Self-state update} & \textbf{External effect} & \textbf{Ref.} \\
\midrule
\multicolumn{7}{@{}l}{\textbf{\underline{Environment $\mathcal{M}$}}} \\
f1 & initialise\_model & Setup & d1 (floor plan), $I_R$, d2 (schedules) & $\tau,\sigma$, care-level map & create $\mathcal{L}$, $\mathcal{R}$, initial $\mathcal{S}$ & \citep{Stone2026} \\
f2 & advance\_clock & Time integration & $\Delta t$ & $\tau_t$ via \eqref{eq:ac-clock} & --- & --- \\
f3 & manage\_shift & Event-driven & $\tau_t;\,T_1,T_2,T_3$ & $\sigma_t$, shift-boundary flags & create $\mathcal{S}$ (shift start); remove $\mathcal{S}$ via f13 (shift end) & \citep{Stone2026} \\
f4 & aggregate\_contacts & Aggregation & closed sessions (all persons) & $C_t$ via \eqref{eq:ac-matrix} & append finalized records to the contact log (T6b) & \citep{Stone2026} \\
\midrule
\multicolumn{7}{@{}l}{\textbf{\underline{Resident $\mathcal{R}$}}} \\
f5 & initialise\_resident & Setup & d1 (allocated room), care level & $c_i$, mobility, allocations & --- & \citep{Stone2026} \\
f6 & advance\_resident\_activity & Statechart + pedestrian & care-level schedule (d2); $\tau_t$ & position, activity, sleep/assisted/med. flags & \textbf{movement delegated to Pedestrian Library}; update $\mathcal{L}$ occupancy & \citep{Helbing2000} \\
f7 & detect\_contacts & Proximity query & all-person positions; $\rho$ & persons-in-range, open sessions & --- (opens transient session) & \citep{Stone2026} \\
f8 & age\_contacts & Duration update & $\Delta t,\,\delta$ & open-session duration; close + verdict via \eqref{eq:ac-contact} & --- & \citep{Stone2026} \\
\midrule
\multicolumn{7}{@{}l}{\textbf{\underline{Staff $\mathcal{S}$}}} \\
f9 & initialise\_staff & Setup & $\sigma_t,\,\rho$; assigned wing & working flag, wing, residents-responsible & --- & \citep{Stone2026} \\
f10 & advance\_staff\_activity & Priority statechart + pedestrian & task set (d2); $\tau_t$; queue timeout & position, task/priority state, working wing & \textbf{movement delegated to Pedestrian Library}; update $\mathcal{L}$ occupancy & \citep{Helbing2000} \\
f11 & detect\_contacts & Proximity query & all-person positions; $\rho$ & persons-in-range, open sessions & --- (opens transient session) & \citep{Stone2026} \\
f12 & age\_contacts & Duration update & $\Delta t,\,\delta$ & open-session duration; close + verdict via \eqref{eq:ac-contact} & --- & \citep{Stone2026} \\
f13 & depart\_staff & Event-driven & shift end ($\tau_t$) & depart flag & remove self from $\mathcal{S}$ & \citep{Stone2026} \\
\midrule
\multicolumn{7}{@{}l}{\textbf{\underline{Floor Plan $\mathcal{L}$}}} \\
f5' & initialise\_location & Setup & d1 (floor-plan registry) & name, type, geometry, associations & --- & \citep{Stone2026} \\
\bottomrule
\end{longtable}
\egroup

\emph{(The Location init is labelled f5\('\) to avoid colliding with the Resident init f5;
it is a distinct function of a distinct agent type.)}

\textbf{Behavioural equations.} The clock and the contact aggregation are

\begin{equation}
\tau_t = \tau_{t-1} + \Delta t \pmod{86400}
\label{eq:ac-clock}
\end{equation}

\begin{equation}
\mathrm{isContact} := \mathrm{duration}\ge\delta,\qquad
C_t[\,\mathrm{role}(a),\mathrm{role}(b)\,]\mathrel{+}= \mathbf{1}[\mathrm{isContact}]
\quad\text{for each session closed at }t .
\label{eq:ac-contact}
\end{equation}

\begin{equation}
C_t[i,j]=\#\{\text{closed contact sessions at }\le t:
\mathrm{role}(\text{contacter}){=}i\,\land\,\mathrm{role}(\text{contactee}){=}j\,\land\,\mathrm{isContact}\}
\label{eq:ac-matrix}
\end{equation}

\textbf{Contact lifecycle (f7\ensuremath{\rightarrow}f8, mirrored by f11\ensuremath{\rightarrow}f12).} Each step, a person queries peers
within \(\rho\) (f7/f11); peers \emph{newly} in range open a session recording onset time.
Open sessions are aged each step (f8/f12: duration \(+\!=\Delta t\)). When a peer leaves
range, the session closes, the verdict \eqref{eq:ac-contact} is evaluated, and the closed
session is handed to the Environment. f4 (Environment) aggregates all freshly-closed
sessions into the matrix \(C_t\) \eqref{eq:ac-matrix} and the record log (T6b). This is
the epidemiological core; there is no separate ``contact-record agent'' --- contacts are
data produced by detection + aging and consumed by aggregation.

\begin{center}\rule{0.5\linewidth}{0.5pt}\end{center}

\subsubsection{T5 --- Associated Data}

\bgroup
\setlength{\tabcolsep}{4pt}
\renewcommand{\arraystretch}{1.12}
\begin{longtable}{@{}>{\centering\arraybackslash}p{0.7cm}
                    >{\raggedright\arraybackslash}p{3.6cm}
                    >{\raggedright\arraybackslash}p{2.0cm}
                    >{\raggedright\arraybackslash}p{1.6cm}
                    >{\raggedright\arraybackslash}p{2.4cm}
                    >{\raggedright\arraybackslash}p{3.6cm}
                    >{\raggedright\arraybackslash}p{2.4cm}@{}}
\toprule
\textbf{ID} & \textbf{Name} & \textbf{Type} & \textbf{Temporal} & \textbf{Collection} & \textbf{Pre-processing} & \textbf{Availability} \\
\midrule
d1 & Floor-plan / location registry & Empirical & Static & model distribution & geometry \ensuremath{\rightarrow} typed nodes/polygons; assigned location type & \textbf{Unavailable} (backing layout not distributed) \\
d2 & Activity schedules \& task templates (resident care-level activities; staff daily tasks with durations) & Empirical / Synthetic & Daily cycle & staff activity-diary survey + template objects & mapped to statechart tasks; care-level stratification; priority assignment & \textbf{Unavailable} (survey data collected for this study, not released) \\
d3 & Staff shift roster (shift starts, per-shift headcounts, wing assignments) & Empirical & Daily cycle & facility parameters & shift-boundary table & \textbf{Partial} (scalar params in T6a; full roster unavailable) \\
\bottomrule
\end{longtable}
\egroup

\textbf{All three are marked Unavailable / Partial.} The backing datasets were not
distributed with the model and could not be located (the activity-diary survey was
collected specifically for this study). This is \textbackslash textbf\{reproduction
blocker \#2\} (see below): the provenance chain is incomplete, so even the reproducible
custom logic cannot be fully instantiated. The IDs d1--d3 nevertheless exist, so every
empirical reference in T6a/T8 resolves to a T5 record (\ensuremath{\rightarrow} r19).

\begin{center}\rule{0.5\linewidth}{0.5pt}\end{center}

\subsubsection{T6 --- Input/Output}

\paragraph{(a) Inputs}

\bgroup
\setlength{\tabcolsep}{4pt}
\renewcommand{\arraystretch}{1.12}
\begin{longtable}{@{}>{\raggedright\arraybackslash}p{1.8cm}
                    >{\raggedright\arraybackslash}p{4.3cm}
                    >{\raggedright\arraybackslash}p{2.0cm}
                    >{\raggedright\arraybackslash}p{1.9cm}
                    >{\raggedright\arraybackslash}p{3.5cm}
                    >{\raggedright\arraybackslash}p{2.0cm}@{}}
\toprule
\textbf{Symbol} & \textbf{Value / Distribution} & \textbf{Data source} & \textbf{Derivation} & \textbf{Algorithm} & \textbf{Ref.} \\
\midrule
\multicolumn{6}{@{}l}{\textbf{\underline{Environment $\mathcal{M}$}}} \\
$I_R$ & 60 & Author & Assumed & --- & --- \\
$\rho$ & 1.5 or 3 (scenario) & Author & Assumed & --- & --- \\
$\delta$ & 3 s & Author & Assumed & --- & --- \\
$\Delta t$ & 1.0 s & Author & Direct & --- & --- \\
sim start & 6:15 & Author & Assumed & --- & --- \\
$T_1,T_2,T_3$ & 6:45 / 14:30 / 22:25 & d3 (partial) & Assumed & --- & --- \\
$M_1,M_2,M_3$ & 12 / 12 / 6 & Author & Assumed & --- & --- \\
staff per wing & 2 & Author & Assumed & --- & --- \\
queue timeout & 40 min & Author & Assumed & --- & --- \\
floor plan & (layout) & d1 (unavailable) & Estimated & --- & --- \\
activity/task schedules & (templates) & d2 (unavailable) & Estimated & --- & --- \\
\midrule
\multicolumn{6}{@{}l}{\textbf{\underline{Resident $\mathcal{R}$}}} \\
care-level mix & per facility (high/med/low) & Author & Assumed & --- & --- \\
\midrule
\multicolumn{6}{@{}l}{\textbf{\underline{Staff $\mathcal{S}$}}} \\
(init per shift) & $M_1,M_2,M_3$ per wing & Author & Direct & --- & --- \\
\bottomrule
\end{longtable}
\egroup

\paragraph{(b) Outputs}

\bgroup
\setlength{\tabcolsep}{4pt}
\renewcommand{\arraystretch}{1.12}
\begin{longtable}{@{}>{\centering\arraybackslash}p{0.8cm}
                    >{\raggedright\arraybackslash}p{2.8cm}
                    >{\raggedright\arraybackslash}p{1.8cm}
                    >{\raggedright\arraybackslash}p{2.2cm}
                    >{\raggedright\arraybackslash}p{1.8cm}
                    >{\raggedright\arraybackslash}p{4.4cm}
                    >{\raggedright\arraybackslash}p{2.5cm}@{}}
\toprule
\textbf{ID} & \textbf{Output} & \textbf{Symbol} & \textbf{Type} & \textbf{Derivation} & \textbf{Formula} & \textbf{Freq.} \\
\midrule
o1 & Contact matrix & $C$ & Matrix[Integer] & Computed & \eqref{eq:ac-matrix} & $-1$ (terminal) \\
o2 & Contact-record log & $\{\kappa\}$ & Table & Direct & each closed session (contacter/contactee role, onset, duration, isContact) & $\Delta t$ (streaming) \\
\bottomrule
\end{longtable}
\egroup

\begin{center}\rule{0.5\linewidth}{0.5pt}\end{center}

\subsubsection{T7 --- Schedule}

\paragraph{(a) Execution}

\bgroup
\setlength{\tabcolsep}{4pt}
\renewcommand{\arraystretch}{1.12}
\begin{longtable}{@{}>{\centering\arraybackslash}p{0.9cm}
                    >{\centering\arraybackslash}p{1.1cm}
                    >{\centering\arraybackslash}p{0.9cm}
                    >{\raggedright\arraybackslash}p{3.3cm}
                    >{\raggedright\arraybackslash}p{4.6cm}
                    >{\raggedright\arraybackslash}p{4.6cm}@{}}
\toprule
\textbf{Step} & \textbf{Agent} & \textbf{ID} & \textbf{Function} & \textbf{Exec. mode} & \textbf{Condition} \\
\midrule
1 & $\mathcal{M}$ & f2 & advance\_clock & Synchronous ($\Delta t$) & every step \\
2 & $\mathcal{M}$ & f3 & manage\_shift & Event-driven (shift table) & $\tau_t\in\{T_1,T_2,T_3\}$ \\
3 & $\mathcal{R}$ & f5 & initialise\_resident & Synchronous & at init / on resident inject \\
4 & $\mathcal{R}$ & f6 & advance\_resident\_activity & Composite: Sequential ($\mathcal{R}$) $\oplus$ Asynchronous (library; velocity, $\rho$) & every step \\
4 & $\mathcal{R}$ & f7 & detect\_contacts & Composite: Sequential ($\mathcal{R}$) $\oplus$ Asynchronous (library) & every step \\
4 & $\mathcal{R}$ & f8 & age\_contacts & Composite: Sequential ($\mathcal{R}$) $\oplus$ Asynchronous (library) & every step \\
5 & $\mathcal{S}$ & f9 & initialise\_staff & Event-driven (shift table) & at shift start / end \\
5 & $\mathcal{S}$ & f13 & depart\_staff & Event-driven (shift table) & at shift end \\
6 & $\mathcal{S}$ & f10 & advance\_staff\_activity & Composite: Sequential ($\mathcal{S}$) $\oplus$ Asynchronous (library; velocity, $\rho$) & every step \\
6 & $\mathcal{S}$ & f11 & detect\_contacts & Composite: Sequential ($\mathcal{S}$) $\oplus$ Asynchronous (library) & every step \\
6 & $\mathcal{S}$ & f12 & age\_contacts & Composite: Sequential ($\mathcal{S}$) $\oplus$ Asynchronous (library) & every step \\
7 & $\mathcal{M}$ & f4 & aggregate\_contacts & Synchronous & every step (after contact cycles) \\
8 & $\mathcal{L}$ & f5$'$ & initialise\_location & Synchronous & at init only \\
\bottomrule
\end{longtable}
\egroup

\textbf{Exec.mode note.} The agent-level updates --- clock, activity/task selection,
contact detection/aging --- are integrated \textbf{synchronously} (breed-sequential:
environment clock \ensuremath{\rightarrow} residents \ensuremath{\rightarrow} staff, each iterated per 1-second step, then f4 aggregates
contacts). The pedestrian \textbf{movement} (position updates for f6/f10) is integrated by
the \textbf{proprietary AnyLogic Pedestrian Library} in a continuous-time
\textbf{Asynchronous} sub-step with no open specification. Per VISA, this is a
\textbf{composite} mode (Sequential \(\oplus\) Asynchronous); composites are built from
base modes, and the proprietary sub-step is the reproduction blocker (see below).

\paragraph{(b) Termination}

\bgroup
\setlength{\tabcolsep}{4pt}
\renewcommand{\arraystretch}{1.12}
\begin{longtable}{@{}>{\centering\arraybackslash}p{0.9cm}
                    >{\raggedright\arraybackslash}p{1.9cm}
                    >{\raggedright\arraybackslash}p{2.9cm}
                    >{\raggedright\arraybackslash}p{3.0cm}
                    >{\raggedright\arraybackslash}p{2.4cm}
                    >{\raggedright\arraybackslash}p{3.6cm}@{}}
\toprule
\textbf{ID} & \textbf{Indicator} & \textbf{Condition} & \textbf{Description} & \textbf{Value source/ref} & \textbf{Termination logic} \\
\midrule
c1 & $\tau_t$ & $\tau_t\ge$ simulation horizon & Configured run length (e.g. 24 h) reached & Author assumed & Stop when $c_1$; f4 then emits $C$ + $\{\kappa\}$ \\
\bottomrule
\end{longtable}
\egroup

\begin{center}\rule{0.5\linewidth}{0.5pt}\end{center}

\subsubsection{T8 --- Validation}

Validation uses the HAV framework at three levels. Because full reproduction is
\textbf{blocked} (proprietary library + unavailable data), every entry's \emph{empirical}
status is \textbf{Blocked}; the table remains structurally valid --- each indicator traces
to a T2/T6b quantity, and each benchmark traces to a T5 record --- it simply cannot be
executed.

\bgroup
\setlength{\tabcolsep}{4pt}
\renewcommand{\arraystretch}{1.12}
\begin{longtable}{@{}>{\centering\arraybackslash}p{0.8cm}
                    >{\centering\arraybackslash}p{1.2cm}
                    >{\raggedright\arraybackslash}p{4.6cm}
                    >{\raggedright\arraybackslash}p{2.8cm}
                    >{\raggedright\arraybackslash}p{3.0cm}
                    >{\raggedright\arraybackslash}p{2.6cm}@{}}
\toprule
\textbf{ID} & \textbf{Level} & \textbf{Hypothesis} & \textbf{Indicator} & \textbf{Benchmark} & \textbf{Status} \\
\midrule
v1 & Output & $C$ reproduces the empirical aged-care contact pattern by role$\times$role & $C[i,j]$ cells & source-paper matrices (d-class) & \textbf{Blocked} \\
v2 & Model & Staff--resident contacts concentrate during care activities (meals, meds) and in care zones & temporal/spatial distribution of staff--resident contacts & qualitative expectation (paper) & Blocked \\
v3 & Output & Contact-duration distribution is consistent with the $\ge\delta$ definition & duration histogram & constructed (threshold $\delta$) & Blocked \\
v4 & Model & Total contacts scale with staff:resident ratio and population density & $\#\{$contacts$\}$ vs $I_R,M_\text{shift}$ & sensitivity expectation (paper) & Blocked \\
v5 & Output & Matrices differ substantially between $\rho{=}1.5$ m and $\rho{=}3.0$ m & $C_{1.5}$ vs $C_{3.0}$ & known sensitivity (paper) & Blocked \\
\bottomrule
\end{longtable}
\egroup

\begin{center}\rule{0.5\linewidth}{0.5pt}\end{center}

\subsubsection{Consistency check (19 rules)}

Within-table (r1--r4) and cross-table (r5--r19), per the revised VISA rule set.

{\def\LTcaptype{none} 
\begin{longtable}[]{@{}
  >{\raggedright\arraybackslash}p{(\linewidth - 4\tabcolsep) * \real{0.2842}}
  >{\centering\arraybackslash}p{(\linewidth - 4\tabcolsep) * \real{0.0632}}
  >{\raggedright\arraybackslash}p{(\linewidth - 4\tabcolsep) * \real{0.6526}}@{}}
\toprule\noalign{}
\begin{minipage}[b]{\linewidth}\raggedright
Rule
\end{minipage} & \begin{minipage}[b]{\linewidth}\centering
Res.
\end{minipage} & \begin{minipage}[b]{\linewidth}\raggedright
Note
\end{minipage} \\
\midrule\noalign{}
\endhead
\bottomrule\noalign{}
\endlastfoot
r1 Variable-type coverage \& time-indexing & PASS & every T2 variable carries one of the
four leaf types; every endogenous variable carries the time index \(t\)
(\(\tau_t,\sigma_t,C_t\), resident/staff positions, occupancy, open sessions) \\
r2 Function-ID uniqueness & PASS & f1--f4, f5--f8, f9--f13, f5\('\) are unique within
their owners; every f-ID in T2's Value column exists in T4 \\
r3 Function productivity & PASS & every function writes \ensuremath{\geq}1 endogenous variable or
creates/removes an instance (f1 setup; f2 \(\tau_t\); f3 \(\sigma_t\)+create/remove; f4
\(C_t\); f5 allocations; f6 position/flags; f7 persons-in-range/sessions; f8 session
duration; f9--f12 likewise; f13 remove) \\
r4 Step--Exec.mode consistency & PASS & within each T7a step the agent-update mode is
consistent; the proprietary Asynchronous sub-step is an explicit composite component, not
a contradictory base mode \\
r5 Same-type (peer) sensing & PASS & \(\mathcal{R}\) and \(\mathcal{S}\) diagonals state
\(\mathrm{pos}\) (peer positions within \(\rho\)); single-instance types
(\(\mathcal{M},\mathcal{L}\)) have \(\emptyset\) diagonals; self-observation is
implicit \\
r6 Passive-agent implications & PASS & no Passive agents (ContactRecord/ReferenceData were
removed: records\ensuremath{\rightarrow}Environment aggregate, data\ensuremath{\rightarrow}T5) \\
r7 Observer-row completeness & PASS & every active agent
(\(\mathcal{M},\mathcal{L},\mathcal{R},\mathcal{S}\)) has exactly one T3 row \\
r8 Population-dynamics consistency & PASS & variable sets \(\mathcal{R},\mathcal{S}\) each
own a creation and a removal function (f5 / f9-create + f13); fixed sets
\(\mathcal{M},\mathcal{L}\) own none \\
r9 Active-agent function coverage & PASS & every non-Passive agent owns \ensuremath{\geq}1 function ---
\(\mathcal{M}\)(f1--f4), \(\mathcal{R}\)(f5--f8), \(\mathcal{S}\)(f9--f13),
\(\mathcal{L}\)(f5\('\)) \\
r10 Variable observability & PASS & every variable is self-read or T3-authorized (peer
positions via \(\mathrm{pos}\) within \(\rho\); globals \(\tau_t,\sigma_t\); room
occupancy derived from position; activity/task templates read from d2 at init) \\
r11 Endogenous-variable completeness & PASS & every endogenous variable is written by \ensuremath{\geq}1
existing f-ID --- \(\tau_t\)(f2), \(\sigma_t\)(f3), \(C_t\)(f4), allocations(f5), resident
position/flags(f6), sessions(f7--f8), staff state(f9--f12), occupancy(f6/f10) \\
r12 Self-state-update validity & PASS & every Self-state target is endogenous of the
owning agent \\
r13 External-effect validity & PASS & every External Effect is a valid instance create
(\(\mathcal{L},\mathcal{R},\mathcal{S}\)), remove (\(\mathcal{S}\) via f13), occupancy
update (\(\mathcal{L}\)), or output emission \\
r14 Information-access validation & PASS & every Decision-basis variable is a
self-attribute or T3-authorized (peer positions within \(\rho\); co-room occupancy; d2
templates at init) \\
r15 Input--output coverage (bidirectional) & PASS & every T2 Input variable has a T6a
entry; every T6a symbol is an exog T2 variable; outputs computable from T2 \\
r16 Schedule coverage (bidirectional) & PASS & every T4 function appears in \ensuremath{\geq}1 T7a step;
every T7a step's function ID exists in T4 \\
r17 Termination-indicator source & PASS & c1 references \(\tau_t\) (T2) and triggers
f4/o1,o2 (T6b) \\
r18 Validation-object coverage & PASS & v1--v5 indicators (\(C\), contact log, duration
histogram, contact counts) all trace to T2/T6b \\
r19 Data-reference resolution & PASS & d1 (floor plan), d2 (schedules), d3 (roster) all
exist in T5; all T6a/T8 empirical references resolve to a d-ID or to Author \\
\end{longtable}
}

\textbf{Summary: 19/19 PASS.} An 8-table specification authored from an external,
independently-authored AnyLogic model --- with no structural accommodation --- passes all
nineteen consistency rules. This is the expressiveness evidence for Model 3.

\begin{center}\rule{0.5\linewidth}{0.5pt}\end{center}

\subsubsection{Reproduction blocker}

\texttt{visa-code} is \textbf{not} invoked for this model and no reproduction is
attempted. Two independent, structural barriers block the loop
\texttt{visa-code\ \ensuremath{\rightarrow}\ reproduce}:

\paragraph{Blocker 1 --- proprietary Pedestrian Library (the dominant barrier)}

All movement --- both residents and staff --- is executed by the
\textbf{AnyLogic Pedestrian
Library}, which implements the social-force model of \citep{Helbing2000}. The model's own
logic only emits \textbf{movement commands} (move to a destination node / wait / depart);
the library decides the \textbf{physics} --- priority-based pathfinding, congestion- and
density-dependent velocity, continuous-space crowd navigation, collision avoidance. None
of this is specified in the workspace or in any open document; it lives in the closed
pedestrian-library component.

\textbf{Why this blocks reproduction.} The model's output --- the contact matrix --- is
generated by contact detection, and contacts arise from \textbf{co-location} (who is
within \(\rho\) of whom at each second). Co-location arises from \textbf{movement}.
Reproducing only the model's own command logic (captured in T1--T8) reproduces \emph{what
each agent decides to do} but not \emph{how it moves}, hence not where agents co-locate,
hence not the matrices. The output is dominated by the opaque library.

\paragraph{Blocker 2 --- unavailable backing data}

The model loads its floor plan (d1), activity/task schedules (d2), and shift roster (d3)
from databases / hardcoded tables that were \textbf{not distributed} with the model and
could not be located --- the activity-diary survey was collected specifically for this
study. Without them, even the reproducible custom logic cannot be fully instantiated
(e.g.~the location registry \(\mathcal{L}\) and the care-level activity schedules are
empty).

\paragraph{Why this is a finding, not a VISA failure}

VISA cleanly \textbf{separates} the two kinds of behaviour in this model:

\begin{itemize}
\tightlist
\item
  \textbf{(a) custom Java logic} --- activity scheduling, the priority staff task
  statechart, shift management, contact detection at \(\le\rho\) for \(\ge\delta\), matrix
  aggregation, I/O --- \textbf{fully captured} in T1--T8 and \textbf{reproducible};
\item
  \textbf{(b) AnyLogic Pedestrian Library navigation} --- flagged in T4 (``movement
  delegated to Pedestrian Library'') and T7a (Asynchronous composite sub-step),
  \textbf{not reproducible}.
\end{itemize}

By making (a) explicit and (b) a named, localized dependency, VISA \textbf{exposes that
the proprietary library is the real reproducibility barrier} --- a \emph{transparency}
contribution. A reader of the 8 tables can see exactly which parts of the model are open
and which are not. This is precisely the expressiveness claim Model 3 was chosen to
demonstrate: VISA scales to a large real-world ABM, faithfully, and honestly demarcates
its own limit.
}

\section{The three VISA skills}
\label{sec:skills}

The three LLM-executable skills that operationalize VISA --- authoring,
checking, and code generation. These are released as supplementary material and
maintained on the companion GitHub repository.

{\small 
\subsection{VISA Authoring Skill}

\begin{quote}
\textbf{Purpose.} Guide an LLM agent in \textbf{constructing the eight VISA tables} for an
agent-based simulation (ABM) model, so the output is complete, internally consistent, and
machine-parseable. This skill is the constructive counterpart to \texttt{visa-check},
which verifies the 19 consistency rules.
\end{quote}

\subsubsection{When to invoke}

Invoke this skill when the user provides a model description (natural-language narrative,
an ODD description, equations, or a reference implementation) and asks to
\textbf{produce / convert / document it as a VISA specification}.

\subsubsection{Companion skill}

After authoring (or after each table), invoke \texttt{visa-check} to verify the 19
consistency rules. Author and check in tandem until all rules pass.

\begin{center}\rule{0.5\linewidth}{0.5pt}\end{center}

\subsubsection{Authoring pipeline (mandatory order)}

Build the tables \textbf{in this order}. Each table supplies identifiers that later tables
cross-reference; producing them out of order creates dangling references.

\begin{verbatim}
T1 Agent  ->  T2 Variable  ->  T3 Sensing  ->  T4 Internal Function
         ->  T5 Associated Data  ->  T6 Input/Output  ->  T7 Schedule  ->  T8 Validation
\end{verbatim}

\bgroup
\setlength{\tabcolsep}{4pt}
\renewcommand{\arraystretch}{1.1}
\begin{longtable}{@{}>{\raggedright\arraybackslash}p{3.4cm}>{\raggedright\arraybackslash}p{13.0cm}@{}}
\toprule
After you finish\ldots{} & The next tables can now safely reference\ldots{} \\
\midrule
T2 & every variable Symbol and its Type (exog/endog) \\
T3 & which (observer, observed) pairs exist and which variables each cell exposes \\
T4 & function IDs f1, f2, \ldots{} (consumed by T2 Value, T7 schedule) \\
T5 & data IDs d1, d2, \ldots{} (consumed by T6 Data source, T8 Benchmark data) \\
T6a & exogenous parameter values/distributions (consumed by T2 exogenous "Input") \\
T6b & output indicator symbols (consumed by T7b, T8) \\
T7a & the per-step execution loop; T7b termination indicators \\
T8 & (terminal; consumes T2 variables and T6b indicators as validation objects) \\
\bottomrule
\end{longtable}
\egroup

\begin{center}\rule{0.5\linewidth}{0.5pt}\end{center}

\subsubsection{Notation conventions (apply in every table)}

\bgroup
\setlength{\tabcolsep}{4pt}
\renewcommand{\arraystretch}{1.1}
\begin{longtable}{@{}>{\raggedright\arraybackslash}p{3.5cm}>{\raggedright\arraybackslash}p{6.3cm}>{\raggedright\arraybackslash}p{6.5cm}@{}}
\toprule
Element & Convention & Example \\
\midrule
Agent \textbf{instance} & lowercase italic + subscript & $v_i$ the $i$-th vendor \\
\textbf{Exogenous} param / aggregate & uppercase italic & $N$ population, $D$ demand, $T$ horizon \\
\textbf{Endogenous} variable & lowercase italic + subscript \textbf{+ time index $t$} & $q_{i,t}$ order qty, $p_{i,t}$ price, $x_{i,t}$ position \\
\textbf{Vector / matrix} & boldface & $\mathbf{s}_{i,t}$ sales history, $A_t$ adjacency \\
Reserved indices & $i,j,k$ instances; $t$ time step & --- \\
\textbf{Fixed} count & $N_\mathcal{X}$ & $N_\mathcal{W}=1$ \\
\textbf{Variable} count & $n_\mathcal{X}$ & $n_\mathcal{V}$ (vendors enter/exit) \\
Optimal value & superscript $*$ & $q^*$ \\
Temporal lag & subscript $t-1$ & $s_{i,t-1}$ \\
\bottomrule
\end{longtable}
\egroup

\begin{quote}
\textbf{Time index on endogenous variables (r1).} Every endogenous variable evolves over
the simulation, so its \textbf{Symbol} in T2 carries the time index \texttt{\$t\$} (e.g.
\texttt{\$x\_\{i,t\}\$}, \texttt{\$p\_\{i,t\}\$}); exogenous variables do not. The index
is written explicitly in T2 (where each variable is defined) and may be suppressed in T4
cells, equations, and prose for readability.
\end{quote}

\subsubsection{Identifier conventions (stable IDs, reusable across tables)}

\bgroup
\setlength{\tabcolsep}{4pt}
\renewcommand{\arraystretch}{1.1}
\begin{longtable}{@{}>{\raggedright\arraybackslash}p{3.5cm}>{\raggedright\arraybackslash}p{6.3cm}>{\raggedright\arraybackslash}p{6.5cm}@{}}
\toprule
Prefix & Meaning & Lives in \\
\midrule
\texttt{d} & associated data record (d1, d2, \ldots{}) & T5 \\
\texttt{c} & termination condition (c1, c2, \ldots{}) & T7b \\
\texttt{v} & validation entry (v1, v2, \ldots{}) & T8 \\
\texttt{r} & consistency rule (r1--r19) & rules table (handled by \texttt{visa-check}) \\
\bottomrule
\end{longtable}
\egroup

\begin{center}\rule{0.5\linewidth}{0.5pt}\end{center}

\subsubsection{Table-by-table schema}

\paragraph{T1 --- Agent}

\emph{Enumerates the model's ontology: agent types, roles, counts.}

\bgroup
\setlength{\tabcolsep}{4pt}
\renewcommand{\arraystretch}{1.1}
\begin{longtable}{@{}>{\raggedright\arraybackslash}p{3.4cm}>{\raggedright\arraybackslash}p{13.0cm}@{}}
\toprule
Column & Required content \\
\midrule
\textbf{Set} & Calligraphic symbol ($\mathcal{E},\mathcal{S},\mathcal{W},\mathcal{V},\mathcal{C},\dots$) \\
\textbf{Instances} & Instance symbol(s), e.g. $e$, $v_1,v_2,\ldots$ \\
\textbf{Category} & \textbf{One of:} \texttt{Environment} \textbar{} \texttt{Space} \textbar{} \texttt{Decision-maker} \textbar{} \texttt{Passive} (definitions below) \\
\textbf{Description} & One-line role \\
\textbf{Quantity} & $N_\mathcal{X}=k$ (fixed) or $n_\mathcal{X}$ (variable) \\
\bottomrule
\end{longtable}
\egroup

\textbf{Category definitions (controlled vocabulary):} - \textbf{Environment} --- manages
the lifecycle of other agent instances (creation, removal, ordering) and computes
model-level statistics / output aggregates (e.g., a mean price, a Gini coefficient).
Parameters and state that belong to a specific decision-maker are attributed to that
decision-maker, not dumped on the Environment. - \textbf{Space} --- an \textbf{active}
agent that \textbf{owns the spatial/topological geometry} of the world (dimensions,
boundary/wrap conditions, room layout) and maintains the spatial or topological
relationships among the currently-alive agents (a grid answering vision-range queries, a
network exposing neighbors). It typically owns a spatial-query function and is not a
passive container. A metric that belongs to a \emph{sensing} decision-maker (e.g., a
vision radius) is attributed to that decision-maker, not to the Space agent --- the Space
agent only reads it as a query argument. Model space implicitly only when there is no
spatial interaction at all. - \textbf{Decision-maker} --- executes behavioral rules, makes
autonomous choices from internal state + sensed info. - \textbf{Passive} --- static
information carrier; no autonomous behavior. \emph{(Implications enforced by r6: no
endogenous variables, column-only in T3, no functions in T4.)}

\begin{quote}
\textbf{What is --- and is not --- an agent (modeling logic).} Only things that \emph{behave
autonomously} become agents; the host platform's object graph is not the taxonomy. Two
things that platform code often exposes as separate classes are \textbf{not} agents: (i) a
simulation \textbf{output datum} (a finalized contact record, an emitted event, a logged
transaction) is \emph{data produced by a function} --- model it as an Environment
aggregate and/or a T6b output, not a T1 agent; (ii) \textbf{reference / lookup data} (a
floor-plan registry, a survey, a parameter table, a template library) is a \emph{dataset}
--- model it in T5 (Associated Data), not as a Passive agent. Likewise, a parameter read
by several decision-maker types is \textbf{declared on each} of those types in T2 (its
value listed once in T6a), never parked in the Environment.
\end{quote}

\textbf{Quantity rule:} \(N\) = predetermined fixed count; \(n\) = count that may change
during the run (entry/exit).

\begin{center}\rule{0.5\linewidth}{0.5pt}\end{center}

\paragraph{T2 --- Variable}

\emph{Complete state-space specification: every variable of every agent type.}

\bgroup
\setlength{\tabcolsep}{4pt}
\renewcommand{\arraystretch}{1.1}
\begin{longtable}{@{}>{\raggedright\arraybackslash}p{3.4cm}>{\raggedright\arraybackslash}p{13.0cm}@{}}
\toprule
Column & Required content \\
\midrule
\textbf{Symbol} & Notation per conventions above \\
\textbf{Type} & \textbf{One of four leaf types} (see below) \\
\textbf{Data Type} & Programming type: \texttt{Integer}, \texttt{Float}, \texttt{Boolean}, \texttt{List[Integer]}, \texttt{Matrix[Integer]}, \ldots{} \\
\textbf{Value} & \texttt{Input} (exogenous \ensuremath{\rightarrow} ref T6a) \textbf{or} function ID(s) \texttt{f4}, \texttt{f1, f2} (endogenous \ensuremath{\rightarrow} ref T4) \\
\textbf{Unit} & Physical/monetary unit, or \texttt{--} if dimensionless \\
\textbf{Desc.} & Short description \\
\bottomrule
\end{longtable}
\egroup

Group rows by agent type using \texttt{\textbackslash{}multicolumn} headers (matching T1
order).

\textbf{Type --- controlled vocabulary (two-level):} - \textbf{Exogenous} (value supplied
from outside; no internal function writes it) - \texttt{Exog.-homo.} --- identical across
all instances - \texttt{Exog.-hetero.} --- varies per instance (e.g., drawn from a
distribution) - \textbf{Endogenous} (value set within the model by a function) -
\texttt{Endog.-dec.} --- directly set by the agent's own decision function -
\texttt{Endog.} (non-decision) --- computed consequence, not directly controlled by a
decision

\begin{quote}
\textbf{Single-instance relaxation (notation, not a rule):} for any agent type with
\(N{=}1\), the homo/hetero distinction is vacuous; its exogenous variables need not
specify it. This is a notation convenience, not a consistency rule.
\end{quote}

\begin{quote}
\textbf{Value provenance:} exogenous \ensuremath{\rightarrow} \texttt{Input} (T6a gives the value); endogenous \ensuremath{\rightarrow}
the f-ID(s) that compute it.
\end{quote}

\begin{center}\rule{0.5\linewidth}{0.5pt}\end{center}

\paragraph{T3 --- Sensing}

\emph{Information-access matrix: which observer reads which variables from which observed
agent.}

\begin{itemize}
\tightlist
\item
  \textbf{Rows} = observer agents (every \textbf{non-Passive} type, including Environment
  and Space).
\item
  \textbf{Columns} = observed agents (\textbf{all} types). Split with a vertical bar:
  active agents (Env/Space/Decision-maker) left, Passive agents right. Passive agents are
  \textbf{columns only} (never rows).
\item
  \textbf{Cell values (controlled notation):}

  \begin{itemize}
  \tightlist
  \item
    \texttt{*} --- access to \textbf{all} attributes of the observed agent (off-diagonal
    only)
  \item
    \texttt{\ensuremath{\emptyset}} --- no access
  \item
    list of specific variable Symbols, e.g.~\texttt{p\_i,\ \textbackslash{}pi\_i,\ w\_i}
  \item
    \textbf{Diagonal} (an instance observing \emph{other} instances of its own type): list
    the observable \textbf{peer} attributes directly, e.g.~\texttt{p\_j} (no braces); use
    \texttt{\ensuremath{\emptyset}} if same-type agents observe nothing of each other

    \begin{itemize}
    \tightlist
    \item
      Every agent reads its \textbf{own} full state by default --- self-observation is
      implicit and \textbf{never} appears on the diagonal, which lists only peer
      attributes
    \item
      \texttt{\ensuremath{\emptyset}} for any single-instance type (\(N{=}1\)), since it has no peers
    \end{itemize}
  \end{itemize}
\end{itemize}

Space agents appear as \textbf{columns} too: they mediate topology (e.g., a vendor reads
the adjacency matrix \(A_t\) from the Space agent to find its neighbors).

\begin{center}\rule{0.5\linewidth}{0.5pt}\end{center}

\paragraph{T4 --- Internal Function}

\emph{Behavioral logic of every agent: what each function reads, updates, and affects
externally.}

\bgroup
\setlength{\tabcolsep}{4pt}
\renewcommand{\arraystretch}{1.1}
\begin{longtable}{@{}>{\raggedright\arraybackslash}p{3.4cm}>{\raggedright\arraybackslash}p{13.0cm}@{}}
\toprule
Column & Required content \\
\midrule
\textbf{Function} & snake\_case name, e.g. \texttt{decide\_order} \\
\textbf{Method} & Short method label, e.g. \texttt{Optimization}, \texttt{Mean}, \texttt{Multinomial logit}, \texttt{Preferential attachment}, \texttt{Conditional} \\
\textbf{Decision Basis} & Variables consumed (from T2 / sensed from T3). Must be self-attributes or T3-authorized observations (r14). \\
\textbf{Self-state Update} & Endogenous variables of \textbf{this} agent modified, with assignment, e.g. $q_i \leftarrow$ Eq.~(3). Must be endogenous (r12). \\
\textbf{External Effect} & Endogenous variables of \textbf{other} agents modified, or \texttt{---}. Must be endogenous vars of other agents (r13), or an instance create/remove of a variable-quantity type (r8); never another's exogenous var. \\
\textbf{Ref.} & \texttt{\textbackslash{}citep\{...\}} motivating the Method, or \texttt{---} for purely mechanical computation (arithmetic, mean, transfer). \\
\bottomrule
\end{longtable}
\egroup

Group rows by agent type. Every function exists to update endogenous variables --- its own
(Self-state Update) \textbf{or} others' (External Effect).

\begin{quote}
\textbf{Keep cells concise --- formalize, then reference.} This is \emph{why} symbols are
defined first. If a Self-state / External-effect update is more than a short phrase, write
it as a \textbf{numbered equation} below the table and put only the written variables + an
equation reference in the cell
(e.g.~\texttt{\$x\_\{i,t\},\ y\_\{i,t\}\$\ via\ Eq.\textasciitilde{}(3)}). Avoid inventing
inline shorthand (e.g.~ad-hoc ``Rule A'') that is never defined; every piece of logic must
be either a named, defined equation or a direct assignment. The Self-state Update /
External Effect columns then record \emph{which} endogenous variable is written and
\emph{which equation} governs it --- not prose.
\end{quote}

\begin{quote}
\textbf{Table presentation.} Within any table partitioned by agent (T2, T4, T6a), precede
each agent's block with a full-width separator row (a horizontal rule above, then the bold
agent name) so the partition is visually unambiguous. Allocate column widths by content
--- wide columns for prose (Description, Decision basis, updates), narrow for
symbols/units/flags --- rather than dividing evenly.
\end{quote}

\begin{center}\rule{0.5\linewidth}{0.5pt}\end{center}

\paragraph{T5 --- Associated Data}

\emph{Complete provenance chain for every external data source.}

\bgroup
\setlength{\tabcolsep}{4pt}
\renewcommand{\arraystretch}{1.1}
\begin{longtable}{@{}>{\raggedright\arraybackslash}p{3.4cm}>{\raggedright\arraybackslash}p{13.0cm}@{}}
\toprule
Column & Controlled vocabulary / content \\
\midrule
\textbf{Title} & Descriptive name \\
\textbf{Type} & \texttt{Empirical} \textbar{} \texttt{Literature} \textbar{} \texttt{Generated} \\
\textbf{Temporal} & \texttt{Static} \textbar{} \texttt{Dynamic} \\
\textbf{Source} & Named source + year, or \texttt{Author} \\
\textbf{Collection} & \texttt{Survey} \textbar{} \texttt{Administrative} \textbar{} \texttt{Sensor} \textbar{} \texttt{Experimental} \textbar{} \texttt{Computational} \\
\textbf{Pre-processing} & \texttt{None} \textbar{} \texttt{Selected} \textbar{} \texttt{Aggregated} \textbar{} \texttt{Transformed} \\
\textbf{\#Rec.} & Record count \\
\textbf{Avail.} & \texttt{Open} \textbar{} \texttt{Restricted} \textbar{} \texttt{Private} \\
\bottomrule
\end{longtable}
\egroup

\textbf{Definitions (use precisely):} - \textbf{Type:} \emph{Empirical} = real-world
observation/measurement; \emph{Literature} = peer-reviewed values/datasets;
\emph{Generated} = synthetic / sampled / auxiliary-sim output. - \textbf{Temporal:}
\emph{Static} = loaded once; \emph{Dynamic} = reloaded/updated during the run. -
\textbf{Collection:} \emph{Survey}, \emph{Administrative} (institutional/gov records),
\emph{Sensor} (automated measurement), \emph{Experimental} (controlled experiment),
\emph{Computational} (algorithm/auxiliary sim). - \textbf{Pre-processing:} \emph{None}
(raw), \emph{Selected} (subset extraction), \emph{Aggregated} (summary stats / coarser
grains), \emph{Transformed} (math/type/normalization, same record count). -
\textbf{Availability:} \emph{Open} / \emph{Restricted} (conditional) / \emph{Private}.

\begin{center}\rule{0.5\linewidth}{0.5pt}\end{center}

\paragraph{T6 --- Input and Output}

Two sub-tables.

\subparagraph{(a) Input --- every exogenous parameter}

\bgroup
\setlength{\tabcolsep}{4pt}
\renewcommand{\arraystretch}{1.1}
\begin{longtable}{@{}>{\raggedright\arraybackslash}p{3.4cm}>{\raggedright\arraybackslash}p{13.0cm}@{}}
\toprule
Column & Content \\
\midrule
\textbf{Value / Distribution} & Scalar, tuple, or distribution (e.g. $U(M_a,M_b)$) \\
\textbf{Data source} & \texttt{d1}, \texttt{d2}, \ldots{} (ref T5) \textbf{or} \texttt{Author} \\
\textbf{Derivation} & \texttt{Direct} \textbar{} \texttt{Estimated} \textbar{} \texttt{Computed} \textbar{} \texttt{Assumed} (definitions below) \\
\textbf{Algorithm} & Named procedure (e.g. \texttt{maximum likelihood estimation}) or paper-defined algorithm (e.g. \texttt{Alg.~1}); \texttt{---} if Derivation is Direct/Assumed \\
\textbf{Ref.} & \texttt{\textbackslash{}citep\{...\}} motivating the Algorithm, or \texttt{---} \\
\bottomrule
\end{longtable}
\egroup

\textbf{Derivation --- controlled vocabulary:} - \textbf{Direct} --- value taken verbatim
from the data source, no processing. - \textbf{Estimated} --- value or distribution
parameters \textbf{statistically inferred} from data (fitting, mean/variance estimation).
- \textbf{Computed} --- value produced by a \textbf{deterministic or stochastic procedure}
on known quantities (formula, randomized construction), without statistical inference. -
\textbf{Assumed} --- fixed by model design / expert judgment / literature; not from the
data source.

Group rows by agent type (matching T1/T2 order).

\subparagraph{(b) Output --- every output indicator}

\bgroup
\setlength{\tabcolsep}{4pt}
\renewcommand{\arraystretch}{1.1}
\begin{longtable}{@{}>{\raggedright\arraybackslash}p{3.4cm}>{\raggedright\arraybackslash}p{13.0cm}@{}}
\toprule
Column & Content \\
\midrule
\textbf{Indicator} & Descriptive name \\
\textbf{Formula} & Computation from T2 variables \\
\textbf{Data Type} & Reuses T2 type vocabulary (\texttt{Integer}, \texttt{Float}, \texttt{List[Integer]}, \ldots{}) \\
\textbf{Unit} & Unit or \texttt{--} \\
\textbf{Frequency} & Integer $k\geq1$ (sample every $k$ steps; $k{=}1$ = every step) \textbf{or} \texttt{-1} (terminal step only) \\
\textbf{Desc.} & Short description \\
\bottomrule
\end{longtable}
\egroup

\begin{center}\rule{0.5\linewidth}{0.5pt}\end{center}

\paragraph{T7 --- Schedule}

Two sub-tables.

\subparagraph{(a) Execution schedule --- the per-step loop}

\bgroup
\setlength{\tabcolsep}{4pt}
\renewcommand{\arraystretch}{1.1}
\begin{longtable}{@{}>{\raggedright\arraybackslash}p{3.4cm}>{\raggedright\arraybackslash}p{13.0cm}@{}}
\toprule
Column & Content \\
\midrule
\textbf{Agent} & Agent-type Set acting \\
\textbf{ID} & Function ID from T4 (\texttt{f1}, \ldots{}) \\
\textbf{Function} & Function name (mirrors T4) \\
\textbf{Exec. mode} & One of four base modes (+ optional inline parameters); see below \\
\textbf{Condition} & Trigger text (e.g. \texttt{Every step}, \texttt{Only when a new vendor is added}) \\
\bottomrule
\end{longtable}
\egroup

\begin{quote}
\textbf{Exec. mode --- controlled vocabulary (four base modes):} - \textbf{Synchronous} ---
all instances observe start-of-step state, update simultaneously (lockstep). -
\textbf{Sequential} --- instances act one at a time in a \textbf{fixed deterministic}
order; later instances see earlier updates. - \textbf{Random-order} --- as Sequential, but
order uniformly randomized and reshuffled each step. - \textbf{Asynchronous} --- not
synchronized to the global step; event/time-driven activation.

Mode-specific parameters are appended in parentheses,
e.g.~\texttt{Sequential\ (by\ \$w\_i\$,\ descending)},
\texttt{Asynchronous\ (Poisson,\ \$\textbackslash{}lambda\$)}. \textbf{Conditional}
activation is \emph{not} a mode --- it goes in the \textbf{Condition} column.
\textbf{Staged} execution (sense\ensuremath{\rightarrow}decide\ensuremath{\rightarrow}update sub-phases) is a \emph{composite} built
from the four modes, used only to rule out within-step information contamination.
\end{quote}

\begin{quote}
\textbf{r4 constraint:} all functions sharing a Step \textbf{must} share the same Exec.
mode.
\end{quote}

\subparagraph{(b) Termination conditions}

\bgroup
\setlength{\tabcolsep}{4pt}
\renewcommand{\arraystretch}{1.1}
\begin{longtable}{@{}>{\raggedright\arraybackslash}p{3.4cm}>{\raggedright\arraybackslash}p{13.0cm}@{}}
\toprule
Column & Content \\
\midrule
\textbf{Indicator} & A T2 variable \textbf{or} T6b output indicator (r17) \\
\textbf{Condition} & Predicate on the indicator, e.g. $t\geq1000$ \\
\textbf{Description} & Plain-language meaning \\
\textbf{Value Source/Ref} & Where the threshold comes from (e.g. \texttt{Author Assumed}) \\
\textbf{Termination logic} & \texttt{multirow} combining all rows, e.g. \texttt{Stop when $c1 \textbackslash{}lor c2$} \\
\bottomrule
\end{longtable}
\egroup

Initial values are \textbf{not} part of this loop --- they are exogenous inputs via T6a.

\begin{center}\rule{0.5\linewidth}{0.5pt}\end{center}

\paragraph{T8 --- Validation}

\emph{Credibility framework. Instantiates the HAV (Hierarchical ABM Validation) framework:
three levels, data-grounded vs non-data-grounded methods.}

\bgroup
\setlength{\tabcolsep}{4pt}
\renewcommand{\arraystretch}{1.1}
\begin{longtable}{@{}>{\raggedright\arraybackslash}p{3.4cm}>{\raggedright\arraybackslash}p{13.0cm}@{}}
\toprule
Column & Content \\
\midrule
\textbf{Validation object} & What is being validated (an agent behavior, a structural mechanism, an aggregate output) \\
\textbf{Benchmark data} & Empirical reference (ref a d-ID from T5 where applicable) \textbf{or} a theoretical expectation \\
\textbf{Method} & Named statistical/theoretical method, e.g. \texttt{K-S test}, \texttt{Moran's I}, \texttt{Regression} \\
\textbf{Indicator} & Test statistic computed \\
\textbf{Passing cond.} & Concrete accept criterion, e.g. \texttt{$p>0.05$} \\
\textbf{Ref} & \texttt{\textbackslash{}citep\{...\}} for prior work using the same method, or \texttt{---} \\
\bottomrule
\end{longtable}
\egroup

Group rows into three multicolumn level groups (matching HAV hierarchy): -
\textbf{Agent level} --- individual agent behavior. - \textbf{Model level} --- structural
mechanisms. - \textbf{Output level} --- aggregate results.

\begin{quote}
Validation objects must be expressed as T2 variables or T6b outputs (r18); the Indicator
column holds the test statistic, which is computed during validation and is not itself a
model variable. Empirical benchmarks must resolve to a T5 entry (r19). Modelers are
encouraged to add rows for more comprehensive validation.
\end{quote}

\begin{center}\rule{0.5\linewidth}{0.5pt}\end{center}

\subsubsection{Output format}

\begin{itemize}
\tightlist
\item
  Emit each table as LaTeX (\texttt{tabular}/\texttt{tabularx}/\texttt{tabular*})
  consistent with the manuscript style, OR as structured Markdown if the user wants a
  draft.
\item
  Preserve all IDs and cross-references (T2 Value \ensuremath{\rightarrow} T4 f-IDs; T6a Data source \ensuremath{\rightarrow} T5 d-IDs;
  T7a ID \ensuremath{\rightarrow} T4; T7b/T8 indicators \ensuremath{\rightarrow} T2/T6b).
\item
  After emitting, \textbf{recommend running `visa-check`} and list which rules are most
  likely at risk given what was just authored.
\end{itemize}

\subsubsection{Common authoring pitfalls}

\begin{enumerate}
\def\labelenumi{\arabic{enumi}.}
\tightlist
\item
  Forgetting to give a Passive agent's variables in T2 --- they may have
  \textbf{exogenous} variables but \textbf{no endogenous} ones (r6).
\item
  Listing an exogenous variable in T4 Self-state Update (violates r12).
\item
  A T4 function reading a variable the agent cannot observe in T3 (violates r14).
\item
  An endogenous variable in T2 whose Value column references no function, or references a
  non-existent f-ID (violates r11 --- orphaned or dangling).
\item
  A function in T4 that never appears in T7a, or a T7a row citing a function not in T4
  (violates r16 --- unscheduled or phantom).
\item
  A Decision-maker agent in T1 with no function in T4 (violates r9).
\item
  A T6a symbol with no matching exogenous variable in T2, or a T2 ``Input'' variable with
  no T6a entry (violates r15).
\item
  A function that reads inputs but modifies no variable (violates r3 --- unproductive).
\item
  An agent marked with variable quantity (\(n\)) but no create/remove function, or a
  fixed-quantity (\(N\)) agent with one (violates r8).
\item
  An empirical data reference in T6a or T8 that points to no T5 record (violates r19).
\end{enumerate}
}

{\small 
\subsection{VISA Checking Skill}

\begin{quote}
\textbf{Purpose.} Verify that a set of eight VISA tables satisfies the
\textbf{19 consistency rules}. For each rule, report \textbf{pass/fail}, the
\textbf{specific cells} implicated in any violation, and a
\textbf{concrete suggested fix}. Turns the rules from a manual checklist into an
executable validator.
\end{quote}

\subsubsection{When to invoke}

Invoke this skill when the user asks to
\textbf{validate / check / verify a VISA specification} (complete or partial), or right
after \texttt{visa-author} finishes a table.

\subsubsection{Inputs}

The eight tables (T1--T8) in any readable form (LaTeX, Markdown, or structured text).
Partial specifications are allowed --- rules whose required tables are absent are reported
as \textbf{BLOCKED (pending input)} rather than failed.

\subsubsection{Output format}

\begin{verbatim}
VISA CONSISTENCY REPORT
=======================
Within-table rules
  r1  [PASS | FAIL | BLOCKED]  <one-line summary>
       (if FAIL) cells: ... | fix: ...
  ...
Cross-table rules
  r5  ...
  ...
Summary: X/19 PASS, Y FAIL, Z BLOCKED.
\end{verbatim}

For each \textbf{FAIL}, name the table, the row/cell, and a one-sentence fix.

\begin{center}\rule{0.5\linewidth}{0.5pt}\end{center}

\subsubsection{Within-table rules}

\paragraph{r1 --- Variable-type coverage and time-indexing (T2)}

\textbf{Inspect:} every row of T2. \textbf{Pass:} each variable's \textbf{Type} is one of
the four leaf types \texttt{Exog.-homo.} \textbar{} \texttt{Exog.-hetero.} \textbar{}
\texttt{Endog.-dec.} \textbar{} \texttt{Endog.} (non-decision). No blank / top-only /
free-text Type. In addition, every \textbf{endogenous} variable's \textbf{Symbol} carries
the time index \texttt{\$t\$} (e.g.~\texttt{\$x\_\{i,t\}\$}, \texttt{\$p\_\{i,t\}\$}),
since it evolves over the simulation; exogenous variables do not. (The index may be
suppressed in T4 cells, equations, and prose for readability, but must appear on each
endogenous symbol in T2.) \textbf{Fix:} assign the correct leaf type. Recall the two-level
test: (1) can an internal function write this variable? No \ensuremath{\rightarrow} Exogenous; Yes \ensuremath{\rightarrow} Endogenous.
(2a) Exogenous: same for all instances \ensuremath{\rightarrow} homo, else hetero. (2b) Endogenous: directly set
by a decision \ensuremath{\rightarrow} dec, else non-decision. Then append the time index \texttt{\$t\$} to the
Symbol of every endogenous variable that lacks it.

\paragraph{r2 --- Function-ID uniqueness (T4)}

\textbf{Inspect:} T4 \textbf{ID} column. \textbf{Pass:} all f-IDs in T4 are unique.
\textbf{Fix:} remove/renumber duplicates.

\paragraph{r3 --- Function productivity (T4)}

\textbf{Inspect:} every T4 function's \textbf{Self-state Update} and
\textbf{External Effect} columns. \textbf{Pass:} each function lists at least one entry
across the two columns --- it modifies \ensuremath{\geq}1 endogenous variable (own or another's) or
creates/removes an instance. A function that reads inputs but writes nothing fails.
\textbf{Fix:} add the missing write, or remove the function if it is redundant.

\paragraph{r4 --- Step--Exec. mode consistency (T7a)}

\textbf{Inspect:} T7a rows grouped by \textbf{Step} number. \textbf{Pass:} all rows
sharing a Step number have the \textbf{same Exec. mode}; functions that must execute
simultaneously share a Step. \textbf{Fix:} either move the mismatched function to its own
Step, or align the Exec. mode.

\begin{center}\rule{0.5\linewidth}{0.5pt}\end{center}

\subsubsection{Cross-table rules}

\paragraph{r5 --- Same-type (peer) sensing (T1 \ensuremath{\rightarrow} T3)}

\textbf{Inspect:} T3 \textbf{diagonal} cells, cross-referenced with T1 Quantity.
\textbf{Pass:} every agent type with variable quantity \(n{>}1\) records on the diagonal
the attributes its instances observe of one another --- the \textbf{peer set}
(\texttt{\$\textbackslash{}emptyset\$} if they observe nothing of each other); a
single-instance type (\(N{=}1\)) has no peers and its diagonal is
\texttt{\$\textbackslash{}emptyset\$}. Self-observation is implicit (every agent reads its
own full state) and is never recorded in a T3 cell. \textbf{Fix:} for each multi-instance
type, state the peer-observed attributes on the diagonal (or
\texttt{\$\textbackslash{}emptyset\$}); for single-instance types, set the diagonal to
\texttt{\$\textbackslash{}emptyset\$}.

\paragraph{r6 --- Passive agent implications (T1 \ensuremath{\rightarrow} T2, T3, T4)}

\textbf{Inspect:} every agent type whose T1 \textbf{Category} is \texttt{Passive}.
\textbf{Pass:} for each Passive type --- (i) it has \textbf{no endogenous} variables in
T2; (ii) it appears \textbf{only as a column} (right of the \texttt{\textbar{}}) in T3,
never as a row; (iii) it has \textbf{no functions} in T4. \textbf{Fix:} if a Passive agent
has endogenous vars/functions, reclassify it as Decision-maker in T1, or remove the
offending entries.

\paragraph{r7 --- Observer row completeness (T1 \ensuremath{\rightarrow} T3)}

\textbf{Inspect:} every \textbf{non-Passive} agent type in T1 vs.~T3 rows. \textbf{Pass:}
each non-Passive type has \textbf{exactly one} row in T3; Passive types have \textbf{no}
rows. \textbf{Fix:} add the missing row, or remove a spurious Passive row.

\paragraph{r8 --- Population-dynamics consistency (T1 \ensuremath{\rightarrow} T2, T4)}

\textbf{Inspect:} each T1 agent type's \textbf{Quantity} (\(N\) vs \(n\)), its T2 count
variable, and T4 functions. \textbf{Pass:} a type whose Quantity is variable (\(n\)) is
created or removed by \ensuremath{\geq}1 T4 function (its count variable is endogenous, written by that
function); a type whose Quantity is fixed (\(N\)) has no function that creates or removes
its instances. \textbf{Fix:} add the missing entry/exit function, or correct the Quantity
to \(N\) if the population is in fact fixed.

\paragraph{r9 --- Active-agent function coverage (T1 \ensuremath{\rightarrow} T4)}

\textbf{Inspect:} every \textbf{non-Passive} agent type in T1 vs.~T4 ownership.
\textbf{Pass:} each active agent type owns \textbf{\ensuremath{\geq}1} internal function in T4 (and
therefore appears in T7a). \textbf{Fix:} add a function for the agent, or reconsider
whether the agent is genuinely active (a static topology with no behavior may be Passive).

\paragraph{r10 --- Variable observability (T2 \ensuremath{\rightarrow} T3)}

\textbf{Inspect:} every variable in T2; check whether it appears in at least one T3 cell
(as itself, or covered by \texttt{*}). \textbf{Pass:} each variable is observable in \ensuremath{\geq}1 T3
cell, \textbf{except} variables never accessed by any function (verify against T4 Decision
Basis). Endogenous-only/internal bookkeeping variables that no function reads from another
agent are exempt. \textbf{Fix:} add the variable to the appropriate T3 cell, or confirm it
is genuinely never sensed.

\paragraph{r11 --- Endogenous variable completeness (T2 \ensuremath{\rightarrow} T4)}

\textbf{Inspect:} every \textbf{endogenous} variable in T2; check T2 \textbf{Value} column
references \ensuremath{\geq}1 function. \textbf{Pass:} each endogenous variable is updated by
\textbf{at least one} T4 function whose f-ID is listed in the Value column
\textbf{and exists in T4} (no dangling references). A variable may be updated by multiple
functions. \textbf{Fix:} either assign the responsible f-ID(s) in T2 and ensure they exist
in T4, or add the missing function in T4.

\paragraph{r12 --- Self-state update validity (T2 \ensuremath{\rightarrow} T4)}

\textbf{Inspect:} every T4 \textbf{Self-state Update} entry. \textbf{Pass:} all variables
listed are \textbf{endogenous} variables \textbf{of that same agent} (per T2).
\textbf{Fix:} remove exogenous variables from Self-state Update; if a function truly needs
to write an exogenous value, reclassify the variable as endogenous.

\paragraph{r13 --- External effect validity (T2 \ensuremath{\rightarrow} T4)}

\textbf{Inspect:} every T4 \textbf{External Effect} entry. \textbf{Pass:} each entry is
either (i) an \textbf{endogenous} variable \textbf{of another agent} (per T2), or (ii) an
instance creation/removal of a variable-quantity type (per r8). Modifying another agent's
\textbf{exogenous} variable is forbidden. \textbf{Fix:} remove the offending external
write, reclassify the target as endogenous, or re-express a forbidden write as a
legitimate instance operation.

\paragraph{r14 --- Information access validation (T3 \ensuremath{\rightarrow} T4)}

\textbf{Inspect:} every T4 function's \textbf{Decision Basis} variables. \textbf{Pass:}
each Decision Basis variable is either (i) an attribute of the \textbf{agent itself}, or
(ii) a variable the agent is \textbf{authorized to observe in T3} (in the cell for that
observer/observed pair). \textbf{Fix:} remove the unobservable variable from Decision
Basis, or grant access in the relevant T3 cell.

\paragraph{r15 --- Input--output coverage (T6 \ensuremath{\leftrightarrow} T2)}

\textbf{Inspect:} T6a Input symbols vs.~T2 exogenous variables marked \texttt{Input}; T6b
Output formulas vs.~T2 variables. \textbf{Pass:} (a) every T6a \textbf{Input} symbol
corresponds to an \textbf{exogenous} variable in T2; (b) every T2 variable marked
\texttt{Input} has a \textbf{matching T6a entry} (no exogenous variable left without a
value --- an unsupplied input is a reproduction blocker); (c) every T6b \textbf{Output}
indicator is computable from T2 variables. \textbf{Fix:} add the missing T2 variable or
T6a entry, or remove the orphaned entry; ensure input symbols point only to exogenous
vars.

\paragraph{r16 --- Schedule coverage (T7 \ensuremath{\leftrightarrow} T4)}

\textbf{Inspect:} every T4 function ID vs.~T7a \textbf{ID} column, and vice versa.
\textbf{Pass:} every function in T4 appears \textbf{at least once} in T7a, \textbf{and}
every function ID in T7a exists in T4. A function may be scheduled multiple times; none
may be unscheduled or phantom. \textbf{Fix:} add the missing function to T7a, or
remove/correct a phantom T7a row.

\paragraph{r17 --- Termination indicator source (T7b \ensuremath{\rightarrow} T2, T6)}

\textbf{Inspect:} every indicator in T7b's \textbf{Indicator} column. \textbf{Pass:} each
is either a \textbf{T2 variable} or a \textbf{T6b output indicator}. The global time index
\texttt{\$t\$} is always available and need not be declared. \textbf{Fix:} add the
indicator to T2 or T6b, or correct the T7b reference.

\paragraph{r18 --- Validation-object coverage (T8 \ensuremath{\rightarrow} T2, T6)}

\textbf{Inspect:} every T8 \textbf{Validation object} (the quantity submitted to
validation). \textbf{Pass:} each validation object is expressed in terms of
\textbf{T2 variables} or \textbf{T6b output indicators}. The T8 \textbf{Indicator} column
holds the test statistic, which is computed during validation and is \textbf{not} itself a
model variable (so it is not checked here). \textbf{Fix:} re-express the validation object
using T2/T6b quantities, or add the missing variable/output.

\paragraph{r19 --- Data-reference resolution (T8, T6a \ensuremath{\rightarrow} T5)}

\textbf{Inspect:} every empirical data reference in T8 (\textbf{Benchmark data}) and in
T6a (\textbf{Data source}). \textbf{Pass:} each referenced d-ID has a corresponding
\textbf{T5} entry. References marked \texttt{Author} (T6a) and purely \textbf{theoretical}
benchmarks (T8) are exempt. \textbf{Fix:} add the data record to T5, or change the
reference to \texttt{Author}/theoretical.

\begin{center}\rule{0.5\linewidth}{0.5pt}\end{center}

\subsubsection{Checking heuristics}

\begin{itemize}
\tightlist
\item
  \textbf{Build a symbol index first.} From T2 collect every (Symbol, agent, Type). From
  T4 collect every (f-ID, owner agent). From T5 collect every d-ID. From T6b collect every
  output symbol. These indices make r2, r11--r15, r17--r19 mechanical lookups.
\item
  \textbf{Check dependency order.} If T4 is missing, r2/r3/r11/r12/r13/r14/r16 are
  \textbf{BLOCKED}. Report BLOCKED, do not guess.
\item
  \textbf{Be specific.} Always name the exact cell (e.g., ``T2 row
  \texttt{Vendor\ \ensuremath{\rightarrow}\ q\_i}, Value column'') so the modeler can fix it directly.
\item
  \textbf{Suggest, don't rewrite.} Propose the minimal fix; let the modeler accept it.
\end{itemize}

\subsubsection{\texorpdfstring{Relationship to
\texttt{visa-author}}{Relationship to visa-author}}

Run this skill after each table is drafted. The 19 rules partition naturally by the table
they primarily concern, so a partial check after, say, T4 can already exercise r2, r3,
r11--r13. Iterate until the summary reads \textbf{19/19 PASS}.
}

{\small 
\subsection{VISA Code-Generation Skill}

\begin{quote}
\textbf{Purpose.} Convert a complete, rule-passing set of eight VISA tables (Markdown, as
produced by \texttt{visa-author}) into a runnable agent-based simulation in
\textbf{Python} or \textbf{Java}. This skill closes the loop from specification to
executable implementation.
\end{quote}

\subsubsection{When to invoke}

Invoke when a reproducer has the eight VISA tables (md) and wants to
\textbf{generate / scaffold / implement} the simulation code. Best run after
\texttt{visa-check} reports 19/19 PASS; on partial specs it emits explicit \texttt{TODO}
stubs for the gaps.

\subsubsection{Inputs}

\begin{itemize}
\tightlist
\item
  The eight VISA tables in Markdown (T1--T8).
\item
  Target language: \textbf{Python} (default) or \textbf{Java}.
\item
  Optional framework: Python \ensuremath{\rightarrow} plain or \href{https://mesa.readthedocs.io}{Mesa}; Java \ensuremath{\rightarrow}
  plain or Repast Simphony.
\item
  Optional random seed.
\end{itemize}

\subsubsection{Companion skills}

This is the third VISA skill, complementing \texttt{visa-author} (build tables) and
\texttt{visa-check} (verify rules). Together: \textbf{author \ensuremath{\rightarrow} check \ensuremath{\rightarrow} code}.

\subsubsection{Output: project layout}

\begin{verbatim}
project/
|-- README.md            # how to run + VISA->code map
|-- config.*             # T6a parameters (+ Derivation metadata)
|-- data/                # T5 datasets (or download stubs)
|   `-- loaders.*        # one loader per T5 record
|-- agents/              # T1 + T2 + T4
|   |-- environment.*
|   |-- space.*
|   |-- vendor.*
|   `-- ...
|-- sensing.*            # T3 access-control layer
|-- model.*              # T7 schedule + T7b termination
|-- collectors.*         # T6b outputs (Frequency-aware)
|-- validation.*         # T8 post-run analyses
`-- main.*               # entry point
\end{verbatim}

\subsubsection{Architecture mapping}

\bgroup
\setlength{\tabcolsep}{4pt}
\renewcommand{\arraystretch}{1.1}
\begin{longtable}{@{}>{\raggedright\arraybackslash}p{3.4cm}>{\raggedright\arraybackslash}p{13.0cm}@{}}
\toprule
VISA element & Code construct \\
\midrule
T1 \textbf{Category} & \texttt{Environment}/\texttt{Space} \ensuremath{\rightarrow} infrastructure base; \texttt{Decision-maker} \ensuremath{\rightarrow} active base with \texttt{step}; \texttt{Passive} \ensuremath{\rightarrow} data-only class (no behavior) \\
T1 \textbf{Quantity} $N$ / $n$ & fixed list vs. dynamic collection supporting add/remove \\
T2 \textbf{variable} & instance attribute / field \\
T2 \textbf{Data Type} & language type (see table below) \\
T2 \textbf{exogenous} & constructor parameter (value supplied from T6a) \\
T2 \textbf{endogenous} & computed attribute, written by the owning T4 function \\
T3 \textbf{sensing} & getter methods + an access-control layer enforcing the matrix \\
T4 \textbf{function} (f-ID) & one \textbf{method} on the owning agent class \\
T4 \textbf{Decision Basis} & the variables the method reads (self attributes + sensed) \\
T4 \textbf{Self-state Update} & \texttt{self.x = \ldots{}} writes inside the method \\
T4 \textbf{External Effect} & calls to \textbf{other} agents' setters \\
T4 \textbf{Method} / \textbf{Ref} & the algorithm body; cite Ref in a comment \\
T5 \textbf{data record} & a loader (CSV/JSON) \\
T6a \textbf{input} & config / params object \\
T6b \textbf{output} & collector, sampled at the recorded \textbf{Frequency} \\
T7a \textbf{schedule} & \texttt{Model.step()}: ordered function calls per step \\
T7a \textbf{Exec.mode} & see patterns below \\
T7b \textbf{termination} & \texttt{while}-loop condition re-checked each step \\
T8 \textbf{validation} & post-run analysis script \\
\bottomrule
\end{longtable}
\egroup

\paragraph{Data Type \ensuremath{\rightarrow} language type}

\bgroup
\setlength{\tabcolsep}{4pt}
\renewcommand{\arraystretch}{1.1}
\begin{longtable}{@{}>{\raggedright\arraybackslash}p{3.5cm}>{\raggedright\arraybackslash}p{6.3cm}>{\raggedright\arraybackslash}p{6.5cm}@{}}
\toprule
VISA Data Type & Python & Java \\
\midrule
Float & \texttt{float} & \texttt{double} \\
Boolean & \texttt{bool} & \texttt{boolean} \\
List[Integer] & \texttt{list[int]} & \texttt{ArrayList<Integer>} \\
Matrix[Integer] & \texttt{numpy.ndarray} (2-D, int) & \texttt{int[][]} \\
List[Float] & \texttt{list[float]} & \texttt{ArrayList<Double>} \\
\emph{(extend analogously)} & \ldots{} & \ldots{} \\
\bottomrule
\end{longtable}
\egroup

\subsubsection{Exec.mode implementation patterns (critical for reproducibility)}

The execution mode is the single biggest source of irreproducibility. Generate the pattern
\textbf{exactly} as T7a records it.

\begin{itemize}
\tightlist
\item
  \textbf{Synchronous} --- snapshot then swap. Each instance reads all Decision-Basis
  variables into locals, computes its update, and only \textbf{after all} instances have
  computed do they write back simultaneously.
\item
  \textbf{Sequential (order\ldots{})} --- process instances one at a time in the stated fixed
  order (e.g., by ID; by \(w_i\) descending); each write is immediately visible to later
  instances.
\item
  \textbf{Random-order} --- as Sequential, but \texttt{shuffle} the instance list with the
  seeded RNG at the start of every step.
\item
  \textbf{Asynchronous (dist\ldots{})} --- event-driven: each instance draws a next-activation
  time (e.g., Poisson(\(\lambda\))); a priority queue fires activations; the global step
  is bookkeeping.
\end{itemize}

\begin{quote}
Rule \textbf{r4} (same Step \ensuremath{\Rightarrow} same Exec.mode) means a whole Step uses one pattern ---
generate one block per Step.
\end{quote}

\paragraph{Illustrative skeleton (Python, one Step)}

\begin{verbatim}
# Exec.mode = Synchronous  -> snapshot-then-swap
olds = [v.read_basis() for v in vendors]          # snapshot
news = [update_from(o) for o in olds]             # compute, no writes yet
for v, n in zip(vendors, news):                   # swap simultaneously
    v.commit(n)

# Exec.mode = Random-order  -> shuffle, then in place
rng.shuffle(vendors)
for v in vendors:
    v.update_in_place()                            # writes visible to later vendors
\end{verbatim}

\subsubsection{Generation pipeline (ordered)}

\begin{enumerate}
\def\labelenumi{\arabic{enumi}.}
\tightlist
\item
  \textbf{Parse} the eight md tables into a structured representation (agent types,
  variables, functions, schedule, \ldots).
\item
  \textbf{Config} --- emit \texttt{config} from T6a: every exogenous parameter, its
  value/distribution, Derivation, and the loader it needs.
\item
  \textbf{Data loaders} --- for each T5 record, emit a loader honoring
  \textbf{Availability}: \texttt{Open} \ensuremath{\rightarrow} direct path/URL;
  \texttt{Restricted}/\texttt{Private} \ensuremath{\rightarrow} stub that raises a clear ``data unavailable''
  error (see Section 4.3, data bottleneck).
\item
  \textbf{Agent classes} --- for each T1 type: attributes from T2 (typed per the table
  above), a constructor taking the exogenous parameters, and one method per owned T4
  function. Translate each function's \textbf{Method}:

  \begin{itemize}
  \tightlist
  \item
    mechanical (Mean, Arithmetic, Append, Comparison, Transfer, Conditional, Optimization)
    \ensuremath{\rightarrow} direct implementation;
  \item
    literature-grounded (Multinomial logit, Preferential attachment, newsvendor fractile,
    \ldots) \ensuremath{\rightarrow} translate the paper's equation, cite the \textbf{Ref} in a comment;
  \item
    under-specified \ensuremath{\rightarrow} emit a clearly marked \texttt{TODO(ref=\ldots{})} so the reproducer
    confirms the exact form.
  \end{itemize}
\item
  \textbf{Sensing layer} --- emit getters; add an access-control wrapper enforcing T3
  (default: open getters with a comment documenting where T3 restricts access, plus an
  optional strict mode).
\item
  \textbf{Model} --- emit agent collections, a \texttt{step()} executing T7a Steps in
  order (with the Exec.mode pattern per Step), and the termination loop from T7b.
\item
  \textbf{Collectors} --- for each T6b output, emit a recorder honoring \textbf{Frequency}
  (\(k \geq 1\) every \(k\) steps; \(-1\) terminal only).
\item
  \textbf{Validation} --- for each T8 row, emit the post-run computation + statistical
  test (K-S, Moran's \(I\), regression, \ldots) + passing-condition check against the
  benchmark.
\item
  \textbf{Main + README} --- wire everything; the README documents how to run, the RNG
  seed, and a table mapping every generated file back to its VISA source.
\end{enumerate}

\subsubsection{Reproducibility safeguards (emit by default)}

\begin{itemize}
\tightlist
\item
  A \textbf{single seeded RNG} threaded through every stochastic draw (T6a distributions,
  Random-order shuffles, Asynchronous activation, stochastic methods).
\item
  Every T4 method body cites its \textbf{Ref} in a comment.
\item
  A \texttt{-\/-seed} CLI flag and a recorded seed in the run log.
\item
  The T3 sensing matrix reproduced as a docstring/comment on the sensing layer, so a
  reproducer can audit information flow.
\end{itemize}

\subsubsection{Common pitfalls}

\begin{enumerate}
\def\labelenumi{\arabic{enumi}.}
\tightlist
\item
  \textbf{Wrong Exec.mode} --- silently changes results; verify each Step against T7a.
\item
  \textbf{Implicit synchronous assumption} --- if T7a says Synchronous but the code
  updates in place, results differ; always snapshot-then-swap for Synchronous.
\item
  \textbf{Unseeded randomness} --- without a threaded seed, re-runs diverge.
\item
  \textbf{Open access vs. T3} --- generating public getters for everything violates the
  model's rationality bounds; at minimum document the T3 restrictions.
\item
  \textbf{Restricted/Private data (T5)} --- the loader cannot fetch these; emit a clear
  stub so the reproducer knows reproduction is blocked at data acquisition.
\item
  \textbf{Passive agents with methods} --- r6 forbids this; if it appears, the spec failed
  \texttt{visa-check} and should be corrected before code generation.
\end{enumerate}
}

\end{document}